\documentclass[prd,nofootinbib,twocolumn,amsmath,amssymb,eqsecnum,longbibliography,superscriptaddress]{revtex4-1}
\usepackage[utf8]{inputenc}
\usepackage{mathtools}
\usepackage{graphics}
\usepackage{graphicx}
\usepackage{dcolumn}
\usepackage{bm}
\usepackage{dsfont} 
\usepackage{amsmath,amssymb}
\usepackage{hyperref}
\usepackage{tabularx}
\usepackage{epstopdf}
\usepackage{tensor}
\usepackage{mathrsfs}
\usepackage[normalem]{ulem}
\usepackage[usenames]{color}
\hypersetup{
    colorlinks=true,
    linkcolor=blue,
    filecolor=magenta,      
    urlcolor=blue,
    citecolor=blue
}
\usepackage[dvipsnames]{xcolor}
\usepackage{mathrsfs}

\newcommand\be{\begin{equation}}
\newcommand\ba{\begin{eqnarray}}
\newcommand\ee{\end{equation}}
\newcommand\ea{\end{eqnarray}}
\newcommand\bw{\begin{widetext}}
\newcommand\ew{\end{widetext}}

\newcommand{\calA}{\mathcal{A}}

\newcommand{\UVA}{Department of Physics, University of Virginia, P.O.~Box 400714, Charlottesville, VA 22904-4714, USA}

\AtBeginDocument{%
    \newwrite\bibnotes
    \def\bibnotesext{Notes.bib}
    \immediate\openout\bibnotes=\jobname\bibnotesext
    \immediate\write\bibnotes{@CONTROL{REVTEX41Control}}
    \immediate\write\bibnotes{@CONTROL{%
    apsrev41Control,author="08",editor="1",pages="1",title="0",year="1"}}
    \if@filesw
    \immediate\write\@auxout{\string\citation{apsrev41Control}}%
    \fi
}

\begin{document}

\title{Regularizing Parameterized Black Hole Spacetimes with Kerr Symmetries}

 \author{Kent Yagi}
 \affiliation{\UVA}

 \author{Samantha Lomuscio}
 \affiliation{\UVA}
 \affiliation{Applied Physics Laboratory, Johns Hopkins University, Laurel, Maryland 20723-6099, USA}

 \author{Tristen Lowrey}
 \affiliation{\UVA}

 \author{Zack Carson}
 \affiliation{\UVA}

\date{\today}

\begin{abstract}

Parameterized Kerr spacetimes allow us to test the nature of black holes in model-independent ways. Such spacetimes contain several arbitrary functions and, as a matter of practicality, one Taylor expands them about infinity and keeps only to finite orders in the expansion. In this paper, we focus on the parameterized spacetime preserving Killing symmetries of a Kerr spacetime and show that an unphysical divergence may appear in the metric if such a truncation is performed in the series expansion. To remedy this, we redefine the arbitrary functions so that the divergence disappears, at least for several known black hole solutions that can be mapped to the parameterized Kerr spacetime. We propose two restricted classes of the refined parameterized Kerr spacetime that only contain one or two arbitrary functions and yet can reproduce exactly all the example black hole spacetimes considered in this paper. The Petrov class of the parameterized Kerr spacetime is of type I while that for the restricted class with one arbitrary function remains type D. We also compute the ringdown frequencies and the shapes of black hole shadows for the parameterized spacetime and show how they deviate from Kerr. The refined black hole metrics with Kerr symmetries presented here are practically more useful than those proposed in previous literature.  

\end{abstract}

\maketitle

\section{Introduction}

Today, General Relativity (GR) has been extensively tested and has passed every test with flying colors. GR has been studied with e.g. solar system experiments~\cite{Will:2014kxa, TEGP} and binary pulsar observations~\cite{stairs, Wex:2014nva} that probe gravity in the weak and/or non-dynamical field regime. Recent observations of gravitational waves~\cite{TheLIGOScientific:2016src,Yunes:2016jcc,Berti:2018cxi,Berti:2018vdi,LIGOScientific:2019fpa,LIGOScientific:2020tif,LIGOScientific:2021sio}, and black hole (BH) shadows~\cite{Psaltis:2020lvx, Psaltis:2020ctj,EventHorizonTelescope:2022xqj,Vagnozzi:2022moj} can test gravity in the strong and/or dynamical field regime.  

An important consequence of GR is the BH no-hair theorem~\cite{israel,hawking-uniqueness}. This theorem states that isolated, stationary, uncharged BHs that are regular outside the event horizon are uniquely characterized by the Kerr metric. Once we go beyond GR, however, BH solutions differ from Kerr in general. Moreover, even within GR, there are many BH mimickers known, such as boson stars and gravastars (see e.g.~\cite{Cardoso:2019rvt} for a review), that are compact exotic objects.
Properties of BHs and the no-hair theorem have been tested through BH shadows \cite{Psaltis:2020lvx, Psaltis:2020ctj,EventHorizonTelescope:2022xqj,Vagnozzi:2022moj}, orbits of supermassive BH stellar companions \cite{Will_2008, Merritt:2009ex, Sadeghian:2011ub}, BH observations through x-rays~\cite{Kong:2014wha,Bambi:2014sfa,Jiang:2015dla,Jiang:2016bdj,Xu:2018lom}, and gravitational waves from binary BH mergers through inspiral~\cite{Krishnendu:2017shb,LIGOScientific:2020tif,LIGOScientific:2021sio} and ringdown~\cite{Isi:2019aib, Capano:2021etf,Cotesta:2022pci,Finch:2022ynt,Ma:2023vvr,Wang:2023xsy}.

An efficient way to test the nature of BHs is to use a parameterically-deformed spacetime that can capture deviations from Kerr in a theory-agnostic way~
\cite{Collins:2004ex,vigelandhughes,Vigeland:2010xe,vigelandnico,Glampedakis:2005cf,johannsen-metric,Johannsen:2015pca,Carson:2020dez,Papadopoulos:2018nvd,Papadopoulos:2020kxu,Chen:2020aix,Rezzolla:2014mua,Konoplya:2016jvv,Konoplya:2018arm,Konoplya:2020hyk,Junior:2020lya,Delaporte:2022acp,Baines:2023dhq}. For example, the first non-Ricci-flat parameterized Kerr spacetime was constructed by Vigeland, Yunes and Stein~\cite{vigelandnico} 
where they required the spacetime to perturbatively possess the Killing symmetries of the Kerr spacetime (that we refer to as \emph{Kerr symmetries} throughout this paper). This spacetime thus possesses a Carter-like constant and makes the Hamilton-Jacobi equation separable. This work was later extended by Johannsen~\cite{Johannsen:2015pca} who treated deviations from Kerr to be exact (without a perturbative scheme) and was further generalized by two of the authors of this paper~\cite{Carson:2020dez} that allowed an extra arbitrary function of the radial coordinate. A similar analysis was carried out by Papadopoulos and Kokkotas~\cite{Papadopoulos:2018nvd,Papadopoulos:2020kxu}. Konoplya et al.~\cite{Konoplya:2018arm} studied a parameterized Kerr spacetime that admits the separability on both Hamilton-Jacobi and Klein-Gordon equations while Lima Junior \emph{et al}.~\cite{Junior:2020lya} constructed a parameterically-deformed rotating BH spacetime with a separable Hamilton-Jacobi equation through a modified Newman-Janis algorithm.  Chen~\cite{Chen:2020aix} constructed a parameterized Kerr spacetime with Kerr symmetries by relaxing the $Z_2$ symmetry of spacetime while Delporte \emph{et al.}~\cite{Delaporte:2022acp} derived the metric without the circularity of spacetime.

In this paper, we focus on the parameterized rotating BH spacetime with Kerr symmetries constructed by Carson and Yagi (CY)~\cite{Carson:2020dez} that includes the one by Johannsen~\cite{Johannsen:2015pca}. There are several arbitrary functions of the radial coordinate $r$ in this spacetime.
Practically, these functions are Taylor expanded about infinity with Taylor coefficients becoming deviation parameters away from Kerr. Some of these deviation parameters can be removed through the redefinition of the BH mass and spin or constrained by imposing observational bounds from e.g. solar system experiments. When comparing with other observations, we typically keep a few leading parameters to constrain them. Here, we point out that, when one truncates the Taylor series in the arbitrary functions, we sometimes find an unphysical divergence in the metric which may make the use of such metrics problematic to probe the nature of BHs through observations.  

Here, we show such pathological behaviors in the original CY metric, taking a braneworld BH as an example, and demonstrate that a simple redefinition of the arbitrary functions can remove the divergence, at least for several example BH metrics studied in this paper. The refined metric (summarized in Sec.~\ref{sec:summary}) has five arbitrary functions of $r$ in total (same number as the original CY metric in~\cite{Carson:2020dez} while Johannsen's metric has four arbitrary functions in $r$~\cite{Johannsen:2015pca}). We propose two restricted classes (class I with one arbitrary function and class II with two arbitrary functions) that can reproduce several known rotating BH spacetimes in theories beyond GR when the arbitrary functions are specified appropriately.
We then study some of the properties and outcomes of the refined parameterized Kerr spacetime. We first show that the Petrov type of the parameterized Kerr spacetime, including the restricted class II, is of type I, while that of restricted class I is of type D. We next study BH observables, namely the ringdown frequencies and the shapes of BH shadows as specific examples. We show how such observables for the parameterized Kerr spacetime deviate from the Kerr case.

The rest of the paper is organized as follows. In Sec.~\ref{sec:summary}, we provide a summary of the refined parameterized Kerr spacetime. In Sec.~\ref{sec:orig}, we review the original parameterized Kerr BH proposed in~\cite{Carson:2020dez}. We then develop in Sec.~\ref{sec:refined} refined parameterized Kerr BHs by removing pathological behaviors. In Sec.~\ref{sec:Petrov}, we study the Petrov type of the refined parameterized metric. In Sec.~\ref{sec:observables}, we compute the quasinormal mode (QNM) frequencies and the shape of the BH shadow (or photon rings). We make our conclusions in Sec.~\ref{sec:conclusion}. We use the geometric units of $G=c=1$.

\subsection{Summary of Refined Parameterized Kerr Metric}
\label{sec:summary}

We summarize here the refined parameterized BH spacetime. The final form of the metric is given by
\begin{align}
\label{eq:refined_CY}
    ds^2 =&  - \frac{\tilde \Sigma A_5 \left(A_5-a^2 A_2^2 \sin ^2 \theta \right)}{\rho^4} dt^2 \nonumber \\
&+\frac{2 a A_5 (A_5-A_0) \tilde \Sigma \sin ^2\theta}{\rho^4}dt d\phi \nonumber \\
&+ \frac{\tilde \Sigma \sin ^2 \theta  A_5 \left(A_1^2 - a^2 A_5 \sin ^2\theta
   \right)}{\rho^4} d\phi^2 \nonumber \\
&+\tilde \Sigma \left( \frac{dr^2}{A_5}+d\theta^2\right)\,,
\end{align}
with
\begin{align}
 \allowdisplaybreaks
   \rho^4 \equiv &  a^4 A_2^2 A_5 \sin ^4 \theta \nonumber \\
& +a^2 \sin ^2\theta  \left(A_0^2-2
   A_0 A_5-A_1^2 A_2^2\right)  +A_1^2 A_5\,, \\
\label{eq:Sigma_tilde}
\tilde\Sigma \equiv& \Sigma+f(r)+g(\theta)\,,\\
\Sigma =& r^2+ a^2 \cos ^2\theta \,, \\ 
\Delta =& r^2  -2 M r + a^2\,.
\end{align}
Here, $A_i(r)$, $f(r)$ and $g(\theta)$ are arbitrary functions of $r$ and $\theta$ while $M$ is the BH mass and $a$ is the spin parameter. The above metric reduces to Kerr in the limit
\begin{equation}
\label{eq:Kerr_limit}
(A_0,A_1,A_2,A_5,\tilde \Sigma) \to (r^2+a^2,r^2+a^2,1,\Delta,\Sigma)\,.
\end{equation}
The arbitrary functions can further be expanded about infinity as 
\begin{align}
\allowdisplaybreaks
\label{eq:Ai_exp}
A_i(r)&\equiv r^2 \left[ 1 + \frac{a^2}{r^2} +  \sum\limits_{n=1}^\infty\alpha_{in}\left(\frac{M}{r}\right)^n \right]\,, \quad (i=0,1)\,, \\
A_2(r) &\equiv 1 + \sum\limits_{n=1}^\infty\alpha_{2n}\left(\frac{M}{r}\right)^n\,, \\
\label{eq:A5_expansion} 
A_5(r) & \equiv r^2 \left[ 1 - \frac{2M}{r} + \frac{a^2}{r^2} +  \sum\limits_{n=1}^\infty\alpha_{5n}\left(\frac{M}{r}\right)^n \right]\,,  \\ 
f(r)&\equiv r^2\sum\limits_{n=1}^\infty \epsilon_n\left(\frac{M}{r}\right)^n\,,\\
\label{eq:g_exp}
g(\theta) & \equiv  M^2\sum\limits^\infty_{n=0}\gamma_n P_n(\cos\theta)\,,
\end{align}
where $P_n$ is the Legendre polynomial. The above metric reduces to that of Johannsen~\cite{Johannsen:2015pca} (with the rescaled radial functions) in the limit $A_0 \to A_1 A_2$.

We provide below metrics within two restricted classes whose metrics are simpler than the above with less arbitrary functions and yet contain several examples of known BH spacetimes (the mapping between the refined parameterized Kerr metrics and the example BH spacetimes can be found in Appendix~\ref{app:mapping}): 

\begin{itemize}

  \item[1.] restricted class (I): \\
 $A_0 = A_1 = r^2+a^2$, $A_2=1$, $f=0$, $g=0$ \\
arbitrary function: $A_5(r)$ \\
example: braneworld~\cite{Aliev:2005bi}, Hayward~\cite{Hayward:2005gi}, Bardeen~\cite{Bardeen_proceedings}, Ghosh~\cite{Ghosh:2014pba}, Kalb-Ramond~\cite{Kumar:2020hgm}
\begin{align}
\label{eq:restricted_1}
    ds^2 = & - \frac{A_5 - a^2 \sin ^2\theta }{\Sigma } dt^2 \nonumber \\
&-\frac{2a \sin ^2\theta  \left(r^2+a^2-A_5\right)}{\Sigma } dt d\phi \nonumber \\
& + \frac{\left(r^2+a^2\right)^2  -a^2 A_5 \sin ^2\theta}{\Sigma }\sin ^2\theta d\phi^2 \nonumber \\
& + \Sigma \left( \frac{dr^2}{A_5}+d\theta^2\right)\,.
\end{align}

   \item[2.] restricted class (II): \\ $A_0 = A_1=r^2+a^2+f$, $A_2=1$, $g=0$ \\
arbitrary functions: $A_5(r)$, $f(r)$ \\
example: braneworld~\cite{Aliev:2005bi}, Hayward~\cite{Hayward:2005gi}, Bardeen~\cite{Bardeen_proceedings}, Ghosh~\cite{Ghosh:2014pba}, Kalb-Ramond~\cite{Kumar:2020hgm}, Kerr-Sen~\cite{Sen:1992ua}
\begin{align}
\label{eq:restricted_2}
    ds^2 =&   - \frac{A_5 - a^2 \sin ^2\theta }{\tilde \Sigma } dt^2 \nonumber \\
& -\frac{2a \sin ^2\theta  \left(r^2+a^2+f-A_5\right)}{\tilde \Sigma } dt d\phi \nonumber \\
& + \frac{\left(r^2+a^2+f\right)^2  -a^2 A_5 \sin ^2\theta}{\tilde \Sigma }\sin ^2\theta d\phi^2 \nonumber \\
& + \tilde \Sigma \left( \frac{dr^2}{A_5}+d\theta^2\right)\,, \\
\tilde \Sigma = &\Sigma + f\,.
\end{align}

\end{itemize}

\section{Original Parameterized Kerr Spacetime}
\label{sec:orig}

We begin by reviewing the original parameterized BH spacetime with Kerr symmetries developed in~\cite{Carson:2020dez} by two of the authors. The metric that preserves Kerr symmetries (which we call the CY metric hereafter) is given by\footnote{In the original CY metric, arbitrary radial functions were referred to as $A_i(r)$. Because $A_1$ and $A_2$ only enter quadratically, we redefine the radial functions as $(\calA_0, \calA_1, \calA_2, \calA_5) = (A_0, A_1^2, A_2^2, A_5)$.}
\begin{align}
\label{eq:CY_metric_original}
      ds^2 = &- \frac{\tilde\Sigma \left(\Delta -a^2 \calA_2 \sin ^2\theta \right)}{\tilde \rho^4} dt^2 \nonumber \\
& -\frac{2a \tilde{\Sigma } \sin^2\theta  \left[ \left(a^2+r^2\right)\calA_0 - \Delta\right]}{\tilde \rho^4} dt d\phi \nonumber \\
&+ \frac{\tilde{\Sigma } \sin^2\theta \left[\left(a^2+r^2\right)^2\calA_1 -a^2 \Delta  \sin ^2\theta \right]}{\tilde \rho^4} d\phi^2 \nonumber \\
&+\tilde\Sigma \left( \frac{dr^2}{\calA_5\Delta}  + d\theta^2\right)\,,
\end{align}
where $a$ is the object's spin, $\mathcal{A}_i(r)$ is an arbitrary function of $r$ while
\begin{eqnarray}
\tilde \rho^4 &=& a^4 \calA_2 \sin ^4\theta +
   \left(a^2+r^2\right)^2  \calA_1\nonumber \\
   & & + a^2 \left(a^2+r^2\right)  \left(\frac{ a^2+r^2}{\Delta} \left(\calA_0^2-\calA_1
   \calA_2\right)-2 \calA_0  \right) \sin
   ^2\theta\,. \nonumber \\
\end{eqnarray}
The above metric reduces to the Kerr BH when $\mathcal{A}_i \to 1$ and $f(r) \to 0$, while it reduces to the Johannsen metric \cite{Johannsen:2015pca} in the limit $\mathcal A_0^2 \to \mathcal A_1 \mathcal A_2$.

The CY metric \cite{Carson:2020dez} exhibits a pathology in certain situations. The metric itself consists of five functions of the radial coordinate $r$ that capture the deviations from Kerr. When mapping this metric to existing beyond-GR theories, these functions are expanded about $r = \infty$. The expansion coefficients represent the deviation parameters from Kerr. Naturally, these are infinite expansions, so for practical purposes, we truncate the expansion so as to have a finite number of beyond-Kerr deviation parameters. This truncation can introduce pathological behavior, namely a nonphysical divergence, into the spacetime.

To put the above in context, let us now consider a BH in the Randall-Sundrum braneworld model~\cite{Randall:1999ee,Randall:1999vf} as an example. The braneworld scenario is an extra dimension model motivated by e.g. string theory. A rotating BH solution in this model is given by~\cite{Aliev:2005bi}
\begin{eqnarray}
\label{eq:brane}
ds^2 &=& -\left(1 - \frac{2 M r-\beta }{\Sigma }\right) dt^2 -2\frac{a   (2 M r-\beta )}{\Sigma }\sin ^2\theta dt d\phi \nonumber \\
&&+ \sin ^2\theta  \left(\frac{a^2  (2 M r-\beta )}{\Sigma
   }\sin ^2\theta +a^2+r^2\right) d\phi^2 \nonumber \\
&& + \Sigma \left( \frac{dr^2}{\beta +\Delta }  +  d\theta^2 \right)\,.
\end{eqnarray}
Here  $\beta$ is the tidal charge and
the above solution reduces to the Kerr metric in the limit $\beta \to 0$. This metric can be mapped to the CY metric in Eq.~\eqref{eq:CY_metric_original} as
\begin{equation}
\label{eq:Ai_brane}
(\calA_0,\calA_1,\calA_2,\calA_5) = \left(\frac{\Delta}{\Delta+\beta}, \frac{\Delta}{\Delta+\beta}, \frac{\Delta}{\Delta+\beta},   \frac{\Delta+\beta}{\Delta} \right)\,,
\end{equation}
and $f = g = 0$. We can further expand the above functions about infinity to yield
\allowdisplaybreaks
\begin{eqnarray}
\label{eq:A0-brane}
\calA_0 &=& 1-\frac{\beta }{r^2}-\frac{2 \beta 
   M}{r^3}+\frac{\beta  \left(a^2+\beta -4 M^2\right)}{r^4}+ \mathcal{O}\left( \frac{M^5}{r^5} \right)\,, \nonumber \\
\\   
\calA_1 &=& 1-\frac{\beta }{2 r^2}-\frac{\beta 
   M}{r^3}+\frac{\beta  \left(4 a^2+3 \beta -16 M^2\right)}{8 r^4} \nonumber \\
& & + \mathcal{O}\left( \frac{M^5}{r^5} \right)\,, 
   \\
   \calA_2 &=& 1-\frac{\beta }{2 r^2}-\frac{\beta 
   M}{r^3}+\frac{\beta  \left(4 a^2+3 \beta -16 M^2\right)}{8 r^4} \nonumber \\
&& + \mathcal{O}\left( \frac{M^5}{r^5} \right)\,, 
   \\
 \label{eq:A5-brane}
 \calA_5 &=& 1+\frac{\beta
   }{r^2}+\frac{2 \beta  M}{r^3}+ \frac{4 \beta  M^2-a^2 \beta }{r^4}+ \mathcal{O}\left( \frac{M^5}{r^5} \right)\,. 
\end{eqnarray}

\begin{figure*}[t]
\includegraphics[width=8.5cm]{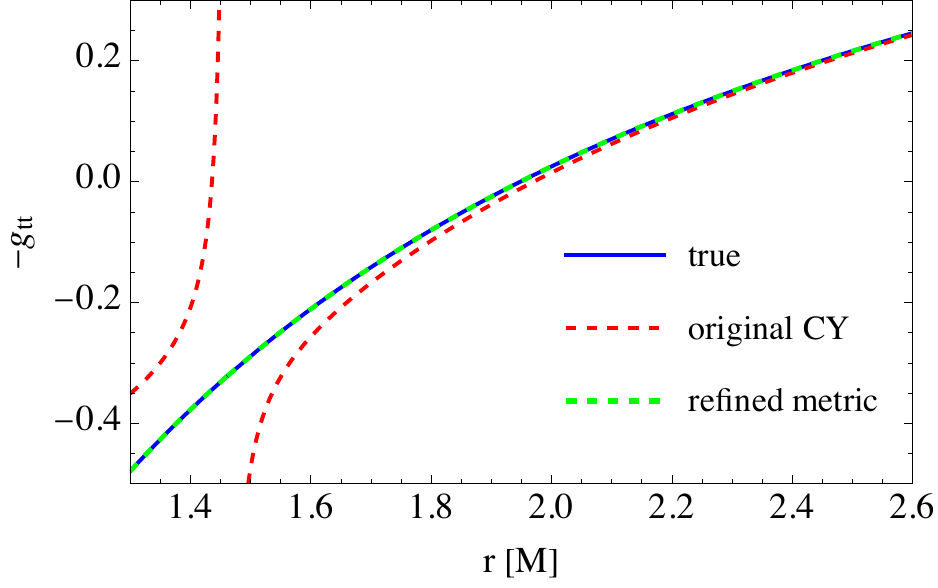}
\includegraphics[width=8.5cm]{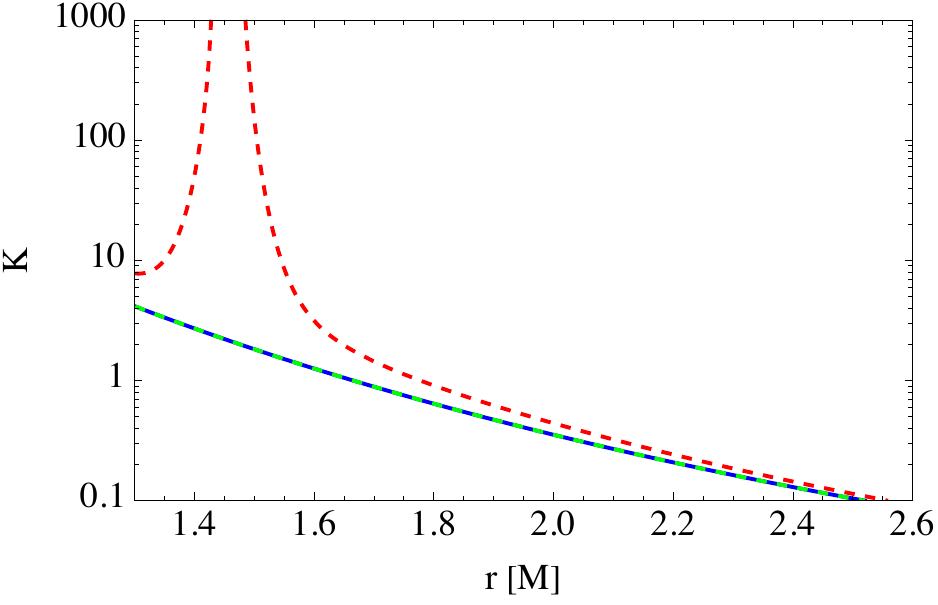}
\caption{\label{fig:gtt_comparison_original} (Left) Comparison of $-g_{tt}$ for the exact braneworld BH (blue solid), the original CY metric with $\mathcal{A}_i$ given in Eqs.~\eqref{eq:A0-brane}--\eqref{eq:A5-brane} (red dashed), and the refined parameterized BH metric with $A_i$ in Eq.~\eqref{eq:Ai_mapping_brane} (green dashed). We chose the parameters as $a = 0.9M$, $\beta = 0.1 M^2$, and $\theta = \pi/2$. The left edge of $r=1.3M$ corresponds to the event horizon. Observe that the original CY metric has an unphysical divergence which is remedied in the refined metric. (Right) Similar to the left panel but for the Kretschmann invariant, showing that the unphysical divergence in the original CY metric is not a gauge artifact.
}
\end{figure*} 

\begin{figure}[h]
\includegraphics[width=8.5cm]{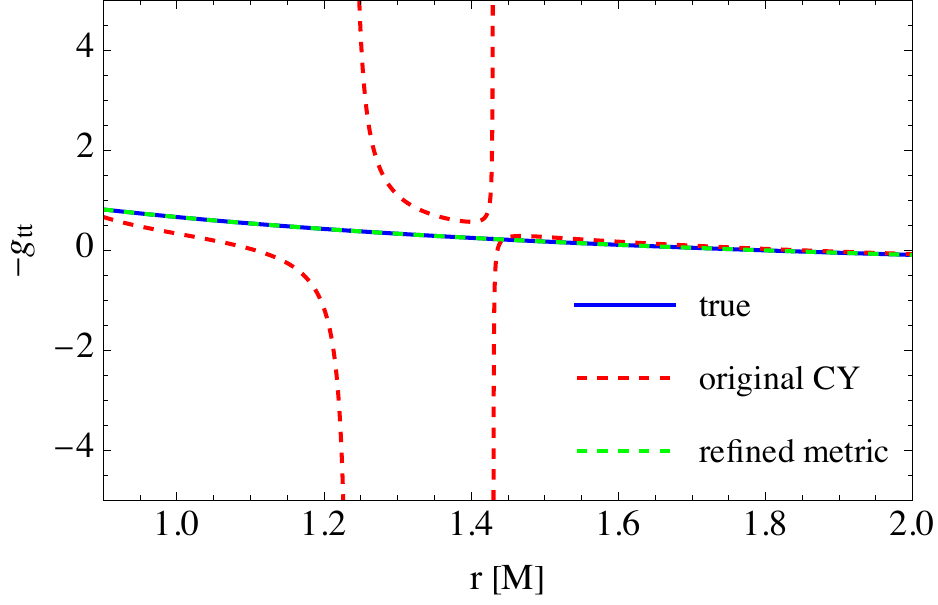}
\caption{\label{fig:Kerr-Sen_refined} Similar to the left panel of Fig.~\ref{fig:gtt_comparison_original} but for a Kerr-Sen BH with the parameters $a = 0.9M$, $b = 0.1 M$ and $\theta = \pi/2$. 
}
\end{figure}

The left panel of Fig.~\ref{fig:gtt_comparison_original} compares $-g_{tt}$ of the braneworld BH with the corresponding CY metric for a certain choice of parameters. Notice that the latter diverges at $r \sim 1.45M$, which is close to the event horizon location for Kerr in GR. To check whether this divergence is not an artifact of bad coordinates, we also show the profile for the Kretschmann invariant $K = R_{\alpha\beta\mu\nu} R^{\alpha\beta\mu\nu}$ in the right panel of Fig.~\ref{fig:gtt_comparison_original} that is a gauge invariant quantity. Observe that it also shows a divergence at the same location as $-g_{tt}$, which means that the divergence cannot be eliminated by a coordinate transformation. Though since this divergence is absent in the true braneworld BH metric, this is a pathology in the original CY metric. We also note that such a divergence arises in other example BH spacetimes if we use the truncated expansion of $\mathcal{A}_i$, such as Kerr-Sen as shown in Fig.~\ref{fig:Kerr-Sen_refined}. Here, $\mathcal{A}_i$ are expanded about $r=\infty$ and truncated at $\mathcal{O}(M^4/r^4)$, similar to the braneworld example. 

What is the origin of this divergence? To address this, we take a look at $g_{tt}$ in Eq.~\eqref{eq:CY_metric_original} and use the full expression for $\mathcal{A}_i$ for the braneworld BH in Eq.~\eqref{eq:Ai_brane}. One can show that both the numerator and denominator are proportional to $\Delta^2$ which cancels out. However, when we use the approximate expression for $\mathcal{A}_i$ in Eqs.~\eqref{eq:A0-brane}--\eqref{eq:A5-brane}, this cancellation is lost and the denominator vanishes when $\Delta \approx 0$, namely the location of the Kerr horizon.

\section{Regularizing the Parameterized Kerr Spacetime}
\label{sec:refined}

We now show that a simple redefinition of the arbitrary radial functions remedies the pathological behavior in the original parameterized Kerr spacetime mentioned in the previous section. Another approach is to assume deviations from Kerr to be small and treat them perturbatively as done in~\cite{vigelandnico}. 
We have successfully removed the divergence in the original CY metric for braneworld and Kerr-Sen BHs. In fact, we could reproduce the former exactly even within the small deviation approximation, while there was some noticeable difference between the true and the new refined metric for the latter. 
We will detail this approach in Appendix~\ref{sec:small_dev} and present the results in Fig.~\ref{fig:small_dev}.

\subsection{Metric}

To remedy the pathologies discussed in the previous section, we perform a simple rescaling of the radial functions.
The idea is to factor $\Delta$ out of the $\mathcal{A}_i(r)$ functions themselves to cancel out the $\Delta$ in the denominator of the metric components explicitly. This will therefore eliminate the divergence at the Kerr horizon. We found that the following redefinition of the radial functions from $\mathcal{A}_i$ to $A_i$ works at least for the example BH metrics considered in this paper:
\begin{align} 
\label{eq:refined_mapping}
&(\calA_0,\calA_1,\calA_2,\calA_5) \nonumber \\
& = \left(\frac{\Delta A_0}{ \left(r^2+a^2\right)A_5}, \frac{\Delta A_1^2}{ \left(r^2+a^2\right)^2 A_5}, \frac{\Delta A_2^2}{A_5}, \frac{A_5}{\Delta } \right)\,.
\end{align}
The $(r^2+a^2)$ dependence in the redefinition is to absorb the same factor in the original CY metric. Further rescaling by $A_5$ is motivated by the following observation. We found that with the simple rescaling by $\Delta$ alone, the new metric components have $A_5$ in the denominator, thus the metric after this $\Delta$ scaling diverges at $A_5=0$, corresponding to the location of the true event horizon. 

The refined metric with these new radial functions $A_i$ is given in Eq.~\eqref{eq:refined_CY} with the Kerr limit shown in Eq.~\eqref{eq:Kerr_limit}. Notice that the new metric is slightly simpler than the original CY metric. Similar to the latter, one can expand $A_i$ about $r=\infty$ as in Eqs.~\eqref{eq:Ai_exp}--\eqref{eq:g_exp}. 

Let us now apply the refined metric to the braneworld BH example. The mapping for the $A_i$ functions is given by
\begin{equation}
\label{eq:Ai_mapping_brane}
(A_0,A_1,A_2,A_5)=\left(r^2+a^2,\; r^2+a^2,\; 1,\; \Delta +\beta \right)\,.
\end{equation}
Observe that the only function that is different from Kerr is $A_5$, which is a simple quadratic function in $r$. This means that the asymptotic series expansion in Eq.~\eqref{eq:A5_expansion} truncates at a finite order and we can recover the branworld BH metric exactly. This is explicitly demonstrated in Fig.~\ref{fig:gtt_comparison_original} for braneworld and Fig.~\ref{fig:Kerr-Sen_refined} for Kerr-Sen. Observe that the new parameterized metric not only removes the artificial divergence but also has a perfect agreement with the true metric.

The mapping of the new radial functions for each example BH metric is shown in Appendix~\ref{app:mapping}.  Similar to the braneworld case, there is only one non-vanishing Taylor coefficient in the series expansion for the arbitrary functions in Eqs.~\eqref{eq:Ai_exp}--\eqref{eq:g_exp} for Kalb-Ramond and Kerr-Sen BHs. Thus, the series truncates at a relatively low order for these BHs and one can recover the original metrics exactly. For Hayward, Bardeen, and Ghosh BHs, the series does not truncate at a finite order. However, the functions $A_0$, $A_1$, and $A_2$ are the same as the Kerr expressions for these example BHs. When imposing these conditions, the refined metric in Eq.~\eqref{eq:refined_CY}  reduces to the one in Eq.~\eqref{eq:restricted_1}. Then, the metric has no unphysical divergence. Thus, we managed to remove the divergence and have successfully ``regularized'' the metric for all the example BHs listed in Appendix~\ref{app:mapping}.

Based on the mapping in Appendix~\ref{app:mapping}, we propose restricted classes of the new parameterized metric that should be easier to handle than the full metric with a lower number of free functions. The first class is obtained by taking the Kerr limit in all of the arbitrary functions except for $A_5$. This metric is given in Eq.~\eqref{eq:restricted_1} and includes braneworld, Hayward, Bardeen, Ghosh, and Kalb-Ramond BHs.  
The second class imposes the conditions $A_0 = A_1$, $A_2=1$, and $g=0$, and the metric is given in Eq.~\eqref{eq:restricted_2}. This restricted metric can describe e.g. Kerr-Sen BH~\cite{Sen:1992ua}, together with all example BHs mentioned for the first restricted class.

Finally, let us present the asymptotic behavior of the metric for the refined parameterized BH spacetime with the expansion in Eqs.~\eqref{eq:Ai_exp}--\eqref{eq:g_exp}. First, $g_{tt}$ and $g_{t\phi}$ behave as
\begin{align}
    g_{tt} &= -1 + (2 + 2 \alpha_{11} - \alpha_{51}  -\epsilon_1)\frac{M}{r} + \mathcal{O}\left( \frac{M^2}{r^2} \right)\,, \\
\label{eq:gtph_asymp}
g_{t\phi} &= -(2+\alpha_{01} - \alpha_{51} ) a  \frac{M}{r} \sin^2\theta + \mathcal{O}\left( \frac{M^2}{r^2} \right)\,.
\end{align} 
We can further redefine the mass $M$ and the spin $a$ to set $\alpha_{01} = \alpha_{51}$ and $\alpha_{11} = (\epsilon_{1}+\alpha_{51})/2$ without loss of generality. Then, the asymptotic behavior of the metric components becomes
\begin{align}
g_{tt} =& -1 + 2\frac{M}{r} + \frac{1}{4}  [8\alpha_{12} + \epsilon_1^2 - 4 \epsilon_2-2 \alpha_{51} (\epsilon_1+4) \nonumber\\
&+\alpha_{51}^2-4 \alpha_{52}-4g] \frac{M^2}{r^2} + \mathcal{O}\left( \frac{M^3}{r^3} \right)\,, \\
g_{rr} =& 1 + \left(2 - \alpha_{51} + \epsilon_1 \right)\frac{M}{r}+ \mathcal{O}\left( \frac{M^2}{r^2} \right)\,, \\
g_{\theta\theta} =& r^2 \left[1 + \epsilon_1\frac{M}{r}+ \mathcal{O}\left( \frac{M^2}{r^2} \right) \right]\,, \\
g_{\phi\phi} = & r^2 \sin^2\theta \left[1 + \epsilon_1\frac{M}{r}+ \mathcal{O}\left( \frac{M^2}{r^2} \right) \right]\,.
\end{align}
One could compare this with the asymptotic behavior of the metric for a non-rotating object within the parameterized post-Newtonian (PPN) framework~\cite{will-living} and use solar system bounds to constrain some of the parameters. However, we do not impose such constraints in this paper because Birkhoff's theorem does not hold in general for non-GR theories, so there is no guarantee the BH spacetime can describe the exterior spacetime of stars.

\subsection{Relation to Other Parameterized Kerr Spacetimes}
 
We next discuss the relation between the refined metric presented in this paper and some of the parameterized BH spacetimes found in previous literature. 

\begin{enumerate}
    \item Cardoso et al.~\cite{Cardoso:2014rha}

 Cardoso et al.~\cite{Cardoso:2014rha} derived a parameterized BH metric which is a generalization of the one constructed by Johannsen and Psaltis~\cite{johannsen-metric}. The latter took the seed non-rotating metric
\begin{eqnarray}
    ds^2 &=& -\bar f(r) [1+\bar h(r)] dt^2 + [1+\bar h(r)] \frac{dr^2}{\bar f(r)} \nonumber \\
&&+ r^2 (d\theta^2 + \sin^2\theta d\phi^2)\,, 
\end{eqnarray}
with $\bar f \equiv 1-2M/r$
and applied a Newman-Janis algorithm~\cite{Newman:1965tw,Drake:1998gf,Erbin:2016lzq} to turn it into a rotating metric. Cardoso et al.~\cite{Cardoso:2014rha} generalized the seed non-rotating metric to
\begin{eqnarray}
    ds^2& = &-\bar f(r) [1+\bar h_t(r)] dt^2 + [1+\bar h_r(r)] \frac{dr^2}{\bar f(r)} \nonumber \\
&&+ r^2 (d\theta^2 + \sin^2\theta d\phi^2)\,, 
\end{eqnarray}
and found the following rotating metric via the Newman-Janis transformation:
\begin{align}
\label{eq:Cardoso}
    ds^2 =& -F(1+h_t)dt^2 -2a[H-F(1+h_t)]\sin^2\theta dtd\phi \nonumber \\
& +\{\Sigma + [2H-F(1+h_t)]a^2 \sin^2\theta\}\sin^2\theta d\phi^2 \nonumber \\
&+\Sigma \left(\frac{1+h_r}{\Delta + h_r a^2 \sin^2\theta}dr^2 + d\theta^2\right)\,,
\end{align} 
where 
\begin{equation}
    F \equiv 1-\frac{2Mr}{\Sigma}\,, \quad H \equiv \sqrt{(1+h_t)(1+h_r)}\,,
\end{equation}
and $h_i = h_i(r,\theta)$. 

We found that the refined parameterized metric for the first restricted class in Eq.~\eqref{eq:restricted_1} is a special case of the one found by Cardoso et al.~\cite{Cardoso:2014rha} in Eq.~\eqref{eq:Cardoso}.
If we take the non-rotating limit of the second restricted metric in Eq.~\eqref{eq:restricted_1}, we find
\begin{eqnarray}
    ds^2 = - \frac{A_5}{r^2}dt^2 + \frac{r^2}{A_5}dr^2 + r^2 (d\theta^2 + \sin^2\theta d\phi^2)\,. \nonumber\\
\end{eqnarray}
Therefore, if Eq.~\eqref{eq:restricted_1} 
can be constructed via the Newman-Janis algorithm, the seed metric should have the form $ 1+\bar h_r = (1+\bar h_t)^{-1}$.
Indeed, we found that the following relations turn Eq.~\eqref{eq:Cardoso} into Eq.~\eqref{eq:restricted_1}:
\begin{equation}
1+h_r = \frac{1}{1+ h_t}\,,
 \qquad   
h_t = -\frac{\Delta - A_5}{\Delta -a^2 \sin ^2\theta }\,. 
\end{equation}
    
\item Papadopoulos and Kokkotas~\cite{Papadopoulos:2018nvd}

The new radial functions $A_i$ and $f$ in Eq.~\eqref{eq:refined_CY} are similar to those in Papadopoulos and Kokkotas~\cite{Papadopoulos:2018nvd}. The relation between $A_i$ here and $A_i^\mathrm{PK}$ in~\cite{Papadopoulos:2018nvd} is
\begin{align}
&(A_1^\mathrm{PK},A_2^\mathrm{PK},A_3^\mathrm{PK},A_4^\mathrm{PK},A_5^\mathrm{PK})   \nonumber \\
&=\left(f+r^2, A_5, -\frac{a^2 A_2^2}{A_5}, -\frac{a A_0}{A_5}, -\frac{A_1^2}{A_5} \right)\,.
\end{align} 
Because of the difference in the scaling of $A_5$ appearing on the right-hand side of the above mapping, $A_i^\mathrm{PK}$ for BH solutions beyond GR (for example braneworld) in general contains a denominator that is a function of $r$. This leads to an infinite series when expanded about $r=\infty$ so the metric in Eq.~\eqref{eq:refined_CY} is more accurate (and can reduce to the exact metric in some cases) than that in~\cite{Papadopoulos:2018nvd} for the example metrics considered in this paper when we use the expanded $A_i$ or $A_i^\mathrm{PK}$.

\item Baines and Visser~\cite{Baines:2023dhq}

Another generalized Kerr spacetime metric was devised by Baines and Visser~\cite{Baines:2023dhq} that preserves Kerr symmetries and keeps the timelike Hamilton-Jacobi and massive Klein-Gordon equations separable. The line element is
\ba
\label{eq:Baines}
ds^2 = &-& \frac{\tilde{\Delta} e^{-2\Phi} - a^2 \sin^2{\theta}}{\Xi^2 + a^2 \cos^2{\theta}} dt^2 + \frac{\Xi^2 + a^2 \cos^2{\theta}}{\tilde{\Delta}} dr^2 \nonumber \\
&+& \left[\Xi^2 + a^2 \cos^2{\theta}\right] d\theta^2 \nonumber \\
&-&2a\frac{ \Xi^2 - \tilde{\Delta} e^{-2\Phi} + a^2 }{\Xi^2 + a^2 \cos^2{\theta}} \sin^2{\theta} dtd\phi \nonumber \\
&+& \frac{\left(\Xi^2 + a^2 \right)^2 - e^{-2\Phi} \tilde{\Delta} a^2 \sin^2{\theta}}{\Xi^2 + a^2 \cos^2{\theta}}  \sin^2{\theta} d\phi^2 \,, \nonumber\\
\ea
with arbitrary deviation functions $\tilde{\Delta}(r)$, $\Phi(r)$, and $\Xi(r)$.
We found the mapping between the refined parameterized metric in Eq.~\eqref{eq:refined_CY} and that in Eq.~\eqref{eq:Baines} as
\ba
&&(A_0,A_1,A_2,A_5,f) \nonumber \\
&=& \left(e^{2\Phi} \left(a^2 + \Xi^2 \right), e^{\Phi} \left(a^2 + \Xi^2 \right), e^{\Phi}, \tilde{\Delta}, -r^2 + \Xi^2 \right). \nonumber\\
\ea
Thus, the metric developed in~\cite{Baines:2023dhq} is a subclass of the refined CY metric.

\end{enumerate}

\section{Petrov Type}
\label{sec:Petrov}

In this section, we study the Petrov type of the refined parameterized BH spacetime.
We first focus on the first restricted class with the metric in Eq.~\eqref{eq:restricted_1}. We will then consider a more general parameterized metric. 

\subsection{Determinig Petrov Type}

The Petrov type of spacetime can be determined by constructing a null tetrad $l^\mu$, $n^\mu$, $m^\mu$ and $\bar m^{\mu}$ (a bar refers to complex conjugate) and computing the Newman-Penrose Weyl scalars. These null vectors satisfy the normalization $l^\mu n_\mu = -1$ and $m^\mu \bar m_\mu = 1$ with all the other contractions to vanish. Below, we follow~\cite{stephani,campanelli} to identify the Petrov type.

When the following condition is satisfied, the spacetime is said to be algebraically special (having at least one degenerate principal null direction):
\be
I^3 = 27 J^2\,.
\label{eq:IJ}
\ee
Here
\ba
\label{I-eq}
I &\equiv& 
3 \Psi_2^2-4\Psi_1\Psi_3 + \Psi_4 \Psi_0\,, \\ 
\label{J-eq}
J &\equiv & 
-\Psi_2^3 + 2 \Psi_1 \Psi_3 \Psi_2 + \Psi_0 \Psi_4 \Psi_2 - \Psi_4 \Psi_1^2 - \Psi_0 \Psi_3^2\,, \nonumber \\
\ea
where $\Psi_i$ are the Newman-Penrose Weyl scalars:
\allowdisplaybreaks
\begin{align}
\Psi_{0} &=C_{\alpha \beta \gamma \delta} l^{\alpha} m^{\beta} l^{\gamma} m^{\delta}\,, \\
\Psi_{1} &=C_{\alpha \beta \gamma \delta} l^{\alpha} n^{\beta} l^{\gamma} m^{\delta}\,, \\
\Psi_{2} &=C_{\alpha \beta \gamma \delta} l^{\alpha} m^{\beta} \bar{m}^{\gamma} n^{\delta}\,, \\
\Psi_{3} &=C_{\alpha \beta \gamma \delta} l^{\alpha} n^{\beta} \bar{m}^{\gamma} n^{\delta}\,, \\
\Psi_{4} &=C_{\alpha \beta \gamma \delta} n^{\alpha} \bar{m}^{\beta} n^{\gamma} \bar{m}^{\delta}\,,
\end{align} 
for a Weyl tensor $C_{\alpha\beta\gamma\delta}$.
In particular, for Petrov type D (and II), $I$ and $J$ are nontrivial. On the other hand, the spacetime is type I if Eq.~\eqref{eq:IJ} is not satisfied.

To determine the Petrov type further, we compute the following scalar quantities:
\allowdisplaybreaks
\ba
K & \equiv & \Psi_1 \Psi_4^2 - 3 \Psi_4 \Psi_3 \Psi_2 + 2 \Psi_3^3\,, \\
\label{L}
L & \equiv & \Psi_2 \Psi_4 - \Psi_3^2\,, \\
\label{N}
N & \equiv & \Psi_4^3 \Psi_0 - 4 \Psi_4^2 \Psi_1 \Psi_3 + 6 \Psi_4 \Psi_2 \Psi_3^2 - 3 \Psi_3^4\,,
\ea
 with $\Psi_4 \neq 0$.
In particular, type D (and III) satisfies the following condition with $N \neq 0$:
\be
K=0, \quad N-9L^2=0\,,
\label{conditionD}
\ee
where $K$ and $N-9L^2$ are invariant under a tetrad rotation. 

To summarize, the spacetime is type D if Eqs.~\eqref{eq:IJ} and~\eqref{conditionD} are satisfied for non-vanishing $I$ and $J$. On the other hand, the spacetime is type I if Eq.~\eqref{eq:IJ} is not satisfied.

\subsection{Application to the Refined Parameterized Metric}

Let us now apply the above formalism to the refined parameterized BH. Let us first consider restricted class I in Eq.~\eqref{eq:restricted_1}. We begin by finding the null tetrad for this spacetime. For Kerr, a commonly-used null tetrad was derived by Kinnersley~\cite{Kinnersley:1969zza}: 
\begin{align}
    l^\mu_\mathrm{(Kin)} = & \left( \frac{r^2 + a^2}{\Delta}, 1,0,\frac{a}{\Delta} \right)\,, \\ 
n^\mu_\mathrm{(Kin)} = &\left(  \frac{r^2 + a^2}{2\Sigma}, - \frac{\Delta}{2\Sigma}, 0, \frac{a}{2\Sigma} \right)\,,  \\ 
m^\mu_\mathrm{(Kin)} = &\frac{1}{\sqrt{2}(r+ia\cos\theta)} \left(  ia\sin\theta,0,1, \frac{i}{\sin\theta}\right)\,. 
\end{align}
We will rotate these to construct a new tetrad such that $\Psi_4 \neq0$:
\begin{align}
    l^\mu =& l^\mu_\mathrm{(Kin)}\,, \quad m^\mu = m^\mu_\mathrm{(Kin)} + l^\mu_\mathrm{(Kin)}\,,\nonumber \\
 \quad n^\mu =& n^\mu_\mathrm{(Kin)} +l^\mu_\mathrm{(Kin)} +  m^\mu_\mathrm{(Kin)}+ \bar m^\mu_\mathrm{(Kin)}\,.
\end{align}
The tetrad for the refined parameterized metric can be obtained by simply performing the replacement 
\begin{eqnarray}
    2Mr \to r^2 + a^2 - A_5\,.
\end{eqnarray}
This is because the metric in Eq.~\eqref{eq:restricted_1} can be obtained by performing the same transformation to the Kerr metric. Having this tetrad at hand, one can compute the Weyl scalars to yield
\bw
\begin{align}
    \Psi_0 =& \Psi_1 = 0\,, \\
\Psi_2 =& - \frac{1}{12\Sigma^3}\left[ 12 \left(r+ia\mu \right)^2 A_5 -6\,\Sigma\,(r+ia\mu)\,A_5'+{\Sigma}^{2
}A_5'' -2\mu^2 \left( {\mu}^{2}-6 \right) {a}^{4} \right. \nonumber \\
& \left.
+12\,ir \left( {\mu}^{2}-2 \right) \mu\,{a}^{3} +4r^2 \left( 5\,{\mu}^{2}-3 \right) {a}^{2}-12\,i\mu\,a {r}^{3} -2\,{r}^{4}
\right]\,, \\
\Psi_3 = & - \frac{1}{4\Sigma^2 (r+ia\mu)} \left[ 12 \left(r+ia\mu \right) A_5 -6\Sigma A_5' +\Sigma(r-ia\mu) A_5'' +2\,i
 \left( {\mu}^{2}-6 \right) \mu\,{a}^{3} \right. \nonumber \\
& \left. + 2r\left(  5\,{\mu}^{2}-6 \right) {a}^{2}-10\,i  {r}^{2}  \mu\,a-2\,{r}^{3}
  \right]\,, \\
\Psi_4 =& -\frac{1}{2\Sigma (r-ia\mu)^2} \left[ 12\,A_5 - 6(r-ia\mu) A_5' +(r-ia\mu)^2 A_5'' +2 \left( 
 {\mu}^{2} -6 \right) {a}^{2}   -8\,i
  r \mu\,a-2  {r
}^{2}
\right]\,,
\end{align}
\ew
where $\mu = \cos\theta$ and a prime denotes the derivative with respect to $r$.

Now we are ready to determine the Petrov type of the refined parameterized metric in the restricted class. First, we found
\bw
\begin{align}
    I =& \frac{1}{48\Sigma^6}\left[-12 (r+ia\mu)^2\,A_5 +6\Sigma (r+ia\mu) A_5' -\Sigma A_5'' -12\,i{a}^{3}{\mu}^{3}r+2\,{a}^{4}{\mu}^{4}+24\,i{a}^{3}\mu\,r \right. \nonumber \\
&\left.
+12\,ia\mu\,{r}^{3}-12\,{a}^{4}{\mu}^{2}-20\,{\mu}^{2}{a}^{2}{r}^{2} a\mu\,r+12\,{a}^{2}{r}^{2}+2\,{r}^{
4}\right] ^{2}\,, \\
J=& - \frac{1}{1728\Sigma^9}\left[-12 (r+ia\mu)^2\,A_5 +6\Sigma (r+ia\mu) A_5' -\Sigma A_5'' -12\,i{a}^{3}{\mu}^{3}r+2\,{a}^{4}{\mu}^{4}+24\,i{a}^{3}\mu\,r \right. \nonumber \\
&\left.
+12\,ia\mu\,{r}^{3}-12\,{a}^{4}{\mu}^{2}-20\,{\mu}^{2}{a}^{2}{r}^{2} a\mu\,r+12\,{a}^{2}{r}^{2}+2\,{r}^{
4}\right] ^{3}\,.
\end{align}
\ew
This leads to $I^3 - 27 J^2 = 0$. Next, we found $K=0$ while $L$ and $N$ are non-vanishing and the latter two satisfying $N - 9 L^2 = 0$. This concludes the refined parameterized metric in Eq.~\eqref{eq:restricted_1} is Petrov type D.  Further, this means the braneworld, Hayward, Bardeen, Ghosh, and Kalb-Ramond rotating BHs are all type D. This is consistent with the recent work in~\cite{Guo:2023wtx} that identified a general, stationary, axisymmetric and asymptotically-flat spacetime under Petrov type D which includes the refined metric here under the first restricted class. This is also consistent with the finding by Walker and Penrose~\cite{walkerpenrose1970} that any spacetimes with Petrov type D allow integrability of the geodesic equations.

We now discuss the Petrov type of a more general class of the refined metric. For example, let us consider the second restricted class in Eq.~\eqref{eq:restricted_2}. Similar to the first restricted class, one can construct the null tetrad based on the Kennersley tetrad for Kerr by applying the following replacement (apply the first replacement and then the second one):
\begin{eqnarray}
    2Mr \to r^2 + a^2 - A_5\,, \qquad r \to \sqrt{r^2 + f}\,.
\end{eqnarray}
We then compute the Weyl scalars $\Psi_i$. We found that, in general, Eq.~\eqref{eq:IJ} is not satisfied. This means that the second restricted class in Eq.~\eqref{eq:restricted_2} and the general refined metric in Eq.~\eqref{eq:refined_CY} are type I.  This further means that the Kerr-Sen BH is type I.

\section{Black Hole Observations}
\label{sec:observables}             
We now explore how the refined parameterized Kerr spacetime modifies observables from the Kerr case. In particular, we study the QNM ringdown frequencies and the shape of BH shadows.

\subsection{Gravitational-wave Ringdown}
\label{sec:ringdown}

Following~\cite{Silva:2019scu,Glampedakis:2017dvb,Glampedakis:2019dqh}, we estimate the real and imaginary part of QNM ringdown frequencies $\omega_R$ and $\omega_I$ for the refined parameterized Kerr spacetime through the post-Kerr formalism. Such a formalism makes use of the eikonal approximation, in which $\omega_R$ and $\omega_I$ are associated with the angular frequency $\Omega_0$ and the Lyapunov exponent $\gamma_0$ (corresponding to the divergence rate of photon orbits grazing the light ring) at the light ring $r_0$ as
\ba
\label{eq:damp_freq_1}  \omega_R &=& 2\Omega_0 = 2(\Omega_K + \delta \Omega_0), \\
\label{eq:damp_freq_2}  \omega_I &=& -\frac{1}{2}|\gamma_0| = -\frac{1}{2}|\gamma_K + \delta \gamma_0|.
\ea
Here
\be
\Omega_K = \pm \frac{M^{1/2}}{r_K^{3/2} \pm a M^{1/2}}
\ee
is the angular frequency of the Kerr light ring at
\be
r_K = 2M\left\{1+\cos{\left[\frac{2}{3} \cos^{-1}{\left(\mp \frac{a}{M}\right)}\right]}\right\},
\ee
where the upper (lower) sign corresponds to prograde (retrograde) orbit,
while
\ba
\delta \Omega_0 = & \mp & \left(\frac{M}{r_K}\right)^{1/2} \left[h_{\phi\phi} \pm \left(\frac{r_K}{M}\right)^{1/2} (r_K +3M) h_{t\phi} \right. \nonumber \\
&+& \left. \left(3r_K^2 + a^2\right) h_{tt}\right] / \left[(r_K - M) \left(3r_K^2 + a^2\right)\right] \nonumber \\
\ea
is the correction to $\Omega_K$ with $h_{\mu\nu}$ representing the metric deviation away from Kerr on the equatorial plane at $\theta=\pi/2$. Moreover, 
\be
\gamma_K = 2\sqrt{3M} \frac{\Delta_K \Omega_K}{r_K^{3/2}(r_K - M)}
\ee
is the Lyapunov exponent for Kerr with $\Delta_K = r_K^2 - 2Mr_K + a^2$, while $\delta\gamma_0$ is the non-Kerr correction given in Eq. (18) of~\cite{Glampedakis:2017dvb}.

\begin{figure*}[t]
\includegraphics[width=8.5cm]{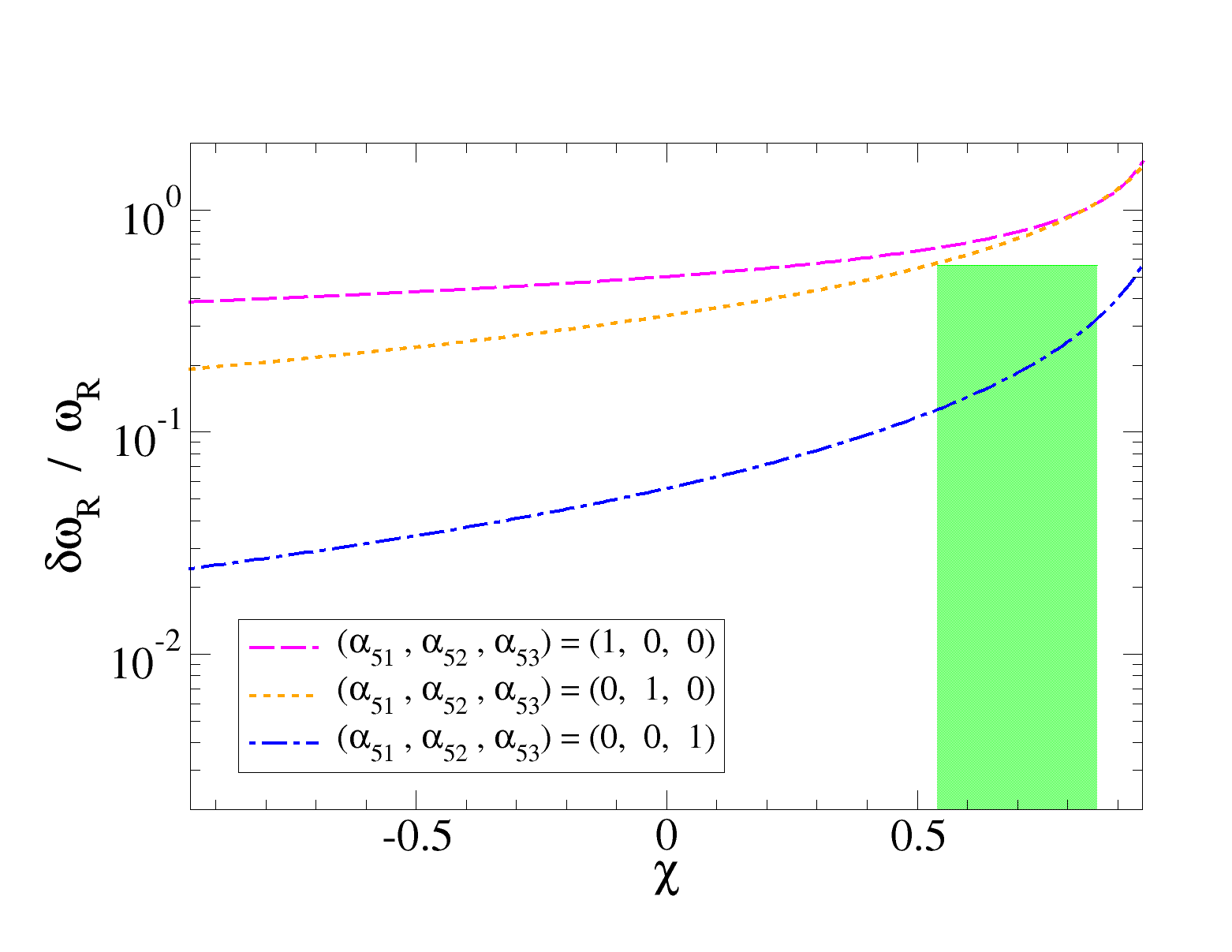}
\includegraphics[width=8.5cm]{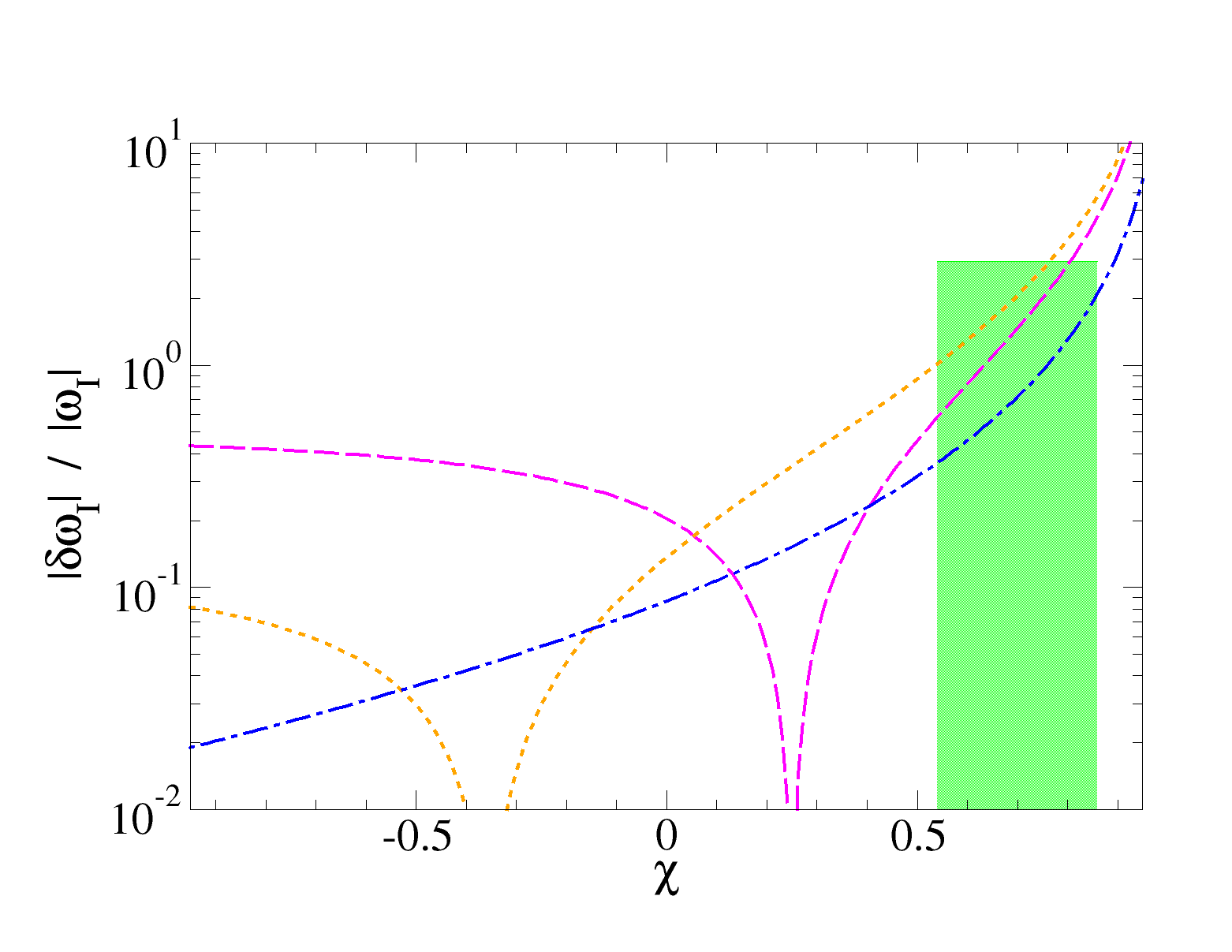}
\caption{\label{fig:qnm_dev_1}
(Left) The fractional difference $\delta\omega_R/\omega_R$ for three different parameter combinations of restricted class I.
 (Right) Similar to the left panel but for the absolute value of the imaginary components. We used the expressions that fully account for the BH spin and did not work in the small spin approximation. We also present the bound from GW$200129\_065458$~\cite{LIGOScientific:2021sio} (green shade). 
}
\end{figure*}

We next apply the refined CY metric with the specific investigation in the refined classes outlined in Eqs.~\eqref{eq:restricted_1} and~\eqref{eq:restricted_2}. We can define small-deviation functions like
\begin{equation}
    A_i = A_i^\mathrm{Kerr} + \varepsilon \delta A_i\,, \quad f = \varepsilon \delta f\,,
\end{equation}
where $\varepsilon$ is a  book-keeping parameter to count the order of deviation from Kerr. Notice in the case of the restricted classes, only $A_5$ and $f$ are arbitrary deviation functions, and we can truncate their respective series at infinity for convenience.  
Assuming deviations from Kerr are small and expanding the restricted metric on $\epsilon$ about 0, keeping to $\mathcal{O}(\epsilon)$ yields
\be
g_{\mu\nu} = g_{\mu\nu}^\mathrm{Kerr} + \epsilon h_{\mu\nu},
\ee
where 
\allowdisplaybreaks
\ba
\label{eq:htt_PK}
h_{tt} &=& -\frac{M}{r} \left[\alpha_{51} -\epsilon_1+ (\alpha_{52}+2\epsilon_1 - \epsilon_{2})\frac{M}{r} 
\right. \nonumber  \\
& & \left. + (\alpha_{53} + 2\epsilon_2 - \epsilon_3)\frac{M^2}{r^2} +\mathcal{O}\left( \frac{M^3}{r^3} \right) \right]\,, \\  
h_{t\phi} &=& \frac{aM}{r} \left[\alpha_{51} -\epsilon_1+ \left(\alpha_{52} + 2\epsilon_1 - \epsilon_2 \right)\frac{M}{r} 
\right. \nonumber  \\
& & \left. + \left(\alpha_{53} + 2\epsilon_2 - \epsilon_3\right) \frac{M^2}{r^2} + \mathcal{O}\left( \frac{M^3}{r^3} \right)\right],\\
h_{rr} &=& -\frac{Mr^3}{\Delta^2} \left[\alpha_{51} -\epsilon_1 + (\alpha_{52} + 2\epsilon_1 - \epsilon_2)\frac{M}{r} 
\right. \nonumber  \\
& & \left. + (\alpha_{53} - \epsilon_1 \chi^2 - 2 \epsilon_2 - \epsilon_3)\frac{M^2}{r^2} + \mathcal{O}\left( \frac{M^3}{r^3} \right) \right], \nonumber \\
\\
h_{\theta\theta} &=& Mr \left[\epsilon_1 + \epsilon_2 \frac{M}{r} + \epsilon_3 \frac{M^2}{r^2} + \mathcal{O} \left(\frac{M^3}{r^3} \right)\right],\\
\label{eq:hphph_PK}
h_{\phi\phi} &=& Mr \left[\epsilon_1 + \epsilon_2 \frac{M}{r} - (\alpha_{51} \chi^2 - \epsilon_1 \chi^2 - \epsilon_3) \frac{M^2}{r^2} 
\right. \nonumber  \\
& & \left. + \mathcal{O} \left(\frac{M^3}{r^3} \right)\right],
\ea
with $\chi = a/M$.
Similarly, if we expand Eqs.~\eqref{eq:damp_freq_1} and~\eqref{eq:damp_freq_2} on $\varepsilon$ around 0 and keep to first order in $\varepsilon$, the QNM corrections in the refined parameterized spacetime to first order in each deviation parameter, and quadratic in spin are given by 
\bw
\ba
\label{eq:omegaR}
\omega_R &=& \omega_R^\mathrm{Kerr} + \frac{\epsilon}{M} \left(\frac{27\alpha_{51} + 9\alpha_{52} + 3\alpha_{53} - 18\epsilon_{1} - 6\epsilon_{2} - 4\epsilon_{3}}{81\sqrt{3}} + \frac{2(54\alpha_{51} + 27\alpha_{52} + 12\alpha_{53} - 27\epsilon_{1} - 15\epsilon_{2} - 17\epsilon_{3})}{729}\chi \right. \nonumber \\ 
&& \left. + \frac{297\alpha_{51} + 177\alpha_{52} + 95\alpha_{53} - 126\epsilon_{1} - 84\epsilon_{2} - 142\epsilon_{3}}{1458\sqrt{3}}\chi^2 \right) + \mathcal{O}(\epsilon^2,\chi^3),\\
\label{eq:omegaI}
\omega_I &=& \omega_I^\mathrm{Kerr} + \frac{\epsilon}{M} \left(\frac{33\alpha_{51} - 11\alpha_{52} - 14\alpha_{53} - 64\epsilon_{1} - 20\epsilon_{2} + 14\epsilon_{3}}{972\sqrt{3}} - \frac{240\alpha_{51} + 122\alpha_{52} + 74\alpha_{53} - 14\epsilon_{1} + 14\epsilon_{2} - 83\epsilon_{3}}{4374}\chi
\right. \nonumber \\ 
&& \left. - 
\frac{3162\alpha_{51} + 1712\alpha_{52} + 1031\alpha_{53} - 350\epsilon_{1} - 13\epsilon_{2} - 1250\epsilon_{3}}{26244\sqrt{3}}\chi^2 \right)+ \mathcal{O}(\epsilon^2,\chi^3).
\ea
\ew
Notice that one can easily reduce to restricted class I by $\epsilon_i = 0$.

\begin{figure*}[t]
\includegraphics[width=8.5cm]{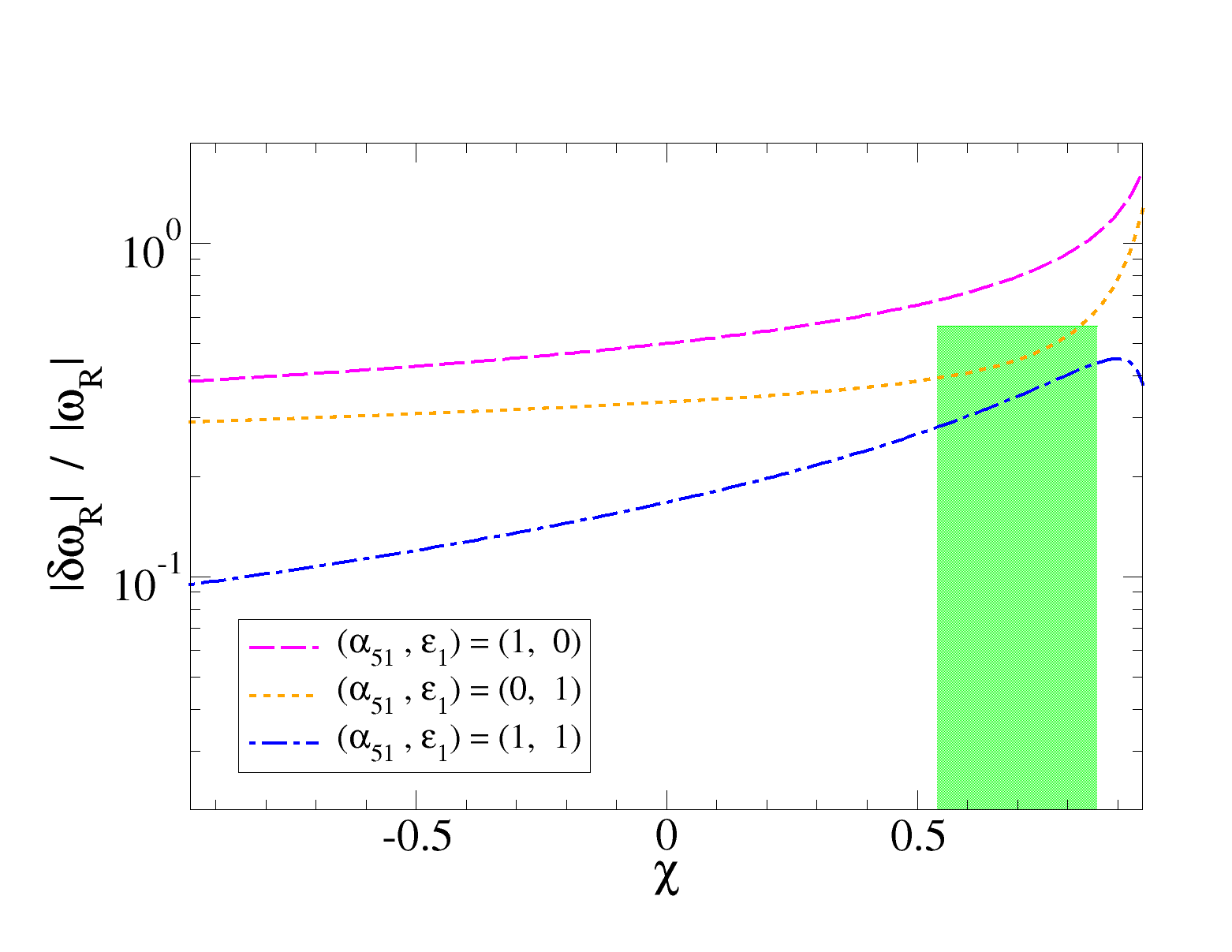}
\includegraphics[width=8.5cm]{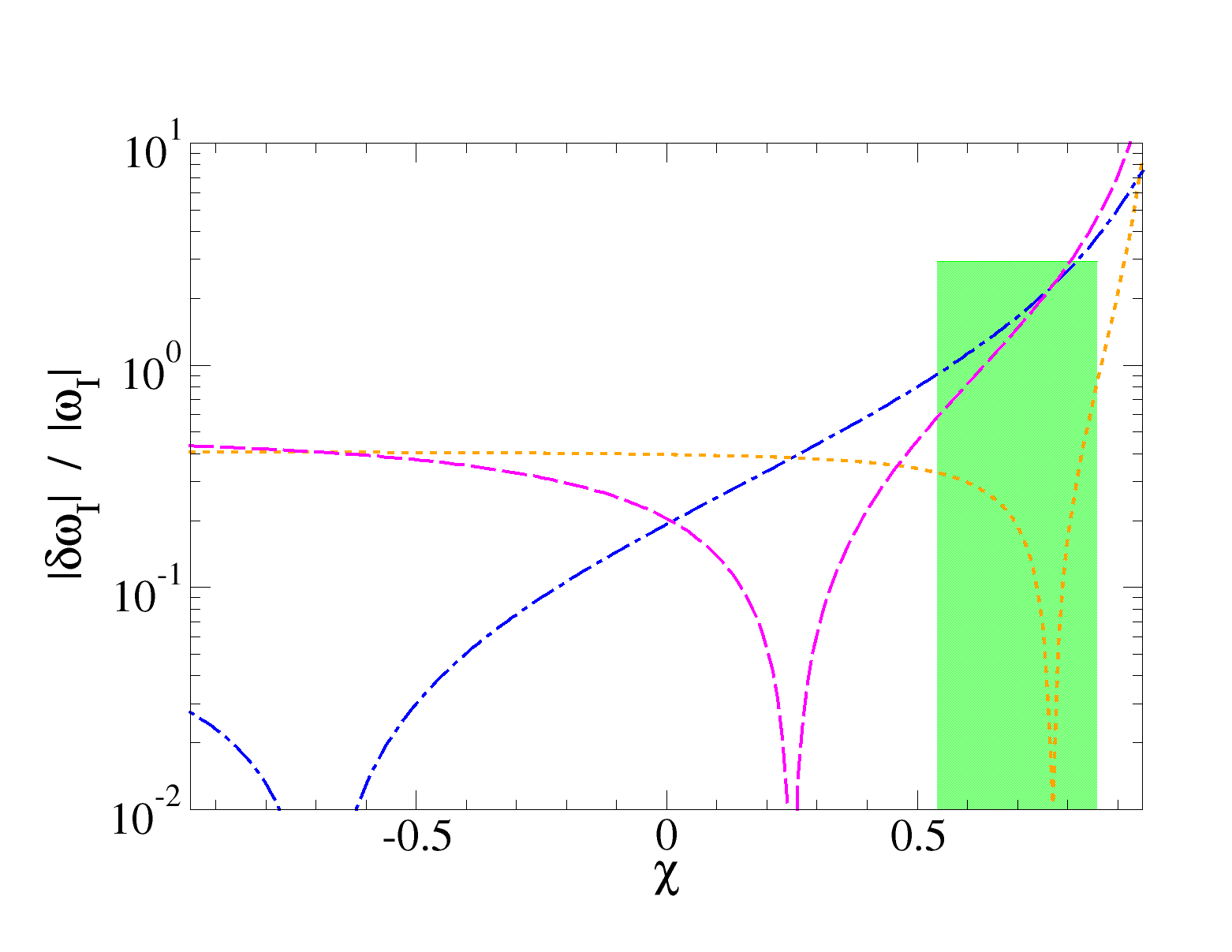}
\caption{\label{fig:qnm_dev_2}
Similar to Fig.~\ref{fig:qnm_dev_1} but for restricted class II. 
}
\end{figure*}

Figure~\ref{fig:qnm_dev_1} shows the fractional difference between the corrections 
$\delta\omega_R$ made by restricted class I  for three example parameter combinations: $(\alpha_{51},\alpha_{52},\alpha_{53})=(1,0,0),(0,1,0),(0,0,1)$\footnote{As mentioned below Eq.~\eqref{eq:gtph_asymp}, $\alpha_{51}$ can be absorbed into the definition of the BH mass and spin, though here, we consider the effect of this parameter as it is non-vanishing in some of the example metrics in Appendix~\ref{app:mapping}.}. Unlike Eqs.~\eqref{eq:omegaR} and~\eqref{eq:omegaI} that are obtained under the expansion about $\chi = 0$, for our calculations, we used the expressions for $\delta \omega_R$ and $\delta \omega_I$ that are valid to full order in $\chi$ (though we do not present such expressions here as they are too lengthy). For the real frequency (left), notice that the deviation becomes larger as we turn on deviation parameters entering at lower orders. For the imaginary frequency (right), we see some of the curves cross zero as we increase $\chi$. We compare these results with the upper bound found from GW$200129\_065458$ that puts the most stringent bound out of all the gravitational-wave events found so far~\cite{LIGOScientific:2021sio}. The allowed parameter space from this event is shown by the green shaded region\footnote{The range of the spin is taken from Table XIII of~\cite{LIGOScientific:2021sio}.}. Notice that curves for the parameter combination of (1,0,0) and (0,1,0) go outside of the allowed region in the real frequency plot and are thus ruled out, while there are other combinations that are still viable. For the imaginary part, the constraint is slightly weaker, but the analytic results from the post-Kerr formulation may not be accurate as we comment later.

Figure~\ref{fig:qnm_dev_2} is similar to Fig.~\ref{fig:qnm_dev_1} but for restricted class II. Motivated by the Kerr-Sen BH in Appendix~\ref{sec:KS}, we vary $(\alpha_{51},\epsilon_1)$. The fractional deviation for the parameter combination of $(\alpha_{51},\epsilon_1)=(1,1)$ is a linear combination of that for (1,0) and (0,1). For the real frequency (left), the one for (0,1) is negative and thus the deviation for (1,1) is the smallest. A similar behavior is seen for the imaginary frequency where e.g. the curve for (1,1) goes to 0 when the curves for (1,0) and (0,1) cross. Comparing these with the bound from GW$200129\_065458$, we see that the parameter choice of (1,0) is ruled out from the real frequency bound while there are other combinations that are still allowed.

\begin{figure}[t]
\includegraphics[width=8.5cm]{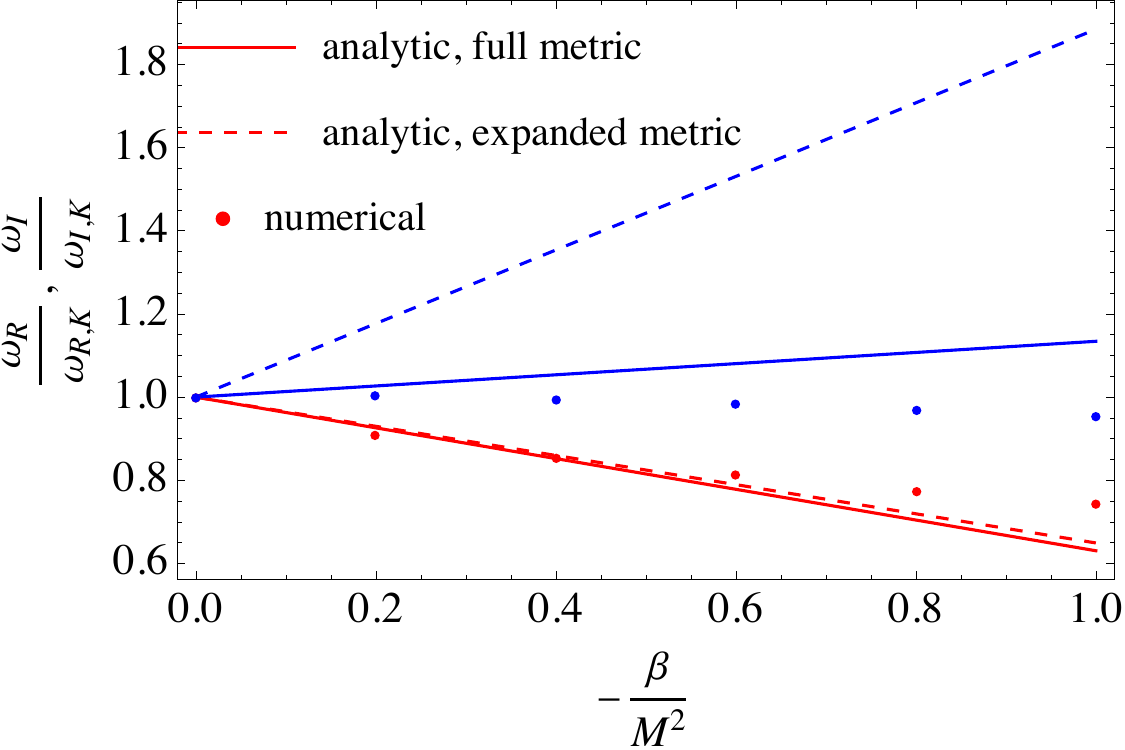}
\caption{\label{fig:brane_QNM}
Real (red) and imaginary (blue) frequencies (normalized by the Kerr values) for braneworld BHs as a function of the dimensionless tidal charge with the dimensionless spin value of $\chi = 0.67$. We compare our analytic estimate via the post-Kerr formulation with numerical results in~\cite{Mishra:2021waw}. For the former, we use the full braneworld BH metric (solid lines) and the series expanded one in Eqs.~\eqref{eq:htt_PK}--\eqref{eq:hphph_PK} (dashed lines). Observe that the analytic estimate is in good agreement with the numerical one for the real frequency when $|\beta|$ is small while the former is not an accurate approximation to the latter for the imaginary frequency. 
}
\end{figure}

We now study the validity of the post-Kerr formalism by taking the braneworld BH as an example, whose QNM frequencies have been computed numerically in~\cite{Mishra:2021waw}. Figure~\ref{fig:brane_QNM} compares the normalized analytic QNM frequencies within the post-Kerr formalism against the numerical ones. For the former, we show the results using the exact braneworld BH metric as well as using the series-expanded metric in Eqs.~\eqref{eq:htt_PK}--\eqref{eq:hphph_PK} with all the deviation parameters set to 0 except for $\alpha_{52} = \beta/M^2$. For the real frequency, we see that the analytic estimates are qualitatively in agreement with the numerical ones, especially when the dimensionless tidal charge $|\beta|$ is small. This is expected as we work in the small-deviation approximation for the post-Kerr formalism. For the imaginary frequency, on the other hand, the agreement between the analytic and numerical estimates is rather poor\footnote{We have checked that our post-Kerr calculation can correctly reproduce Fig.~4 of~\cite{Glampedakis:2017dvb} for the imaginary QNM frequency for the Johannsen metric~\cite{Johannsen:2015pca} and can produce results that are qualitatively in agreement with the imaginary QNM frequency for the axial perturbation in Einstein-dilaton Gauss-Bonnet gravity~\cite{Bryant:2021xdh}.}. In particular, the deviation from Kerr is positive (mostly negative) for the analytic (numerical) estimates (see Appendix~\ref{app:QNM_brane_imaginary} for further details). This suggests that, while the real QNM frequencies for the parameterized Kerr spacetime found via the post-Kerr formalism may be reliable, the imaginary QNM frequency results may not be accurate.

\subsection{Black Hole Shadows}

Let us now compute the photon rings, or the shape of BH shadows, for the refined parameterized BH (the shape of BH shadows for parameterized BHs and some of the example spacetimes in Appendix~\ref{app:mapping} have been studied in e.g.~\cite{Johannsen:2013vgc,Carson:2020dez,Wei:2013kza,Atamurotov:2013sca,Ghasemi-Nodehi:2015raa,Abdujabbarov:2016hnw,Younsi:2016azx,Mizuno:2018lxz,Psaltis:2020ctj,Narang:2020bgo,Konoplya:2021slg,Kumar:2020hgm,Zubair:2023cor,Tsukamoto:2014tja,Tsukamoto:2017fxq,Shaikh:2019fpu}). We begin by following ~\cite{Johannsen:2013vgc,Carson:2020dez}, starting with the Hamilton-Jacobi function and Hamilton-Jacobi equations respectively
\allowdisplaybreaks
\ba
    S &\equiv& -\frac{1}{2}\mu\tau - Et + L_z\phi + S_r(r) + S_\theta(\theta)\,, \\
    -\frac{\partial S}{\partial \tau}&=&\frac{1}{2} g^\mathrm{\alpha \beta} \frac{\partial S}{\partial x^\alpha} \frac{\partial S}{\partial x^\beta}\,,
\ea
for particle mass $\mu$, proper time $\tau$, orbital energy E, angular momentum $L_z$, and generalized radial and polar functions $S_r(r)$ and $S_\theta(\theta)$, to yield
\ba
   && -\mu ^2 \left(a^2 \cos ^2\theta+f+g+r^2\right) \nonumber \\ &=& \frac{1}{\Delta}\left[2 a^3 A_0 E L_z - \frac{\Delta}{A_5} \left(a^2 A_2^2 L_z^2 +A_1^2 E^2\right) \right. \nonumber \\
    && \left. +a^2 A_4^2 \Delta E^2 \sin ^2\theta -2 a A_0 \Delta E L_z + 2a A_0 E L_z r^2 \right. \nonumber \\
    && \left. + A_3^2 \Delta L_z^2 \csc ^2\theta +A_5\Delta \left(\frac{\partial S_r}{\partial r}\right)^2 + \Delta\left(\frac{\partial S_\theta}{\partial\theta}\right)^2 \right]. \nonumber \\
\ea
It is now helpful to separate the Hamilton-Jacobi equations using the separation constant
\begin{align}
\allowdisplaybreaks
C =& -\mu ^2 \left(f + r^2 \right)  - \frac{1}{A_5}\left[-a^2 L_z^2 A_2^2 + 2aE L_z A_0  \right. \nonumber \\
& \left. - E^2 A_1^2 + A_5^2 \left(\frac{\partial S_r}{\partial r}\right)^2 \right]\,,
\\
C =& a^2 E^2 \sin ^2\theta +\mu ^2 \left(g + a^2 \cos ^2\theta \right)-2 a E L_z \nonumber \\
&+L_z^2 \csc ^2\theta+\left(\frac{\partial S_\theta}{\partial\theta}\right)^2 .
\end{align}
If one defines the Carter-like constant of motion $Q \equiv C-(L_z-aE)^2$, a solution for $S_r(r)$ takes the form
\ba
S_r(r) &=& \pm \int dr  \sqrt{\frac{R(r)}{\Delta A_5(r)}}, \\
R(r)&\equiv& \frac{\Delta}{A_5}
[a^2  L_z^2 A_2^2 \nonumber \\
&&- 2 a  E L_z A_0 + E^2 A_1^2 ] - a^2 E^2 + 2 a  E L_z \nonumber \\
&& -  f\mu ^2 -  L_z^2 - Q - \mu ^2 r^2,
\ea
where a difference in sign represents particles with prograde and retrograde motion. 

At this point, we can compute the generalized momenta $p_\alpha$ utilizing the expression
\be
p_\alpha = \frac{\partial S}{\partial x^\alpha},
\ee
where we are particularly interested in the radial momenta given in both covariant and contravariant form: 
\be
\label{eq:pr}
p_r = \pm \sqrt{\frac{R(r)}{\Delta A_5(r)}}, \quad
p^r = \pm \frac{1}{\tilde{\Sigma}} \sqrt{\frac{A_5(r) R(r)}{\Delta }}.
\ee
Following~\cite{Johannsen:2013vgc}, we now turn our attention to the impact parameters $x$ and $y$~\cite{Bardeen_1973}
\ba
x= -\frac{\xi}{\sin{\iota}}, \qquad y= \pm \sqrt{\eta + a^2 \cos^2{\iota} - \xi^2\cot^2{\iota}}\,, \nonumber \\
\ea
which describe the image plane from an observer's point of view at infinity and where $\iota$ is the inclination angle, $\xi \equiv L_z/E$, and $\eta \equiv Q/E^2$. Notice the new invariant parameters $\xi$ and $\eta$ have been constructed entirely out of constants of motion.

The photon rings of interest are described by the solutions to these new parameters $\xi$ and $\eta$. As these constants of motion are conserved along null geodesics, they can be solved in the special case of circular orbits for simplicity. Here, the radial photon momentum $p^r$ found in Eq.~\eqref{eq:pr}, as well as its radial derivative, must vanish. Because $\tilde{\Sigma}$ and $A_5(r)$ are both non-negative (at least when deviations from Kerr are small), this results in the system of equations
\ba
R(r)=0, \qquad \frac{dR(r)}{dr}=0,
\ea
with the full re-parameterized expression for $R(r)$ of an orbiting photon $(\mu=0)$ given by
\ba
R(r)=&&\frac{1}{A_5}
\left[A_1^2(r)\Delta - 2a A_0(r)\Delta\xi + a^2 A_2^2(r)\Delta\xi^2 \right] \nonumber \\
&&- a^2 \Delta + 2 a \Delta \xi - \Delta \eta - \Delta \xi^2.
\ea

\begin{figure*}[t]
\includegraphics[width=8.5cm]{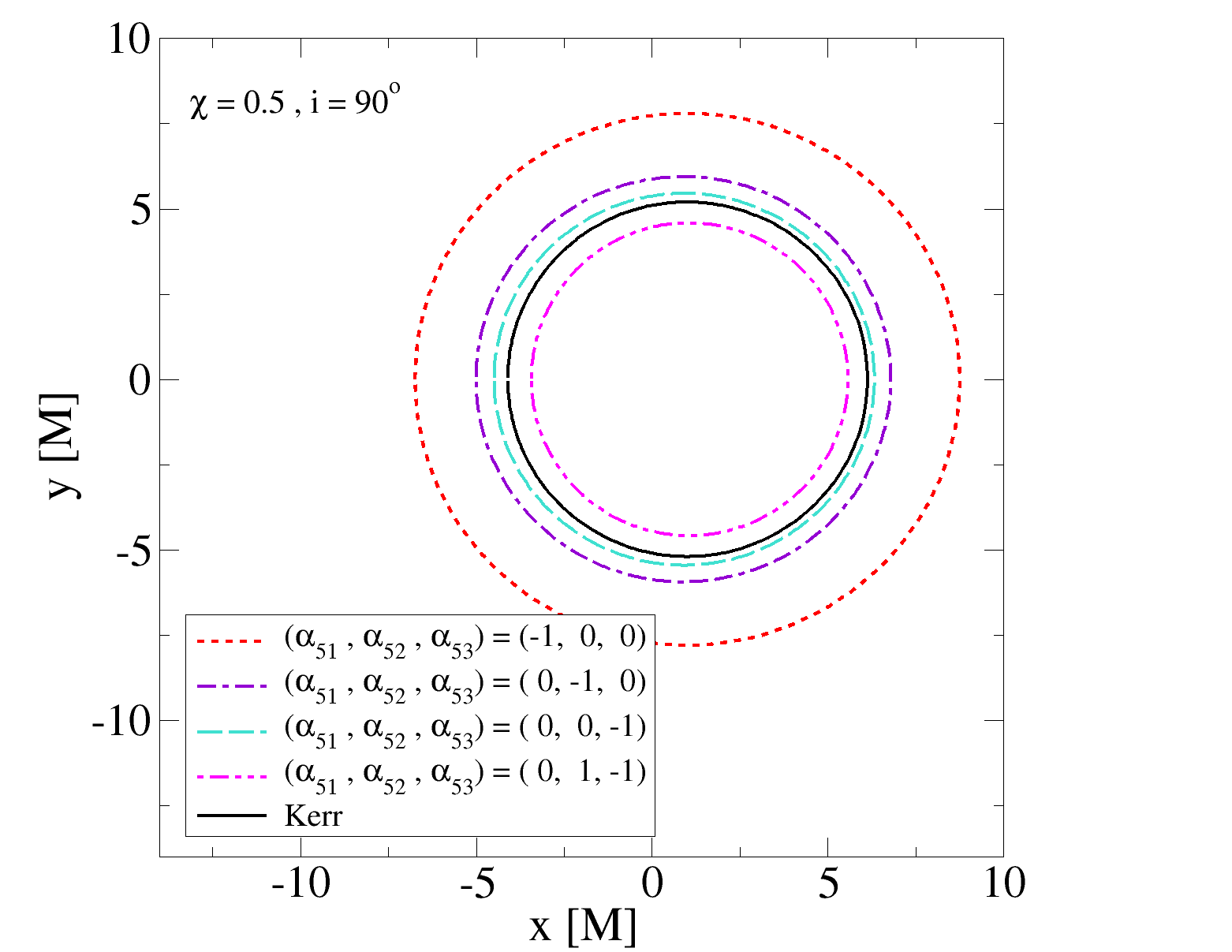}
\includegraphics[width=8.5cm]{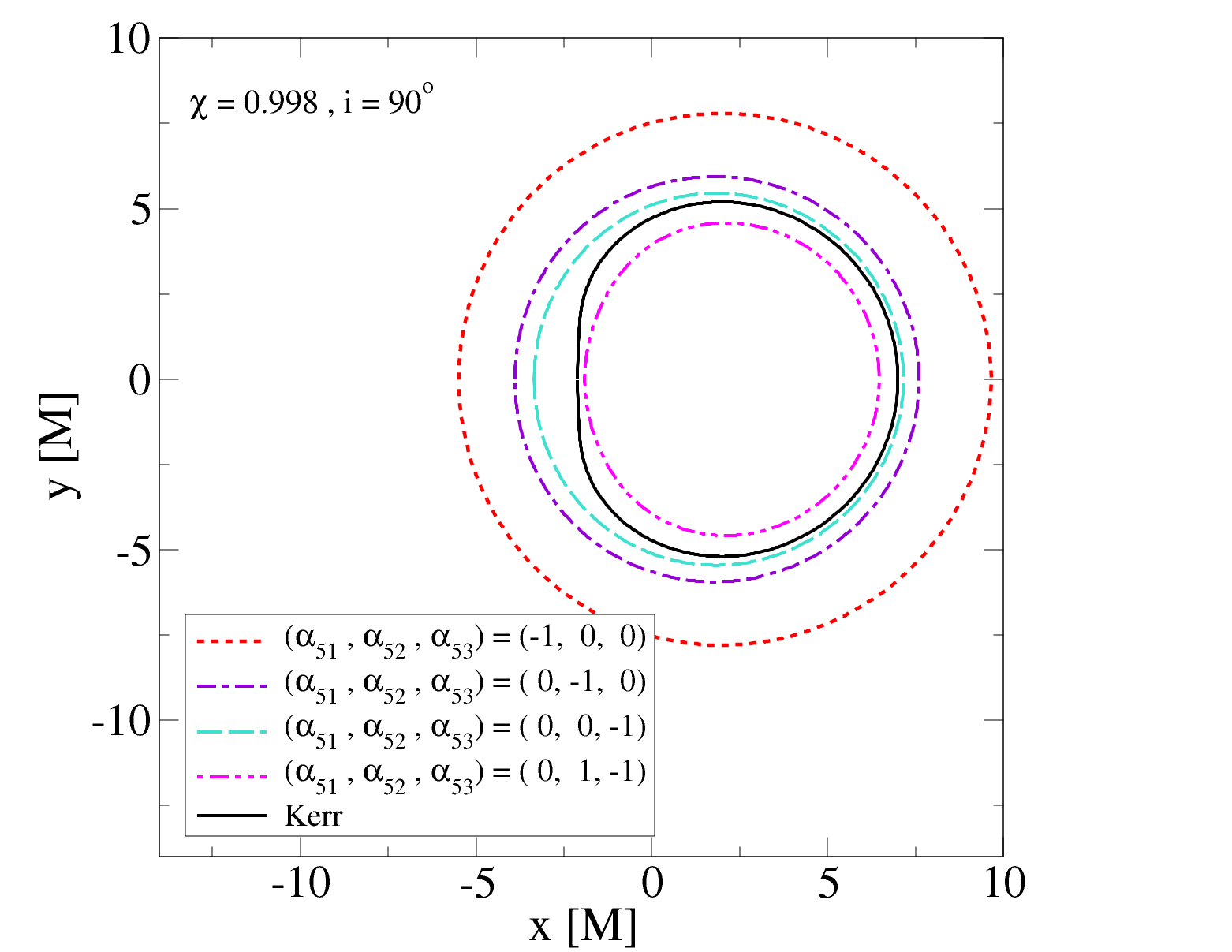}
\includegraphics[width=8.5cm]{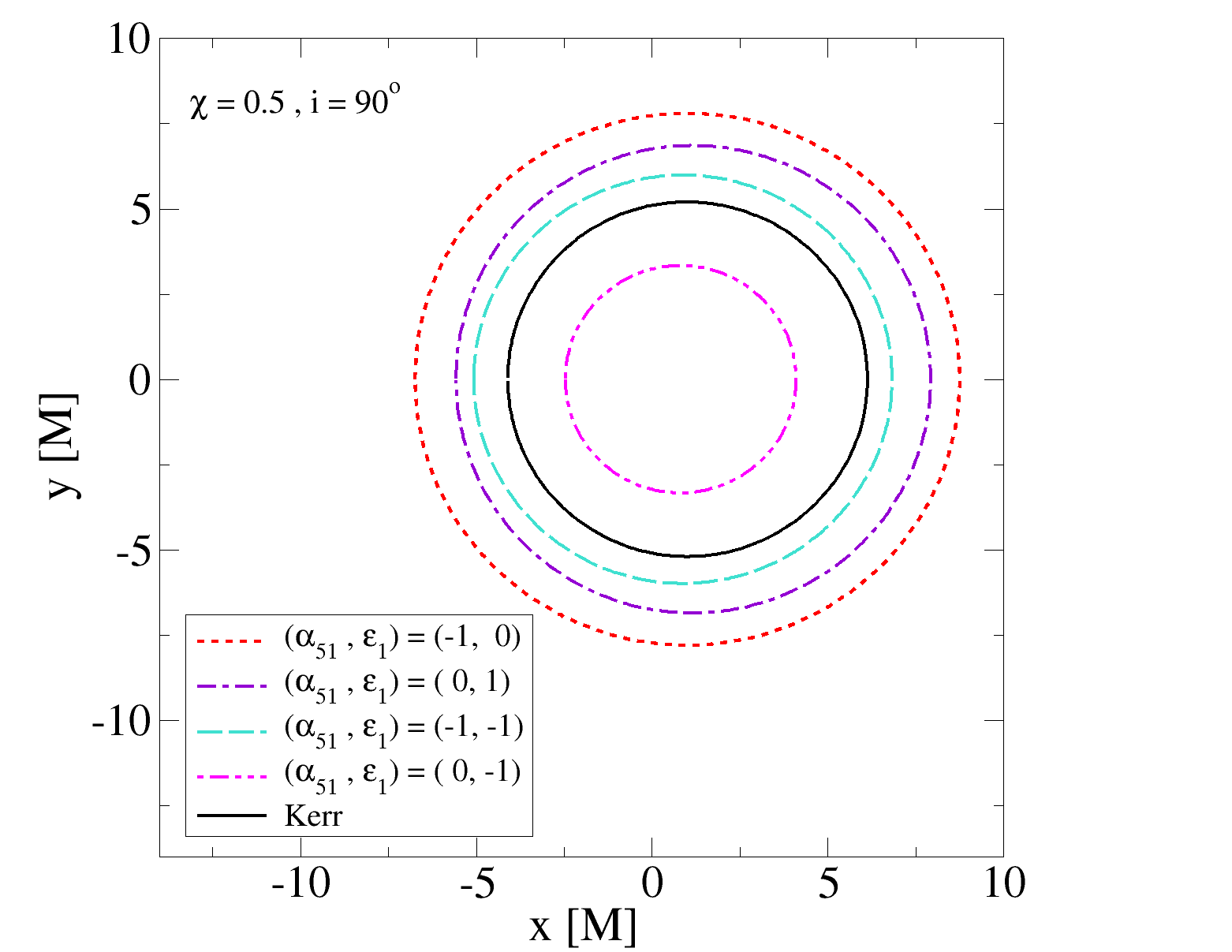}
\includegraphics[width=8.5cm]{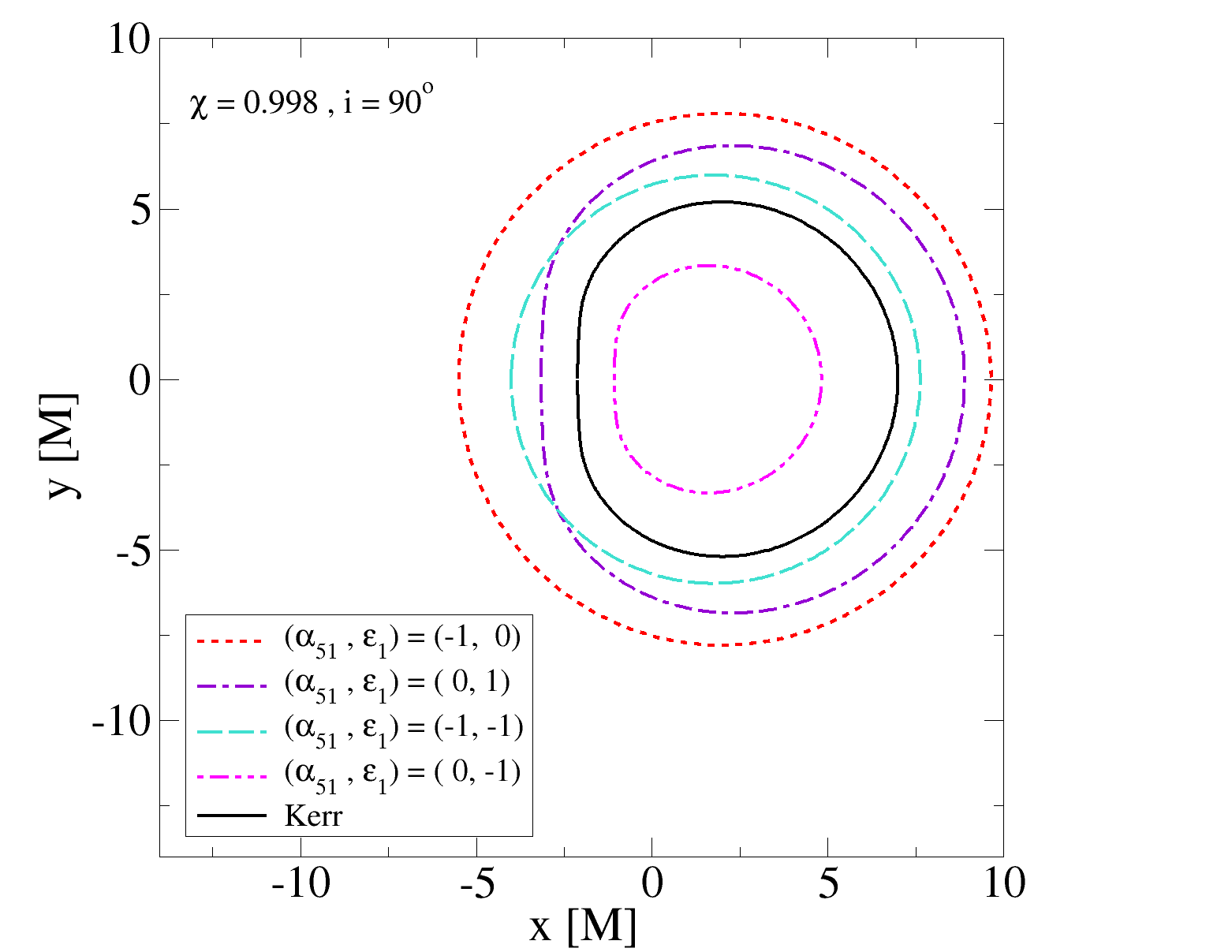}
\caption{\label{fig:bh_shadows}
Shapes of BH shadows (photon rings) about a medium spin $\chi=0.5$ (left) and a high spin $\chi=0.998$ (right), plotted for example parameter combinations corresponding to restricted class I (top) and restricted class II (bottom) of the refined parameterized spacetime. The inclination is fixed at the extreme case of $i=90^\circ$ in every scenario for demonstration purposes.}
\end{figure*}

Figure~\ref{fig:bh_shadows} shows the shape of BH shadows for the refined parameterized Kerr spacetime with several parameter combinations for two spin choices: $\chi$ = 0.5 and 0.998. For restricted class I (top panels), we have kept the series in $A_5/r^2$ in Eq.~\eqref{eq:A5_expansion} up to $\mathcal{O}(M^3/r^3)$ and chosen combinations: $(\alpha_{51},\alpha_{52},\alpha_{53})= (-1, 0, 0), (0, -1, 0), (0, 0, -1), (0, 1, -1)$, together with the Kerr case. Notice we have chosen parameters to mostly 0 or $-1$. This is because the spacetime becomes a naked singularity for certain combinations of $\alpha_{5i}$. For example, if we keep only $\alpha_{51}$ and $\alpha_{52}$, the event horizon exists only when $|\chi| \leq  \sqrt{1-\alpha_{51}+(\alpha_{51}^2/4)-\alpha_{52}}$, so for example, $|\chi| \leq 0.5$ when $(\alpha_{51},\alpha_{52},\alpha_{53})= (1, 0, 0)$ or (0,1,0). 
Of those plotted, we observe a pattern where singularly attributing $-1$ to each sequential parameter increases the radius of the shadow, with $\alpha_{51}\neq 0$ ($\alpha_{53}\neq 0$) giving the largest (smallest). This is simply due to lower order terms in the expansion of $A_5$ at infinity giving larger contributions as expected. Additionally, the combination that contains the positive parameter, $(\alpha_{51},\alpha_{52},\alpha_{53})=(0, 1, -1)$, produces a curve smaller than Kerr, implying a reductive nature to a positive parameter choice, with $\alpha_{52}$ making a larger contribution than $\alpha_{53}$. 
Notice also that, unlike the case of a Kerr BH where the shape of the shadows distort as the spin gets closer to the Kerr maximal value of 1, the shadow shapes of the parameterized Kerr spacetime remain mostly spherical. This is because $\chi = 1$ does not correspond to the extremal limit anymore when $\alpha_{5i} \neq 0$.

For restricted class II (bottom panels), we focus on varying $\alpha_{51}$ and $\epsilon_1$ only and have chosen combinations $(\alpha_{51},\epsilon_1)=(-1,0), (0,1), (-1,-1),(0,-1)$. We notice the parameter $\epsilon_1$ increases the size of the image with positive choices (and shrinks the image with negative choices), which is the opposite of what happens when varying $\alpha_{51}$. Moreover, allowing $\alpha_{51}$ and $\epsilon_1$ to exist simultaneously reveals contributions from $\alpha_{51}$ are larger than those from $\epsilon_1$. Furthermore, $\epsilon_1$ retains the shape of Kerr and only deviates in size, as opposed to $\alpha_{51}$ (and $\alpha_{52},$ and $\alpha_{53}$ for class I) which contributes to a deviation in both size and shape. Namely, for $(\alpha_{51},\epsilon_1) = (0,\pm1)$, the shape of the shadows distort like Kerr for a nearly extremal spin because the event horizon location is not modified and $|\chi|=1$ remains to be the extremal spin. For both classes, as concluded previously in~\cite{Carson:2020dez}, deviations from Kerr are easily distinguishable from the exact Kerr result when the deviation parameters are sufficiently large.

\section{Conclusions}

\label{sec:conclusion}

In this paper, we first pointed out that the parameterized rotating BH spacetime with Kerr symmetries constructed by Johansenn~\cite{Johannsen:2015pca} and Carson and Yagi~\cite{Carson:2020dez} may suffer from unphysical divergence if the asymptotic series in the asymptotic behavior of arbitrary functions are kept only to finite orders. 
We then showed that a simple redefinition of the arbitrary functions in the metric can cure the pathology, at least for all the example metrics studied in this paper. 
We further proposed two restricted classes of the refined parameterized Kerr metric that only contain one or two arbitrary functions and can capture several known BH solutions in theories beyond GR. We studied the Petrov type of the refined parameterized Kerr spacetime and found that it is type I, including the second restricted class, while it is type D for the first restricted class. We also computed some observables, namely the ringdown frequencies and the shape of BH shadows, and showed how they deviate from the Kerr case. 

There are a few different avenues for future work. It would be interesting to constrain the parameterized Kerr spacetime through BH shadow observations, similar to e.g.~\cite{Vagnozzi:2022moj} that carried out a systematic survey of constraining non-spinning parameterized BH spacetimes. Regarding constraints from ringdown, we compared the ringdown frequencies for the parameterized Kerr spacetime for several example parameter combinations against the bound from GW$200129\_065458$~\cite{LIGOScientific:2021sio}. It would be important to carry out a more detailed survey to identify the regions in the deviation parameter space that can be ruled out from current gravitational-wave observations. One could also use the inspiral information of gravitational waves to constrain parameterized Kerr spacetimes~\cite{Cardenas-Avendano:2019zxd,Carson:2020iik,Shashank:2021giy}. For example, the leading post-Netwonian correction to the gravitational waveform arises from the modification to the gravitational potential, which can be read off from the $(t,t)$ component of the metric. One can follow calculations in~\cite{Carson:2020iik} to derive such corrections for the parameterized Kerr spacetime presented here. One can further account for corrections to the ringdown frequencies and carry out a parameter estimation study as in~\cite{Carson:2020iik} to derive constraints from existing and future gravitational-wave observations. On the other hand, it sounds challenging to find corrections during the merger phase as it requires one to carry out numerical simulations but there are no field equations that parameterized Kerr metrics follow. It would also be interesting to derive gravitational waveforms for extreme mass ratio inspirals where a small compact object orbits the parameterized Kerr spacetime~\cite{Kumar:2023bdf} and study prospects of probing such spacetime with future gravitational-wave observations.

\begin{acknowledgements}
K.Y. and S.L. acknowledge support from NASA Grant No. 80NSSC20M0056 through the Virginia Space Grant Consortium. K.Y. also acknowledges support from a Sloan Foundation Research Fellowship and the Owens Family Foundation.
\end{acknowledgements}

\appendix

\section{Mapping Functions}
\label{app:mapping}

\bw

In this appendix, we present the mapping functions for the radial functions in the parameterized Kerr metric for several example BH solutions\footnote{There are other rotating BH solutions that can be mapped to the refined parameterized metric presented in this paper. Extended Kerr BH in~\cite{Contreras:2021yxe} is one example. However, we do not include it in this appendix since one of the radial functions cannot be expanded in polynomials as in Eq.~\eqref{eq:A5_expansion} because of the exponential dependence.}. $g(\theta) = 0$ for all the examples studied here.  For each example spacetime, we provide (i) the mapping with the original CY metric in Eq.~\eqref{eq:CY_metric_original}, (ii) the mapping with the refined, rescaled metric in Eq.~\eqref{eq:refined_CY}, (iii) non-vanishing Taylor coefficients in Eqs.~\eqref{eq:Ai_exp}--\eqref{eq:g_exp}, and (iv) the mapping with the refined metric under the small deviation approximation in Eq.~\eqref{eq:small dev approx Ai} for $n=0$ (one can easily find the mapping for other $n$ by simply multiplying each function by $\Delta^n$). 

\subsection{Braneworld}

Metrics for BHs in braneworld follow the same form as the Kerr-Newman solution with the electric charge $Q$ with the replacement $Q^2 = \beta$ (a tidal charge that can take either sign)~\cite{Aliev:2005bi}.

\begin{enumerate}
    \item original CY mapping

\begin{align}
(\calA_0,\calA_1,\calA_2,\calA_5,f)=\left( \frac{\Delta}{\Delta + \beta}, \frac{\Delta}{\Delta + \beta}, \frac{\Delta}{\Delta + \beta}, 
 \frac{\Delta + \beta}{\Delta} ,0\right)\,,
\end{align}

    \item refined mapping (rescaling)

\begin{equation}
(A_0,A_1,A_2,A_5,f)=\left(r^2+a^2,\; r^2+a^2,\; 1,\; \Delta +\beta,\; 0 \right)\,,
\end{equation}

    \item Taylor coefficients

\begin{equation}
    \alpha_{52}=\frac{\beta}{M^2}\,,
\end{equation}

    \item refined mapping (small deviation, $n=0$)

\begin{equation}
\label{eq:deltaA_brane}
    (\delta A_0,\delta A_1,\delta A_2,\delta A_5,\delta f)=\left( \frac{\beta}{\Delta}, \frac{\beta}{\Delta}, \frac{\beta}{\Delta}, - \frac{\beta}{\Delta}, 0 \right)\,.
\end{equation}

\end{enumerate}

\subsection{Hayward}

The metric for the Hayward BH is parameterized by  $g$ that controls the regularity of the BH~\cite{Hayward:2005gi}.

\begin{enumerate}
    \item original CY mapping

\begin{align}   (\calA_0,\calA_1,\calA_2,\calA_5,f)=&\left(\Delta \left( r^2 + a^2 - \frac{2 M r^4}{r^3 + g^3} \right)^{-1}, \Delta \left( r^2 + a^2 - \frac{2 M r^4}{r^3 + g^3} \right)^{-1},\right. \nonumber \\
& \left. \Delta \left( r^2 + a^2 - \frac{2 M r^4}{r^3 + g^3} \right)^{-1}, \frac{1}{\Delta} \left( r^2 + a^2 - \frac{2 M r^4}{r^3 + g^3} \right),0 \right)\,,
\end{align}

    \item refined mapping (rescaling)

\begin{equation}
        (A_0,A_1,A_2,A_5,f)=\left(r^2+a^2,\; r^2+a^2,\; 1,\; r^2 + a^2 - \frac{2 M r^4}{r^3+g^3}, \;0 \right)\,,
\end{equation}

    \item Taylor coefficients ($n$ is a positive integer)

\begin{equation}
    \alpha_{5 \; 3n+1}= (-1)^{3n+1} 2 \frac{g^{3n}}{M^{3n
}}\,,
\end{equation}

    \item refined mapping (small deviation, $n=0$) 

\begin{equation}
    (\delta A_0,\delta A_1,\delta A_2,\delta A_5,\delta f)=\left( -\frac{2 g^3 M}{r^2 \Delta}, -\frac{2 g^3 M}{r^2 \Delta}, -\frac{2 g^3 M}{r^2 \Delta}, \frac{2 g^3 M}{r^2 \Delta}, 0\right)\,.
\end{equation}

\end{enumerate}

\subsection{Bardeen}

Similar to the Hayward BH, the metric for the Bardeen BH is also parameterized by  $g$ that controls the regularity of the BH~\cite{Bardeen_proceedings}.

\begin{enumerate}
    \item original CY mapping

\begin{align}   (\calA_0,\calA_1,\calA_2,\calA_5,f)=&\left(\Delta \left( r^2 + a^2 - \frac{2 M r^4}{(r^2 + g^2)^{3/2}} \right)^{-1}, \Delta \left( r^2 + a^2 - \frac{2 M r^4}{(r^2 + g^2)^{3/2}} \right)^{-1},\right. \nonumber \\
& \left. \Delta \left( r^2 + a^2 - \frac{2 M r^4}{(r^2 + g^2)^{3/2}} \right)^{-1}, \frac{1}{\Delta} \left( r^2 + a^2 - \frac{2 M r^4}{(r^2 + g^2)^{3/2}} \right),0 \right)\,,
\end{align}

    \item refined mapping (rescaling)

\begin{equation}
        (A_0,A_1,A_2,A_5,f)=\left(r^2+a^2,\; r^2+a^2,\; 1,\; r^2 + a^2 - \frac{2 M r^4}{(r^2 + g^2)^{3/2}}, \;0 \right)\,,
\end{equation}

    \item Taylor coefficients

\begin{equation}
    (\alpha_{53},\alpha_{55},\alpha_{57},...)=\left(-3\frac{g^2}{M^2},  \frac{15g^4}{4M^4},  -\frac{35g^6}{8M^6},...\right)\,,
\end{equation}

    \item refined mapping (small deviation, $n=0$)

\begin{equation}
    (\delta A_0,\delta A_1,\delta A_2,\delta A_5,\delta f)=\left( -\frac{3 g^2 M}{r \Delta}, -\frac{3 g^2 M}{r \Delta}, -\frac{3 g^2 M}{r \Delta}, \frac{3 g^2 M}{r \Delta}, 0\right)\,.
\end{equation}

\end{enumerate}

\subsection{Ghosh}

The metric for the nonsingular BH found by Ghosh is parameterized by  $k$ that characterizes the mass profile~\cite{Ghosh:2014pba}.

\begin{enumerate}
    \item original CY mapping

\begin{align}
    (\calA_0,\calA_1,\calA_2,\calA_5,f)=&\left(\frac{\Delta}{r^2 +a^2-2M r e^{-k/r}}, \frac{\Delta}{r^2 +a^2-2M r e^{-k/r}}, \frac{\Delta}{r^2 +a^2-2M r e^{-k/r}}, 
 \frac{r^2 +a^2-2M r e^{-k/r}}{\Delta},0 \right)\,,
\end{align}

    \item refined mapping (rescaling)

\begin{equation}
        (A_0,A_1,A_2,A_5,f)=\left(r^2+a^2,\; r^2+a^2,\; 1,\; r^2 + a^2  - 2M r e^{-k/r}, \;0 \right)\,,
\end{equation}

    \item Taylor coefficients

\begin{equation}
    (\alpha_{52},\alpha_{53},\alpha_{54},...)=\left(2\frac{k}{M},  -\frac{k^2}{M^2},  \frac{k^3}{3M^3},...\right)\,,
\end{equation}

    \item refined mapping (small deviation, $n=0$)

\begin{equation}
    (\delta A_0,\delta A_1,\delta A_2,\delta A_5,\delta f)=\left( -\frac{2Mk}{ \Delta},  -\frac{2Mk}{ \Delta},  -\frac{2Mk}{ \Delta},  \frac{2Mk}{ \Delta}, 0\right)\,.
\end{equation}

\end{enumerate}

\subsection{Kalb-Ramond}

The metric for the Kalb-Ramond BH is parameterized by the Kalb-Ramond parameter $s$ and the Lorentz-violating parameter $\Gamma$~\cite{Kumar:2020hgm}.

\begin{enumerate}
    \item original CY mapping

\begin{align}
    (\calA_0,\calA_1,\calA_2,\calA_5,f)=&\left(\frac{\Delta r^{\frac{2}{s}}}{ r^{2 + \frac{2}{s}} + a^2 r^{\frac{2}{s}} - 2 M r^{\frac{2+s}{s}} + r^2 \Gamma}, \frac{\Delta r^{\frac{2}{s}}}{ r^{2 + \frac{2}{s}} + a^2 r^{\frac{2}{s}} - 2 M r^{\frac{2+s}{s}} + r^2 \Gamma}, \right. \nonumber \\
& \left. \frac{\Delta r^{\frac{2}{s}}}{ r^{2 + \frac{2}{s}} + a^2 r^{\frac{2}{s}} - 2 M r^{\frac{2+s}{s}} + r^2 \Gamma}, \frac{ r^{2 + \frac{2}{s}} + a^2 r^{\frac{2}{s}} - 2 M r^{\frac{2+s}{s}} + r^2 \Gamma}{\Delta r^{\frac{2}{s}}}, 0 \right)\,,
\end{align}

    \item refined mapping (rescaling)

\begin{equation}
        (A_0,A_1,A_2,A_5,f)=\left(r^2+a^2,\; r^2+a^2,\; 1,\; \Delta + r^{\frac{2}{s}(s-1)}\Gamma,\; 0 \right)\,,
\end{equation}

    \item Taylor coefficients (e.g. $s=2$)

\begin{equation}
    \alpha_{51}=\frac{\Gamma}{M}\,,
\end{equation}

    \item refined mapping (small deviation, $n=0$)

\begin{equation}
    (\delta A_0,\delta A_1,\delta A_2,\delta A_5,\delta f)=\left( - \frac{r^{2 - \frac{2}{s}} \Gamma}{\Delta}, - \frac{r^{2 - \frac{2}{s}} \Gamma}{\Delta}, - \frac{r^{2 - \frac{2}{s}} \Gamma}{\Delta},  \frac{r^{2 - \frac{2}{s}} \Gamma}{\Delta}, 0\right)\,. 
\end{equation}

\end{enumerate}

\subsection{Kerr-Sen}
\label{sec:KS}

The metric for the Kerr-Sen BH is parameterized by $b$ that is related to the charge~\cite{Sen:1992ua}.

\begin{enumerate}
    \item original CY mapping

\begin{align}
(\calA_0,\calA_1,\calA_2,\calA_5,f)=\left( \frac{(r^2 +a^2 + 2br) \Delta}{(r^2+a^2)(\Delta + 2br)}, \frac{(r^2 +a^2 + 2br)^2 \Delta}{(r^2+a^2)^2(\Delta + 2br)}, \frac{\Delta}{\Delta + 2br}, \frac{\Delta + 2br}{\Delta}, 2br \right)\,,
\end{align}

    \item refined mapping (rescaling)

\begin{equation}
        (A_0,A_1,A_2,A_5,f)=\left(r^2 +a^2 + 2br,\; r^2 +a^2 +2br,\; 1,\; \Delta + 2br,\; 2br \right)\,,
\end{equation}

    \item Taylor coefficients

\begin{equation}
    (\alpha_{01}, \alpha_{11}, \alpha_{51},\epsilon_1)=\left( \frac{2b}{M}, \frac{2b}{M}, \frac{2b}{M}, \frac{2b}{M}\right)\,,
\end{equation}

    \item refined mapping (small deviation, $n=0$)

\begin{equation}
\label{eq:deltaA_KS}  
  (\delta A_0,\delta A_1,\delta A_2,\delta A_5,\delta f)=\left( -\frac{4bMr^2}{(a^2+r^2)\Delta }, \frac{2br(\Delta-2Mr)}{(a^2+r^2)\Delta}, -\frac{2br}{\Delta}, \frac{2br}{\Delta}, 2br \right)\,.
\end{equation}

\end{enumerate}

\ew

\section{Small-deviation Approximation}
\label{sec:small_dev}

Another way to regularize the divergence in the original CY metric is to treat deviations from Kerr to be small and find a perturbative metric.  
Instead of parameterizing $\calA_i$ directly, we split them into Kerr plus correction, expand about a small deviation from Kerr, and keep to linear order in the deviation. Namely, we consider the following:
\begin{equation}
    \label{eq:small dev approx Ai}
        \mathcal{A}_i(r) = 1 + \epsilon \Delta^n \; \delta \mathcal{A}_i(r), \qquad
        f(r) = \epsilon \Delta^n \; \delta f(r)\,,
\end{equation}
where $n$ takes integer values while $\epsilon$ is a book-keeping parameter to count the order of corrections from Kerr. Notice that if we use the full expression for $\delta \mathcal{A}_i$, the results are the same for any $n$. The difference in $n$ becomes important when we use $\delta \mathcal{A}_i$ expanded about $r=\infty$. We will expand the metric about $\epsilon=0$ and only keep to linear order. 
To keep the location of the event horizon to be the non-Kerr one, for $g_{rr}$, we found it more advantageous not to perturb in small $\epsilon$. If we had perturbed also $g_{rr}$, the metric falls into that considered by Vigeland et al.~\cite{Vigeland:2011ji} (see Sec.~VIA of~\cite{Johannsen:2015pca} for the relation between the metrics in Vegeland et al.~\cite{Vigeland:2011ji} and Johannsen~\cite{Johannsen:2015pca}, with the latter same as taking the original CY metric in Eq.~\eqref{eq:CY_metric_original} and taking the limit $\mathcal{A}_0^2 \to \mathcal{A}_1 \mathcal{A}_2$).
We do not explicitly show the expression for the refined metric under the small deviation approximation as the expression itself is quite lengthy and not illuminating. We consider below two example values of $n$ ($n=-1$ and $n=1$) for two BH spacetimes: braneworld and Kerr-Sen. The mapping functions $\delta \calA_i$ and $\delta f$ for these two BHs in the case of $n=0$ can be found in Eqs.~\eqref{eq:deltaA_brane} and~\eqref{eq:deltaA_KS}.

\begin{figure*}[t]
\includegraphics[width=8.5cm]{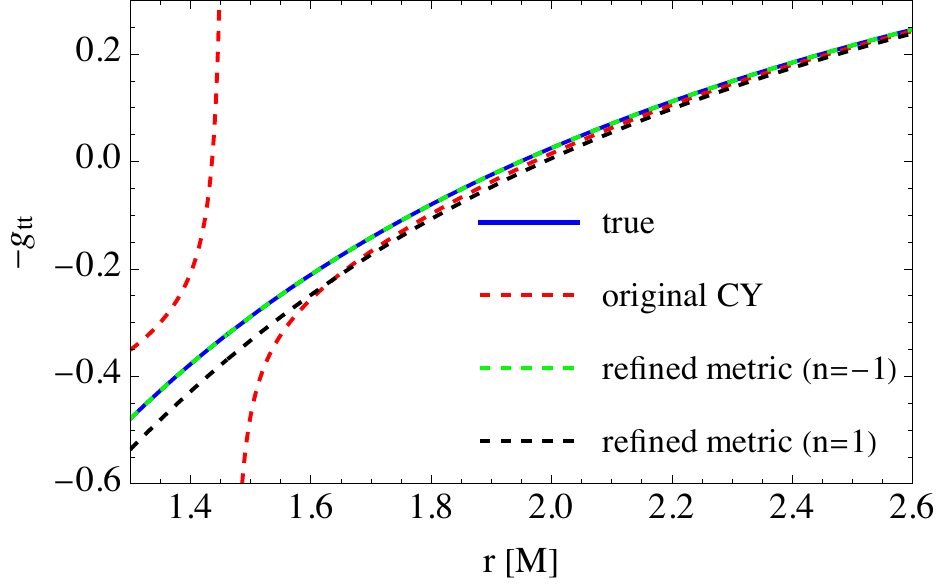}
\includegraphics[width=8.5cm]{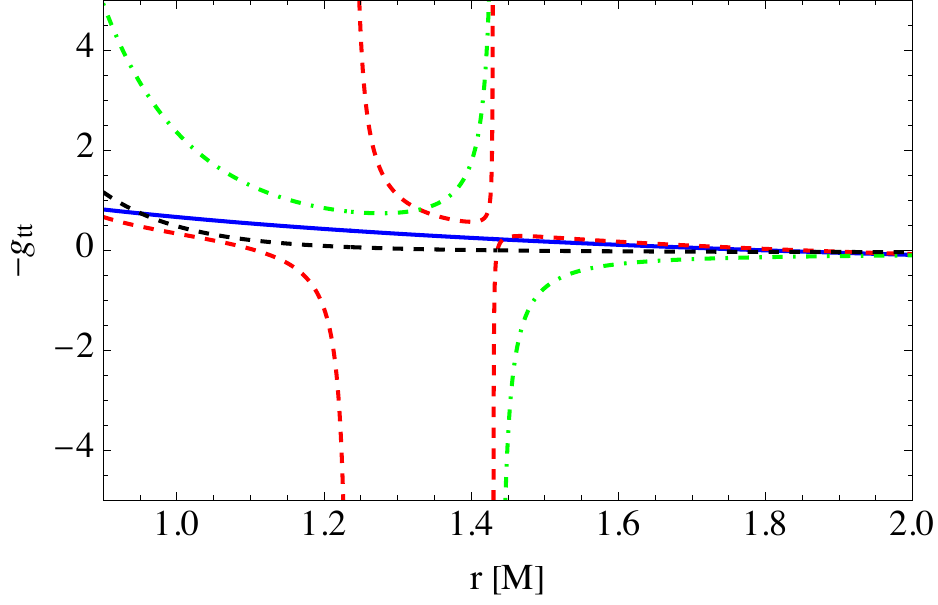}
\caption{\label{fig:small_dev}
(Left) Comparison of $-g_{tt}$ for the braneworld BH with the true one (blue solid), the original CY metric with truncated $\mathcal{A}_i$ (red dashed), and the refined metric under small-deviation approximation for $n=-1$ (green dashed) and $n=1$ (black dashed). We chose the parameters as $a = 0.9M$, $\beta = 0.1 M^2$ and $\theta = \pi/2$.
 (Right) Similar to the left panel but for the Kerr-Sen BH. We chose the parameters as $a = 0.9M$, $b = 0.1 M$ and $\theta = \pi/2$.
}
\end{figure*}

\subsection{$n=-1$}

We first consider the case with $n=-1$. Let us first focus on studying the braneworld BH. The mapping functions are given by 
\begin{equation}
(\delta \calA_0, \delta \calA_1, \delta \calA_2, \delta \calA_5, \delta f) = (-\beta,-\beta,-\beta,\beta,0)\,.
\end{equation}
We can then try to construct the braneworld BH by substituting the above mapping of $\delta \calA_i$ to the CY metric that is expanded and valid to $\mathcal{O}(\epsilon)$. We found that the outcome metric is \emph{exactly} the same as the original braneworld BH metric even though we have kept only to $\mathcal{O}(\epsilon)$ (see the left panel of Fig.~\ref{fig:small_dev}). This is because the correction to Kerr enters linearly in the braneworld BH (except for $g_{rr}$  which we do not expand in small $\epsilon$, see Eq.~\eqref{eq:brane}) and $\delta \mathcal{A}_i$ and $\delta f$ are constants (so the expressions become exact even if one expands about $r=\infty$). Thus, this new parametrization works at least for describing the braneworld BH. 

This approach with $n=-1$ may not work for other metrics. For a Kerr-Sen BH with 
\begin{align}
  &  (\delta \calA_0, \delta \calA_1, \delta \calA_2,\delta \calA_5, \delta f) \nonumber \\
& = \left( -\frac{4bMr^2}{a^2+r^2}, \frac{2br(\Delta-2Mr)}{a^2+r^2}, -2br, 2br, 2br \Delta \right)\,,
\end{align}
we find
\begin{equation}
g_{tt} - g_{tt,K} \propto \frac{1}{\Delta^2}\,,
\end{equation}
where $g_{tt,K}$ is the $(t,t)$ component of the Kerr metric. Thus, $g_{tt}$ diverges at the Kerr horizon where $\Delta = 0$. 
 The result for $-g_{tt}$ is
  shown in the right panel of Fig.~\ref{fig:small_dev} where we expand $\delta \calA_0$ and $\delta \calA_1$ about $r=\infty$ and keep up to $\mathcal{O}(1/r^2)$. Notice there is indeed a divergence close to the Kerr horizon ($r=1.44M$). For the braneworld BH case, a specific combination of $\delta \calA_i$ eliminates $\Delta^2$ in the denominator and prevents the metric from diverging at $\Delta = 0$. Indeed, when $\delta \calA_0 = \delta \calA_1 = \delta \calA_2 = -\delta \calA_5 $, as in the case of the braneworld BH, we find
\begin{equation}
g_{tt} - g_{tt,K} = -\frac{\calA_5
   \Sigma}{r^4} \epsilon + \mathcal{O}(\epsilon^2)\,,
\end{equation}
and thus no divergence. Similar behavior can be seen for the Hayward, Bardeen Ghosh, and Kalb-Ramond BHs.

\subsection{$n=1$}

Next, let us study the case with $n=1$.  
In this case, we find
\begin{equation}
g_{tt} - g_{tt,K} \propto \frac{\beta}{\Delta^0}\,.
\end{equation}

As before, let us first focus on the braneworld example. 
$\delta \calA_i$ for the second approach is given by
\begin{equation}
\label{eq:calA_appr2_full}
(\delta \calA_0, \delta \calA_1, \delta \calA_2, \delta \calA_5) = \left(-\frac{\beta}{\Delta^2},-\frac{\beta}{\Delta^2},-\frac{\beta}{\Delta^2},\frac{\beta}{\Delta^2} \right)\,.
\end{equation}
Substituting this to the CY metric expanded to $\mathcal{O}(\epsilon)$, we find that it matches with the original braneworld BH metric exactly, similar to the $n=-1$ case.
The above $\delta \calA_i$ mapping can be expanded about $r=\infty$ as
\begin{eqnarray}
\label{eq:calA_appr2_exp}
\delta \calA_0 &=& \delta \calA_1 = \delta \calA_2 = -\delta \calA_5 \nonumber \\
&=& -\frac{\beta }{r^4} -\frac{4 \beta  M}{r^5}+\frac{2 \beta 
   \left(a^2-6 M^2\right)}{r^6}+ \mathcal{O}\left( \frac{M^{7}}{r^{7}} \right)\,.\nonumber\\
\end{eqnarray}
The left panel of Fig.~\ref{fig:small_dev} contains $-g_{tt}$ with $n=1$. Notice that although the divergence has been removed, the result is not very accurate near the horizon where the large-$r$ expansion breaks down.

Let us next look at Kerr-Sen. The mapping for the radial functions is given by
\begin{align}
  &  (\delta \calA_0, \delta \calA_1, \delta \calA_2,\delta \calA_5, \delta f)\nonumber \\
& = \left( -\frac{4bMr^2}{(a^2+r^2) \Delta^2}, \frac{2br(\Delta-2Mr)}{(a^2+r^2) \Delta^2}, -\frac{2br}{\Delta^2}, \frac{2br}{\Delta^2}, \frac{2br}{\Delta} \right)\,.
\end{align}
We expand this about $r=\infty$ and keep up to 3rd order from the leading. The right panel of Fig.~\ref{fig:small_dev} also includes $-g_{tt}$ with $n=1$. Notice that, unlike the $n=-1$ case, the divergence is now absent, though the result has some deviation from the true one for smaller $r$. Given these results, we conclude that although the small-deviation approximation works to remove the divergence when $n$ is chosen appropriately, the rescaling approach in Sec.~\ref{sec:refined} seems to work better in terms of reproducing true metrics.

\section{Imaginary QNM frequency for braneworld black hole}
\label{app:QNM_brane_imaginary}

\begin{figure}[t]
\includegraphics[width=8.5cm]{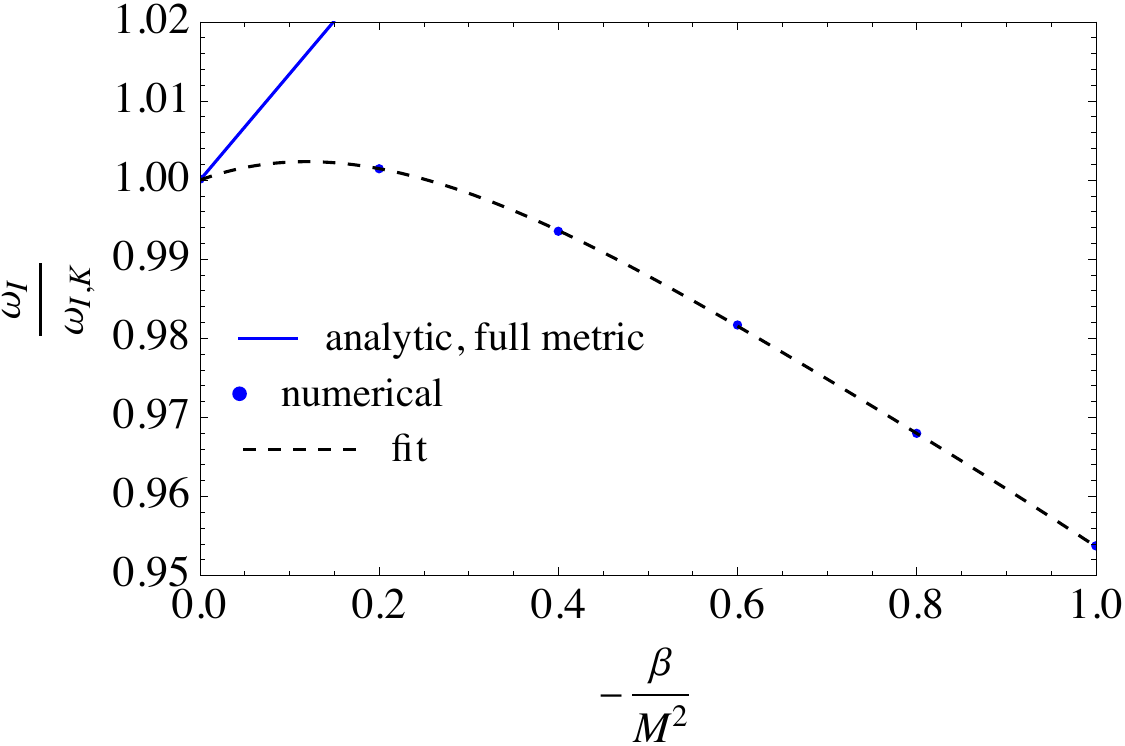}
\caption{\label{fig:QNM_brane_imaginary}
Similar to Fig.~\ref{fig:brane_QNM} but focusing on the imaginary QNMs with the analytic estimate given in Eq.~\eqref{eq:brane_QNM_analytic}. We also present a fit for the numerical QNMs given in Eq.~\eqref{eq:brane_fit}.
}
\end{figure}

Let us study the imaginary QNM frequencies for braneworld BHs in more detail. Figure~\ref{fig:QNM_brane_imaginary} shows the comparison between the analytic estimate given in Sec.~\ref{sec:ringdown} and the numerical values in~\cite{Mishra:2021waw}. The former is given by
\begin{equation}
\label{eq:brane_QNM_analytic}
    \frac{\omega_I}{\omega_{I,K}} = 1 + 0.134 \left( -\frac{\beta}{M^2}\right)\,,
\end{equation}
with the dimensionless spin value of $\chi = 0.67$. On the other hand, we can fit the numerical values in a quartic polynomial in the tidal charge and find
\begin{equation}
\label{eq:brane_fit}
    \frac{\omega_I}{\omega_{I,K}} = 1 +0.0413 x-0.202 x^2+0.171 x^3-0.0558 x^4\,,
\end{equation}
with $x=(-\beta/M^2)$.
Interestingly, the numerical imaginary QNMs increase near $\beta \sim 0$, which is qualitatively consistent with the analytic finding in Eq.~\eqref{eq:brane_QNM_analytic}. Though quantitatively, the analytic and numerical behaviors do not match as can be seen by the difference in the coefficients of the linear terms in Eqs.~\eqref{eq:brane_QNM_analytic} and~\eqref{eq:brane_fit}. We note that a similar behavior is seen for scalar Gauss-Bonnet gravity where the imaginary part of the QNM frequency for BHs is not well captured by the analytic eikonal approximation~\cite{Bryant:2021xdh}. This may indicate a limitation in these analytic approaches on the imaginary QNM frequencies.

\bibliography{master}

\begin{thebibliography}{94}%
\makeatletter
\providecommand \@ifxundefined [1]{%
 \@ifx{#1\undefined}
}%
\providecommand \@ifnum [1]{%
 \ifnum #1\expandafter \@firstoftwo
 \else \expandafter \@secondoftwo
 \fi
}%
\providecommand \@ifx [1]{%
 \ifx #1\expandafter \@firstoftwo
 \else \expandafter \@secondoftwo
 \fi
}%
\providecommand \natexlab [1]{#1}%
\providecommand \enquote  [1]{``#1''}%
\providecommand \bibnamefont  [1]{#1}%
\providecommand \bibfnamefont [1]{#1}%
\providecommand \citenamefont [1]{#1}%
\providecommand \href@noop [0]{\@secondoftwo}%
\providecommand \href [0]{\begingroup \@sanitize@url \@href}%
\providecommand \@href[1]{\@@startlink{#1}\@@href}%
\providecommand \@@href[1]{\endgroup#1\@@endlink}%
\providecommand \@sanitize@url [0]{\catcode `\\12\catcode `\$12\catcode
  `\&12\catcode `\#12\catcode `\^12\catcode `\_12\catcode `\%12\relax}%
\providecommand \@@startlink[1]{}%
\providecommand \@@endlink[0]{}%
\providecommand \url  [0]{\begingroup\@sanitize@url \@url }%
\providecommand \@url [1]{\endgroup\@href {#1}{\urlprefix }}%
\providecommand \urlprefix  [0]{URL }%
\providecommand \Eprint [0]{\href }%
\providecommand \doibase [0]{http://dx.doi.org/}%
\providecommand \selectlanguage [0]{\@gobble}%
\providecommand \bibinfo  [0]{\@secondoftwo}%
\providecommand \bibfield  [0]{\@secondoftwo}%
\providecommand \translation [1]{[#1]}%
\providecommand \BibitemOpen [0]{}%
\providecommand \bibitemStop [0]{}%
\providecommand \bibitemNoStop [0]{.\EOS\space}%
\providecommand \EOS [0]{\spacefactor3000\relax}%
\providecommand \BibitemShut  [1]{\csname bibitem#1\endcsname}%
\let\auto@bib@innerbib\@empty
\bibitem [{\citenamefont {Will}(2014)}]{Will:2014kxa}%
  \BibitemOpen
  \bibfield  {author} {\bibinfo {author} {\bibfnamefont {C.~M.}\ \bibnamefont
  {Will}},\ }\bibfield  {title} {\enquote {\bibinfo {title} {{The Confrontation
  between General Relativity and Experiment}},}\ }\href {\doibase
  10.12942/lrr-2014-4} {\bibfield  {journal} {\bibinfo  {journal} {Living Rev.
  Rel.}\ }\textbf {\bibinfo {volume} {17}},\ \bibinfo {pages} {4} (\bibinfo
  {year} {2014})},\ \Eprint {http://arxiv.org/abs/1403.7377} {arXiv:1403.7377
  [gr-qc]} \BibitemShut {NoStop}%
\bibitem [{\citenamefont {{Will}}(1993)}]{TEGP}%
  \BibitemOpen
  \bibfield  {author} {\bibinfo {author} {\bibfnamefont {C.~M.}\ \bibnamefont
  {{Will}}},\ }\href@noop {} {\emph {\bibinfo {title} {Theory and Experiment in
  Gravitational Physics, by Clifford M.~Will, pp.~396.~ISBN
  0521439736.~Cambridge, UK: Cambridge University Press, March 1993.}}},\
  edited by\ \bibinfo {editor} {\bibnamefont {{Will, C.~M.}}}\ (\bibinfo {year}
  {1993})\BibitemShut {NoStop}%
\bibitem [{\citenamefont {Stairs}(2003)}]{stairs}%
  \BibitemOpen
  \bibfield  {author} {\bibinfo {author} {\bibfnamefont {I.~H.}\ \bibnamefont
  {Stairs}},\ }\bibfield  {title} {\enquote {\bibinfo {title} {{Testing general
  relativity with pulsar timing}},}\ }\href@noop {} {\bibfield  {journal}
  {\bibinfo  {journal} {Living Rev.Rel.}\ }\textbf {\bibinfo {volume} {6}},\
  \bibinfo {pages} {5} (\bibinfo {year} {2003})}\BibitemShut {NoStop}%
\bibitem [{\citenamefont {Wex}(2014)}]{Wex:2014nva}%
  \BibitemOpen
  \bibfield  {author} {\bibinfo {author} {\bibfnamefont {N.}~\bibnamefont
  {Wex}},\ }\bibfield  {title} {\enquote {\bibinfo {title} {{Testing
  Relativistic Gravity with Radio Pulsars}},}\ }\href@noop {} {\  (\bibinfo
  {year} {2014})},\ \Eprint {http://arxiv.org/abs/1402.5594} {arXiv:1402.5594
  [gr-qc]} \BibitemShut {NoStop}%
\bibitem [{\citenamefont {Abbott}\ \emph {et~al.}(2016)\citenamefont {Abbott}
  \emph {et~al.}}]{TheLIGOScientific:2016src}%
  \BibitemOpen
  \bibfield  {author} {\bibinfo {author} {\bibfnamefont {B.~P.}\ \bibnamefont
  {Abbott}} \emph {et~al.} (\bibinfo {collaboration} {Virgo, LIGO
  Scientific}),\ }\bibfield  {title} {\enquote {\bibinfo {title} {{Tests of
  general relativity with GW150914}},}\ }\href {\doibase
  10.1103/PhysRevLett.116.221101} {\bibfield  {journal} {\bibinfo  {journal}
  {Phys. Rev. Lett.}\ }\textbf {\bibinfo {volume} {116}},\ \bibinfo {pages}
  {221101} (\bibinfo {year} {2016})},\ \Eprint
  {http://arxiv.org/abs/1602.03841} {arXiv:1602.03841 [gr-qc]} \BibitemShut
  {NoStop}%
\bibitem [{\citenamefont {Yunes}\ \emph {et~al.}(2016)\citenamefont {Yunes},
  \citenamefont {Yagi},\ and\ \citenamefont {Pretorius}}]{Yunes:2016jcc}%
  \BibitemOpen
  \bibfield  {author} {\bibinfo {author} {\bibfnamefont {N.}~\bibnamefont
  {Yunes}}, \bibinfo {author} {\bibfnamefont {K.}~\bibnamefont {Yagi}}, \ and\
  \bibinfo {author} {\bibfnamefont {F.}~\bibnamefont {Pretorius}},\ }\bibfield
  {title} {\enquote {\bibinfo {title} {{Theoretical Physics Implications of the
  Binary Black-Hole Mergers GW150914 and GW151226}},}\ }\href {\doibase
  10.1103/PhysRevD.94.084002} {\bibfield  {journal} {\bibinfo  {journal} {Phys.
  Rev.}\ }\textbf {\bibinfo {volume} {D94}},\ \bibinfo {pages} {084002}
  (\bibinfo {year} {2016})}\BibitemShut {NoStop}%
\bibitem [{\citenamefont {Berti}\ \emph
  {et~al.}(2018{\natexlab{a}})\citenamefont {Berti}, \citenamefont {Yagi},\
  and\ \citenamefont {Yunes}}]{Berti:2018cxi}%
  \BibitemOpen
  \bibfield  {author} {\bibinfo {author} {\bibfnamefont {E.}~\bibnamefont
  {Berti}}, \bibinfo {author} {\bibfnamefont {K.}~\bibnamefont {Yagi}}, \ and\
  \bibinfo {author} {\bibfnamefont {N.}~\bibnamefont {Yunes}},\ }\bibfield
  {title} {\enquote {\bibinfo {title} {{Extreme Gravity Tests with
  Gravitational Waves from Compact Binary Coalescences: (I)
  Inspiral-Merger}},}\ }\href {\doibase 10.1007/s10714-018-2362-8} {\bibfield
  {journal} {\bibinfo  {journal} {Gen. Rel. Grav.}\ }\textbf {\bibinfo {volume}
  {50}},\ \bibinfo {pages} {46} (\bibinfo {year} {2018}{\natexlab{a}})},\
  \Eprint {http://arxiv.org/abs/1801.03208} {arXiv:1801.03208 [gr-qc]}
  \BibitemShut {NoStop}%
\bibitem [{\citenamefont {Berti}\ \emph
  {et~al.}(2018{\natexlab{b}})\citenamefont {Berti}, \citenamefont {Yagi},
  \citenamefont {Yang},\ and\ \citenamefont {Yunes}}]{Berti:2018vdi}%
  \BibitemOpen
  \bibfield  {author} {\bibinfo {author} {\bibfnamefont {E.}~\bibnamefont
  {Berti}}, \bibinfo {author} {\bibfnamefont {K.}~\bibnamefont {Yagi}},
  \bibinfo {author} {\bibfnamefont {H.}~\bibnamefont {Yang}}, \ and\ \bibinfo
  {author} {\bibfnamefont {N.}~\bibnamefont {Yunes}},\ }\bibfield  {title}
  {\enquote {\bibinfo {title} {{Extreme Gravity Tests with Gravitational Waves
  from Compact Binary Coalescences: (II) Ringdown}},}\ }\href {\doibase
  10.1007/s10714-018-2372-6} {\bibfield  {journal} {\bibinfo  {journal} {Gen.
  Rel. Grav.}\ }\textbf {\bibinfo {volume} {50}},\ \bibinfo {pages} {49}
  (\bibinfo {year} {2018}{\natexlab{b}})},\ \Eprint
  {http://arxiv.org/abs/1801.03587} {arXiv:1801.03587 [gr-qc]} \BibitemShut
  {NoStop}%
\bibitem [{\citenamefont {Abbott}\ \emph {et~al.}(2019)\citenamefont {Abbott}
  \emph {et~al.}}]{LIGOScientific:2019fpa}%
  \BibitemOpen
  \bibfield  {author} {\bibinfo {author} {\bibfnamefont {B.~P.}\ \bibnamefont
  {Abbott}} \emph {et~al.} (\bibinfo {collaboration} {LIGO Scientific,
  Virgo}),\ }\bibfield  {title} {\enquote {\bibinfo {title} {{Tests of General
  Relativity with the Binary Black Hole Signals from the LIGO-Virgo Catalog
  GWTC-1}},}\ }\href {\doibase 10.1103/PhysRevD.100.104036} {\bibfield
  {journal} {\bibinfo  {journal} {Phys. Rev. D}\ }\textbf {\bibinfo {volume}
  {100}},\ \bibinfo {pages} {104036} (\bibinfo {year} {2019})},\ \Eprint
  {http://arxiv.org/abs/1903.04467} {arXiv:1903.04467 [gr-qc]} \BibitemShut
  {NoStop}%
\bibitem [{\citenamefont {Abbott}\ \emph
  {et~al.}(2021{\natexlab{a}})\citenamefont {Abbott} \emph
  {et~al.}}]{LIGOScientific:2020tif}%
  \BibitemOpen
  \bibfield  {author} {\bibinfo {author} {\bibfnamefont {R.}~\bibnamefont
  {Abbott}} \emph {et~al.} (\bibinfo {collaboration} {LIGO Scientific,
  Virgo}),\ }\bibfield  {title} {\enquote {\bibinfo {title} {{Tests of general
  relativity with binary black holes from the second LIGO-Virgo
  gravitational-wave transient catalog}},}\ }\href {\doibase
  10.1103/PhysRevD.103.122002} {\bibfield  {journal} {\bibinfo  {journal}
  {Phys. Rev. D}\ }\textbf {\bibinfo {volume} {103}},\ \bibinfo {pages}
  {122002} (\bibinfo {year} {2021}{\natexlab{a}})},\ \Eprint
  {http://arxiv.org/abs/2010.14529} {arXiv:2010.14529 [gr-qc]} \BibitemShut
  {NoStop}%
\bibitem [{\citenamefont {Abbott}\ \emph
  {et~al.}(2021{\natexlab{b}})\citenamefont {Abbott} \emph
  {et~al.}}]{LIGOScientific:2021sio}%
  \BibitemOpen
  \bibfield  {author} {\bibinfo {author} {\bibfnamefont {R.}~\bibnamefont
  {Abbott}} \emph {et~al.} (\bibinfo {collaboration} {LIGO Scientific, VIRGO,
  KAGRA}),\ }\bibfield  {title} {\enquote {\bibinfo {title} {{Tests of General
  Relativity with GWTC-3}},}\ }\href@noop {} {\  (\bibinfo {year}
  {2021}{\natexlab{b}})},\ \Eprint {http://arxiv.org/abs/2112.06861}
  {arXiv:2112.06861 [gr-qc]} \BibitemShut {NoStop}%
\bibitem [{\citenamefont {Psaltis}\ \emph {et~al.}(2020)\citenamefont {Psaltis}
  \emph {et~al.}}]{Psaltis:2020lvx}%
  \BibitemOpen
  \bibfield  {author} {\bibinfo {author} {\bibfnamefont {D.}~\bibnamefont
  {Psaltis}} \emph {et~al.} (\bibinfo {collaboration} {Event Horizon
  Telescope}),\ }\bibfield  {title} {\enquote {\bibinfo {title} {{Gravitational
  Test Beyond the First Post-Newtonian Order with the Shadow of the M87 Black
  Hole}},}\ }\href {\doibase 10.1103/PhysRevLett.125.141104} {\bibfield
  {journal} {\bibinfo  {journal} {Phys. Rev. Lett.}\ }\textbf {\bibinfo
  {volume} {125}},\ \bibinfo {pages} {141104} (\bibinfo {year} {2020})},\
  \Eprint {http://arxiv.org/abs/2010.01055} {arXiv:2010.01055 [gr-qc]}
  \BibitemShut {NoStop}%
\bibitem [{\citenamefont {Psaltis}\ \emph {et~al.}(2021)\citenamefont
  {Psaltis}, \citenamefont {Talbot}, \citenamefont {Payne},\ and\ \citenamefont
  {Mandel}}]{Psaltis:2020ctj}%
  \BibitemOpen
  \bibfield  {author} {\bibinfo {author} {\bibfnamefont {D.}~\bibnamefont
  {Psaltis}}, \bibinfo {author} {\bibfnamefont {C.}~\bibnamefont {Talbot}},
  \bibinfo {author} {\bibfnamefont {E.}~\bibnamefont {Payne}}, \ and\ \bibinfo
  {author} {\bibfnamefont {I.}~\bibnamefont {Mandel}},\ }\bibfield  {title}
  {\enquote {\bibinfo {title} {{Probing the Black Hole Metric. I. Black Hole
  Shadows and Binary Black-Hole Inspirals}},}\ }\href {\doibase
  10.1103/PhysRevD.103.104036} {\bibfield  {journal} {\bibinfo  {journal}
  {Phys. Rev. D}\ }\textbf {\bibinfo {volume} {103}},\ \bibinfo {pages}
  {104036} (\bibinfo {year} {2021})},\ \Eprint
  {http://arxiv.org/abs/2012.02117} {arXiv:2012.02117 [gr-qc]} \BibitemShut
  {NoStop}%
\bibitem [{\citenamefont {Akiyama}\ \emph {et~al.}(2022)\citenamefont {Akiyama}
  \emph {et~al.}}]{EventHorizonTelescope:2022xqj}%
  \BibitemOpen
  \bibfield  {author} {\bibinfo {author} {\bibfnamefont {K.}~\bibnamefont
  {Akiyama}} \emph {et~al.} (\bibinfo {collaboration} {Event Horizon
  Telescope}),\ }\bibfield  {title} {\enquote {\bibinfo {title} {{First
  Sagittarius A* Event Horizon Telescope Results. VI. Testing the Black Hole
  Metric}},}\ }\href {\doibase 10.3847/2041-8213/ac6756} {\bibfield  {journal}
  {\bibinfo  {journal} {Astrophys. J. Lett.}\ }\textbf {\bibinfo {volume}
  {930}},\ \bibinfo {pages} {L17} (\bibinfo {year} {2022})}\BibitemShut
  {NoStop}%
\bibitem [{\citenamefont {Vagnozzi}\ \emph {et~al.}(2023)\citenamefont
  {Vagnozzi} \emph {et~al.}}]{Vagnozzi:2022moj}%
  \BibitemOpen
  \bibfield  {author} {\bibinfo {author} {\bibfnamefont {S.}~\bibnamefont
  {Vagnozzi}} \emph {et~al.},\ }\bibfield  {title} {\enquote {\bibinfo {title}
  {{Horizon-scale tests of gravity theories and fundamental physics from the
  Event Horizon Telescope image of Sagittarius A}},}\ }\href {\doibase
  10.1088/1361-6382/acd97b} {\bibfield  {journal} {\bibinfo  {journal} {Class.
  Quant. Grav.}\ }\textbf {\bibinfo {volume} {40}},\ \bibinfo {pages} {165007}
  (\bibinfo {year} {2023})},\ \Eprint {http://arxiv.org/abs/2205.07787}
  {arXiv:2205.07787 [gr-qc]} \BibitemShut {NoStop}%
\bibitem [{\citenamefont {Israel}(1967)}]{israel}%
  \BibitemOpen
  \bibfield  {author} {\bibinfo {author} {\bibfnamefont {W.}~\bibnamefont
  {Israel}},\ }\bibfield  {title} {\enquote {\bibinfo {title} {{Event horizons
  in static vacuum space-times}},}\ }\href {\doibase 10.1103/PhysRev.164.1776}
  {\bibfield  {journal} {\bibinfo  {journal} {Phys. Rev.}\ }\textbf {\bibinfo
  {volume} {164}},\ \bibinfo {pages} {1776--1779} (\bibinfo {year}
  {1967})}\BibitemShut {NoStop}%
\bibitem [{\citenamefont {Hawking}(1972)}]{hawking-uniqueness}%
  \BibitemOpen
  \bibfield  {author} {\bibinfo {author} {\bibfnamefont {S.~W.}\ \bibnamefont
  {Hawking}},\ }\bibfield  {title} {\enquote {\bibinfo {title} {{Black holes in
  general relativity}},}\ }\href {\doibase 10.1007/BF01877517} {\bibfield
  {journal} {\bibinfo  {journal} {Commun. Math. Phys.}\ }\textbf {\bibinfo
  {volume} {25}},\ \bibinfo {pages} {152--166} (\bibinfo {year}
  {1972})}\BibitemShut {NoStop}%
\bibitem [{\citenamefont {Cardoso}\ and\ \citenamefont
  {Pani}(2019)}]{Cardoso:2019rvt}%
  \BibitemOpen
  \bibfield  {author} {\bibinfo {author} {\bibfnamefont {V.}~\bibnamefont
  {Cardoso}}\ and\ \bibinfo {author} {\bibfnamefont {P.}~\bibnamefont {Pani}},\
  }\bibfield  {title} {\enquote {\bibinfo {title} {{Testing the nature of dark
  compact objects: a status report}},}\ }\href {\doibase
  10.1007/s41114-019-0020-4} {\bibfield  {journal} {\bibinfo  {journal} {Living
  Rev. Rel.}\ }\textbf {\bibinfo {volume} {22}},\ \bibinfo {pages} {4}
  (\bibinfo {year} {2019})},\ \Eprint {http://arxiv.org/abs/1904.05363}
  {arXiv:1904.05363 [gr-qc]} \BibitemShut {NoStop}%
\bibitem [{\citenamefont {Will}(2008)}]{Will_2008}%
  \BibitemOpen
  \bibfield  {author} {\bibinfo {author} {\bibfnamefont {C.~M.}\ \bibnamefont
  {Will}},\ }\bibfield  {title} {\enquote {\bibinfo {title} {Testing the
  general relativistic “no-hair” theorems using the galactic center black
  hole sagittarius a*},}\ }\href {\doibase 10.1086/528847} {\bibfield
  {journal} {\bibinfo  {journal} {The Astrophysical Journal}\ }\textbf
  {\bibinfo {volume} {674}},\ \bibinfo {pages} {L25} (\bibinfo {year}
  {2008})}\BibitemShut {NoStop}%
\bibitem [{\citenamefont {Merritt}\ \emph {et~al.}(2010)\citenamefont
  {Merritt}, \citenamefont {Alexander}, \citenamefont {Mikkola},\ and\
  \citenamefont {Will}}]{Merritt:2009ex}%
  \BibitemOpen
  \bibfield  {author} {\bibinfo {author} {\bibfnamefont {D.}~\bibnamefont
  {Merritt}}, \bibinfo {author} {\bibfnamefont {T.}~\bibnamefont {Alexander}},
  \bibinfo {author} {\bibfnamefont {S.}~\bibnamefont {Mikkola}}, \ and\
  \bibinfo {author} {\bibfnamefont {C.~M.}\ \bibnamefont {Will}},\ }\bibfield
  {title} {\enquote {\bibinfo {title} {{Testing Properties of the Galactic
  Center Black Hole Using Stellar Orbits}},}\ }\href {\doibase
  10.1103/PhysRevD.81.062002} {\bibfield  {journal} {\bibinfo  {journal} {Phys.
  Rev. D}\ }\textbf {\bibinfo {volume} {81}},\ \bibinfo {pages} {062002}
  (\bibinfo {year} {2010})},\ \Eprint {http://arxiv.org/abs/0911.4718}
  {arXiv:0911.4718 [astro-ph.GA]} \BibitemShut {NoStop}%
\bibitem [{\citenamefont {Sadeghian}\ and\ \citenamefont
  {Will}(2011)}]{Sadeghian:2011ub}%
  \BibitemOpen
  \bibfield  {author} {\bibinfo {author} {\bibfnamefont {L.}~\bibnamefont
  {Sadeghian}}\ and\ \bibinfo {author} {\bibfnamefont {C.~M.}\ \bibnamefont
  {Will}},\ }\bibfield  {title} {\enquote {\bibinfo {title} {{Testing the black
  hole no-hair theorem at the galactic center: Perturbing effects of stars in
  the surrounding cluster}},}\ }\href {\doibase 10.1088/0264-9381/28/22/225029}
  {\bibfield  {journal} {\bibinfo  {journal} {Class. Quant. Grav.}\ }\textbf
  {\bibinfo {volume} {28}},\ \bibinfo {pages} {225029} (\bibinfo {year}
  {2011})},\ \Eprint {http://arxiv.org/abs/1106.5056} {arXiv:1106.5056 [gr-qc]}
  \BibitemShut {NoStop}%
\bibitem [{\citenamefont {Kong}\ \emph {et~al.}(2014)\citenamefont {Kong},
  \citenamefont {Li},\ and\ \citenamefont {Bambi}}]{Kong:2014wha}%
  \BibitemOpen
  \bibfield  {author} {\bibinfo {author} {\bibfnamefont {L.}~\bibnamefont
  {Kong}}, \bibinfo {author} {\bibfnamefont {Z.}~\bibnamefont {Li}}, \ and\
  \bibinfo {author} {\bibfnamefont {C.}~\bibnamefont {Bambi}},\ }\bibfield
  {title} {\enquote {\bibinfo {title} {{Constraints on the spacetime geometry
  around 10 stellar-mass black hole candidates from the disk's thermal
  spectrum}},}\ }\href {\doibase 10.1088/0004-637X/797/2/78} {\bibfield
  {journal} {\bibinfo  {journal} {Astrophys. J.}\ }\textbf {\bibinfo {volume}
  {797}},\ \bibinfo {pages} {78} (\bibinfo {year} {2014})},\ \Eprint
  {http://arxiv.org/abs/1405.1508} {arXiv:1405.1508 [gr-qc]} \BibitemShut
  {NoStop}%
\bibitem [{\citenamefont {Bambi}(2014)}]{Bambi:2014sfa}%
  \BibitemOpen
  \bibfield  {author} {\bibinfo {author} {\bibfnamefont {C.}~\bibnamefont
  {Bambi}},\ }\bibfield  {title} {\enquote {\bibinfo {title} {{Note on the
  Cardoso-Pani-Rico parametrization to test the Kerr black hole hypothesis}},}\
  }\href {\doibase 10.1103/PhysRevD.90.047503} {\bibfield  {journal} {\bibinfo
  {journal} {Phys. Rev. D}\ }\textbf {\bibinfo {volume} {90}},\ \bibinfo
  {pages} {047503} (\bibinfo {year} {2014})},\ \Eprint
  {http://arxiv.org/abs/1408.0690} {arXiv:1408.0690 [gr-qc]} \BibitemShut
  {NoStop}%
\bibitem [{\citenamefont {Jiang}\ \emph {et~al.}(2015)\citenamefont {Jiang},
  \citenamefont {Bambi},\ and\ \citenamefont {Steiner}}]{Jiang:2015dla}%
  \BibitemOpen
  \bibfield  {author} {\bibinfo {author} {\bibfnamefont {J.}~\bibnamefont
  {Jiang}}, \bibinfo {author} {\bibfnamefont {C.}~\bibnamefont {Bambi}}, \ and\
  \bibinfo {author} {\bibfnamefont {J.~F.}\ \bibnamefont {Steiner}},\
  }\bibfield  {title} {\enquote {\bibinfo {title} {{Testing the Kerr Nature of
  Black Hole Candidates using Iron Line Spectra in the CPR Framework}},}\
  }\href {\doibase 10.1088/0004-637X/811/2/130} {\bibfield  {journal} {\bibinfo
   {journal} {Astrophys. J.}\ }\textbf {\bibinfo {volume} {811}},\ \bibinfo
  {pages} {130} (\bibinfo {year} {2015})},\ \Eprint
  {http://arxiv.org/abs/1504.01970} {arXiv:1504.01970 [gr-qc]} \BibitemShut
  {NoStop}%
\bibitem [{\citenamefont {Jiang}\ \emph {et~al.}(2016)\citenamefont {Jiang},
  \citenamefont {Bambi},\ and\ \citenamefont {Steiner}}]{Jiang:2016bdj}%
  \BibitemOpen
  \bibfield  {author} {\bibinfo {author} {\bibfnamefont {J.}~\bibnamefont
  {Jiang}}, \bibinfo {author} {\bibfnamefont {C.}~\bibnamefont {Bambi}}, \ and\
  \bibinfo {author} {\bibfnamefont {J.~F.}\ \bibnamefont {Steiner}},\
  }\bibfield  {title} {\enquote {\bibinfo {title} {{Testing the Kerr nature of
  black hole candidates using iron line reverberation mapping in the
  Cardoso-Pani-Rico framework}},}\ }\href {\doibase 10.1103/PhysRevD.93.123008}
  {\bibfield  {journal} {\bibinfo  {journal} {Phys. Rev. D}\ }\textbf {\bibinfo
  {volume} {93}},\ \bibinfo {pages} {123008} (\bibinfo {year} {2016})},\
  \Eprint {http://arxiv.org/abs/1601.00838} {arXiv:1601.00838 [gr-qc]}
  \BibitemShut {NoStop}%
\bibitem [{\citenamefont {Xu}\ \emph {et~al.}(2018)\citenamefont {Xu},
  \citenamefont {Nampalliwar}, \citenamefont {Abdikamalov}, \citenamefont
  {Ayzenberg}, \citenamefont {Bambi}, \citenamefont {Dauser}, \citenamefont
  {Garcia},\ and\ \citenamefont {Jiang}}]{Xu:2018lom}%
  \BibitemOpen
  \bibfield  {author} {\bibinfo {author} {\bibfnamefont {Y.}~\bibnamefont
  {Xu}}, \bibinfo {author} {\bibfnamefont {S.}~\bibnamefont {Nampalliwar}},
  \bibinfo {author} {\bibfnamefont {A.~B.}\ \bibnamefont {Abdikamalov}},
  \bibinfo {author} {\bibfnamefont {D.}~\bibnamefont {Ayzenberg}}, \bibinfo
  {author} {\bibfnamefont {C.}~\bibnamefont {Bambi}}, \bibinfo {author}
  {\bibfnamefont {T.}~\bibnamefont {Dauser}}, \bibinfo {author} {\bibfnamefont
  {J.~A.}\ \bibnamefont {Garcia}}, \ and\ \bibinfo {author} {\bibfnamefont
  {J.}~\bibnamefont {Jiang}},\ }\bibfield  {title} {\enquote {\bibinfo {title}
  {{A Study of the Strong Gravity Region of the Black Hole in GS
  1354\textendash{}645}},}\ }\href {\doibase 10.3847/1538-4357/aadb9d}
  {\bibfield  {journal} {\bibinfo  {journal} {Astrophys. J.}\ }\textbf
  {\bibinfo {volume} {865}},\ \bibinfo {pages} {134} (\bibinfo {year}
  {2018})},\ \Eprint {http://arxiv.org/abs/1807.10243} {arXiv:1807.10243
  [gr-qc]} \BibitemShut {NoStop}%
\bibitem [{\citenamefont {Krishnendu}\ \emph {et~al.}(2017)\citenamefont
  {Krishnendu}, \citenamefont {Arun},\ and\ \citenamefont
  {Mishra}}]{Krishnendu:2017shb}%
  \BibitemOpen
  \bibfield  {author} {\bibinfo {author} {\bibfnamefont {N.~V.}\ \bibnamefont
  {Krishnendu}}, \bibinfo {author} {\bibfnamefont {K.~G.}\ \bibnamefont
  {Arun}}, \ and\ \bibinfo {author} {\bibfnamefont {C.~K.}\ \bibnamefont
  {Mishra}},\ }\bibfield  {title} {\enquote {\bibinfo {title} {{Testing the
  binary black hole nature of a compact binary coalescence}},}\ }\href
  {\doibase 10.1103/PhysRevLett.119.091101} {\bibfield  {journal} {\bibinfo
  {journal} {Phys. Rev. Lett.}\ }\textbf {\bibinfo {volume} {119}},\ \bibinfo
  {pages} {091101} (\bibinfo {year} {2017})},\ \Eprint
  {http://arxiv.org/abs/1701.06318} {arXiv:1701.06318 [gr-qc]} \BibitemShut
  {NoStop}%
\bibitem [{\citenamefont {Isi}\ \emph {et~al.}(2019)\citenamefont {Isi},
  \citenamefont {Giesler}, \citenamefont {Farr}, \citenamefont {Scheel},\ and\
  \citenamefont {Teukolsky}}]{Isi:2019aib}%
  \BibitemOpen
  \bibfield  {author} {\bibinfo {author} {\bibfnamefont {M.}~\bibnamefont
  {Isi}}, \bibinfo {author} {\bibfnamefont {M.}~\bibnamefont {Giesler}},
  \bibinfo {author} {\bibfnamefont {W.~M.}\ \bibnamefont {Farr}}, \bibinfo
  {author} {\bibfnamefont {M.~A.}\ \bibnamefont {Scheel}}, \ and\ \bibinfo
  {author} {\bibfnamefont {S.~A.}\ \bibnamefont {Teukolsky}},\ }\bibfield
  {title} {\enquote {\bibinfo {title} {{Testing the no-hair theorem with
  GW150914}},}\ }\href {\doibase 10.1103/PhysRevLett.123.111102} {\bibfield
  {journal} {\bibinfo  {journal} {Phys. Rev. Lett.}\ }\textbf {\bibinfo
  {volume} {123}},\ \bibinfo {pages} {111102} (\bibinfo {year} {2019})},\
  \Eprint {http://arxiv.org/abs/1905.00869} {arXiv:1905.00869 [gr-qc]}
  \BibitemShut {NoStop}%
\bibitem [{\citenamefont {Capano}\ \emph {et~al.}(2021)\citenamefont {Capano},
  \citenamefont {Cabero}, \citenamefont {Westerweck}, \citenamefont {Abedi},
  \citenamefont {Kastha}, \citenamefont {Nitz}, \citenamefont {Wang},
  \citenamefont {Nielsen},\ and\ \citenamefont {Krishnan}}]{Capano:2021etf}%
  \BibitemOpen
  \bibfield  {author} {\bibinfo {author} {\bibfnamefont {C.~D.}\ \bibnamefont
  {Capano}}, \bibinfo {author} {\bibfnamefont {M.}~\bibnamefont {Cabero}},
  \bibinfo {author} {\bibfnamefont {J.}~\bibnamefont {Westerweck}}, \bibinfo
  {author} {\bibfnamefont {J.}~\bibnamefont {Abedi}}, \bibinfo {author}
  {\bibfnamefont {S.}~\bibnamefont {Kastha}}, \bibinfo {author} {\bibfnamefont
  {A.~H.}\ \bibnamefont {Nitz}}, \bibinfo {author} {\bibfnamefont {Y.-F.}\
  \bibnamefont {Wang}}, \bibinfo {author} {\bibfnamefont {A.~B.}\ \bibnamefont
  {Nielsen}}, \ and\ \bibinfo {author} {\bibfnamefont {B.}~\bibnamefont
  {Krishnan}},\ }\bibfield  {title} {\enquote {\bibinfo {title} {{A multimode
  quasi-normal spectrum from a perturbed black hole}},}\ }\href@noop {} {\
  (\bibinfo {year} {2021})},\ \Eprint {http://arxiv.org/abs/2105.05238}
  {arXiv:2105.05238 [gr-qc]} \BibitemShut {NoStop}%
\bibitem [{\citenamefont {Cotesta}\ \emph {et~al.}(2022)\citenamefont
  {Cotesta}, \citenamefont {Carullo}, \citenamefont {Berti},\ and\
  \citenamefont {Cardoso}}]{Cotesta:2022pci}%
  \BibitemOpen
  \bibfield  {author} {\bibinfo {author} {\bibfnamefont {R.}~\bibnamefont
  {Cotesta}}, \bibinfo {author} {\bibfnamefont {G.}~\bibnamefont {Carullo}},
  \bibinfo {author} {\bibfnamefont {E.}~\bibnamefont {Berti}}, \ and\ \bibinfo
  {author} {\bibfnamefont {V.}~\bibnamefont {Cardoso}},\ }\bibfield  {title}
  {\enquote {\bibinfo {title} {{Analysis of Ringdown Overtones in GW150914}},}\
  }\href {\doibase 10.1103/PhysRevLett.129.111102} {\bibfield  {journal}
  {\bibinfo  {journal} {Phys. Rev. Lett.}\ }\textbf {\bibinfo {volume} {129}},\
  \bibinfo {pages} {111102} (\bibinfo {year} {2022})},\ \Eprint
  {http://arxiv.org/abs/2201.00822} {arXiv:2201.00822 [gr-qc]} \BibitemShut
  {NoStop}%
\bibitem [{\citenamefont {Finch}\ and\ \citenamefont
  {Moore}(2022)}]{Finch:2022ynt}%
  \BibitemOpen
  \bibfield  {author} {\bibinfo {author} {\bibfnamefont {E.}~\bibnamefont
  {Finch}}\ and\ \bibinfo {author} {\bibfnamefont {C.~J.}\ \bibnamefont
  {Moore}},\ }\bibfield  {title} {\enquote {\bibinfo {title} {{Searching for a
  ringdown overtone in GW150914}},}\ }\href {\doibase
  10.1103/PhysRevD.106.043005} {\bibfield  {journal} {\bibinfo  {journal}
  {Phys. Rev. D}\ }\textbf {\bibinfo {volume} {106}},\ \bibinfo {pages}
  {043005} (\bibinfo {year} {2022})},\ \Eprint
  {http://arxiv.org/abs/2205.07809} {arXiv:2205.07809 [gr-qc]} \BibitemShut
  {NoStop}%
\bibitem [{\citenamefont {Ma}\ \emph {et~al.}(2023)\citenamefont {Ma},
  \citenamefont {Sun},\ and\ \citenamefont {Chen}}]{Ma:2023vvr}%
  \BibitemOpen
  \bibfield  {author} {\bibinfo {author} {\bibfnamefont {S.}~\bibnamefont
  {Ma}}, \bibinfo {author} {\bibfnamefont {L.}~\bibnamefont {Sun}}, \ and\
  \bibinfo {author} {\bibfnamefont {Y.}~\bibnamefont {Chen}},\ }\bibfield
  {title} {\enquote {\bibinfo {title} {{Using rational filters to uncover the
  first ringdown overtone in GW150914}},}\ }\href {\doibase
  10.1103/PhysRevD.107.084010} {\bibfield  {journal} {\bibinfo  {journal}
  {Phys. Rev. D}\ }\textbf {\bibinfo {volume} {107}},\ \bibinfo {pages}
  {084010} (\bibinfo {year} {2023})},\ \Eprint
  {http://arxiv.org/abs/2301.06639} {arXiv:2301.06639 [gr-qc]} \BibitemShut
  {NoStop}%
\bibitem [{\citenamefont {Wang}\ \emph {et~al.}(2023)\citenamefont {Wang},
  \citenamefont {Capano}, \citenamefont {Abedi}, \citenamefont {Kastha},
  \citenamefont {Krishnan}, \citenamefont {Nielsen}, \citenamefont {Nitz},\
  and\ \citenamefont {Westerweck}}]{Wang:2023xsy}%
  \BibitemOpen
  \bibfield  {author} {\bibinfo {author} {\bibfnamefont {Y.-F.}\ \bibnamefont
  {Wang}}, \bibinfo {author} {\bibfnamefont {C.~D.}\ \bibnamefont {Capano}},
  \bibinfo {author} {\bibfnamefont {J.}~\bibnamefont {Abedi}}, \bibinfo
  {author} {\bibfnamefont {S.}~\bibnamefont {Kastha}}, \bibinfo {author}
  {\bibfnamefont {B.}~\bibnamefont {Krishnan}}, \bibinfo {author}
  {\bibfnamefont {A.~B.}\ \bibnamefont {Nielsen}}, \bibinfo {author}
  {\bibfnamefont {A.~H.}\ \bibnamefont {Nitz}}, \ and\ \bibinfo {author}
  {\bibfnamefont {J.}~\bibnamefont {Westerweck}},\ }\bibfield  {title}
  {\enquote {\bibinfo {title} {{A frequency-domain perspective on GW150914
  ringdown overtone}},}\ }\href@noop {} {\  (\bibinfo {year} {2023})},\ \Eprint
  {http://arxiv.org/abs/2310.19645} {arXiv:2310.19645 [gr-qc]} \BibitemShut
  {NoStop}%
\bibitem [{\citenamefont {Collins}\ and\ \citenamefont
  {Hughes}(2004)}]{Collins:2004ex}%
  \BibitemOpen
  \bibfield  {author} {\bibinfo {author} {\bibfnamefont {N.~A.}\ \bibnamefont
  {Collins}}\ and\ \bibinfo {author} {\bibfnamefont {S.~A.}\ \bibnamefont
  {Hughes}},\ }\bibfield  {title} {\enquote {\bibinfo {title} {{Towards a
  formalism for mapping the spacetimes of massive compact objects: Bumpy black
  holes and their orbits}},}\ }\href {\doibase 10.1103/PhysRevD.69.124022}
  {\bibfield  {journal} {\bibinfo  {journal} {Phys. Rev.}\ }\textbf {\bibinfo
  {volume} {D69}},\ \bibinfo {pages} {124022} (\bibinfo {year} {2004})},\
  \Eprint {http://arxiv.org/abs/gr-qc/0402063} {arXiv:gr-qc/0402063}
  \BibitemShut {NoStop}%
\bibitem [{\citenamefont {Vigeland}\ and\ \citenamefont
  {Hughes}(2010)}]{vigelandhughes}%
  \BibitemOpen
  \bibfield  {author} {\bibinfo {author} {\bibfnamefont {S.~J.}\ \bibnamefont
  {Vigeland}}\ and\ \bibinfo {author} {\bibfnamefont {S.~A.}\ \bibnamefont
  {Hughes}},\ }\bibfield  {title} {\enquote {\bibinfo {title} {{Spacetime and
  orbits of bumpy black holes}},}\ }\href {\doibase 10.1103/PhysRevD.81.024030}
  {\bibfield  {journal} {\bibinfo  {journal} {Phys. Rev.}\ }\textbf {\bibinfo
  {volume} {D81}},\ \bibinfo {pages} {024030} (\bibinfo {year} {2010})},\
  \Eprint {http://arxiv.org/abs/0911.1756} {arXiv:0911.1756 [gr-qc]}
  \BibitemShut {NoStop}%
\bibitem [{\citenamefont {Vigeland}(2010)}]{Vigeland:2010xe}%
  \BibitemOpen
  \bibfield  {author} {\bibinfo {author} {\bibfnamefont {S.~J.}\ \bibnamefont
  {Vigeland}},\ }\bibfield  {title} {\enquote {\bibinfo {title} {{Multipole
  moments of bumpy black holes}},}\ }\href {\doibase
  10.1103/PhysRevD.82.104041} {\bibfield  {journal} {\bibinfo  {journal}
  {Phys.Rev.}\ }\textbf {\bibinfo {volume} {D82}},\ \bibinfo {pages} {104041}
  (\bibinfo {year} {2010})},\ \Eprint {http://arxiv.org/abs/1008.1278}
  {arXiv:1008.1278 [gr-qc]} \BibitemShut {NoStop}%
\bibitem [{\citenamefont {Vigeland}\ \emph
  {et~al.}(2011{\natexlab{a}})\citenamefont {Vigeland}, \citenamefont {Yunes},\
  and\ \citenamefont {Stein}}]{vigelandnico}%
  \BibitemOpen
  \bibfield  {author} {\bibinfo {author} {\bibfnamefont {S.}~\bibnamefont
  {Vigeland}}, \bibinfo {author} {\bibfnamefont {N.}~\bibnamefont {Yunes}}, \
  and\ \bibinfo {author} {\bibfnamefont {L.}~\bibnamefont {Stein}},\ }\bibfield
   {title} {\enquote {\bibinfo {title} {{Bumpy Black Holes in Alternate
  Theories of Gravity}},}\ }\href {\doibase 10.1103/PhysRevD.83.104027}
  {\bibfield  {journal} {\bibinfo  {journal} {Phys. Rev.}\ }\textbf {\bibinfo
  {volume} {D83}},\ \bibinfo {pages} {104027} (\bibinfo {year}
  {2011}{\natexlab{a}})},\ \Eprint {http://arxiv.org/abs/1102.3706}
  {arXiv:1102.3706 [gr-qc]} \BibitemShut {NoStop}%
\bibitem [{\citenamefont {Glampedakis}\ and\ \citenamefont
  {Babak}(2006)}]{Glampedakis:2005cf}%
  \BibitemOpen
  \bibfield  {author} {\bibinfo {author} {\bibfnamefont {K.}~\bibnamefont
  {Glampedakis}}\ and\ \bibinfo {author} {\bibfnamefont {S.}~\bibnamefont
  {Babak}},\ }\bibfield  {title} {\enquote {\bibinfo {title} {{Mapping
  spacetimes with LISA: inspiral of a test-body in a `quasi-Kerr' field}},}\
  }\href {\doibase 10.1088/0264-9381/23/12/013} {\bibfield  {journal} {\bibinfo
   {journal} {Class. Quant. Grav.}\ }\textbf {\bibinfo {volume} {23}},\
  \bibinfo {pages} {4167--4188} (\bibinfo {year} {2006})},\ \Eprint
  {http://arxiv.org/abs/gr-qc/0510057} {arXiv:gr-qc/0510057} \BibitemShut
  {NoStop}%
\bibitem [{\citenamefont {Johannsen}\ and\ \citenamefont
  {Psaltis}(2011)}]{johannsen-metric}%
  \BibitemOpen
  \bibfield  {author} {\bibinfo {author} {\bibfnamefont {T.}~\bibnamefont
  {Johannsen}}\ and\ \bibinfo {author} {\bibfnamefont {D.}~\bibnamefont
  {Psaltis}},\ }\bibfield  {title} {\enquote {\bibinfo {title} {{A Metric for
  Rapidly Spinning Black Holes Suitable for Strong-Field Tests of the No-Hair
  Theorem}},}\ }\href {\doibase 10.1103/PhysRevD.83.124015} {\bibfield
  {journal} {\bibinfo  {journal} {Phys. Rev.}\ }\textbf {\bibinfo {volume}
  {D83}},\ \bibinfo {pages} {124015} (\bibinfo {year} {2011})},\ \Eprint
  {http://arxiv.org/abs/1105.3191} {arXiv:1105.3191 [gr-qc]} \BibitemShut
  {NoStop}%
\bibitem [{\citenamefont {Johannsen}(2013{\natexlab{a}})}]{Johannsen:2015pca}%
  \BibitemOpen
  \bibfield  {author} {\bibinfo {author} {\bibfnamefont {T.}~\bibnamefont
  {Johannsen}},\ }\bibfield  {title} {\enquote {\bibinfo {title} {{Regular
  Black Hole Metric with Three Constants of Motion}},}\ }\href {\doibase
  10.1103/PhysRevD.88.044002} {\bibfield  {journal} {\bibinfo  {journal} {Phys.
  Rev.}\ }\textbf {\bibinfo {volume} {D88}},\ \bibinfo {pages} {044002}
  (\bibinfo {year} {2013}{\natexlab{a}})},\ \Eprint
  {http://arxiv.org/abs/1501.02809} {arXiv:1501.02809 [gr-qc]} \BibitemShut
  {NoStop}%
\bibitem [{\citenamefont {Carson}\ and\ \citenamefont
  {Yagi}(2020{\natexlab{a}})}]{Carson:2020dez}%
  \BibitemOpen
  \bibfield  {author} {\bibinfo {author} {\bibfnamefont {Z.}~\bibnamefont
  {Carson}}\ and\ \bibinfo {author} {\bibfnamefont {K.}~\bibnamefont {Yagi}},\
  }\bibfield  {title} {\enquote {\bibinfo {title} {{Asymptotically flat,
  parameterized black hole metric preserving Kerr symmetries}},}\ }\href
  {\doibase 10.1103/PhysRevD.101.084030} {\bibfield  {journal} {\bibinfo
  {journal} {Phys. Rev.}\ }\textbf {\bibinfo {volume} {D101}},\ \bibinfo
  {pages} {084030} (\bibinfo {year} {2020}{\natexlab{a}})},\ \Eprint
  {http://arxiv.org/abs/2002.01028} {arXiv:2002.01028 [gr-qc]} \BibitemShut
  {NoStop}%
\bibitem [{\citenamefont {Papadopoulos}\ and\ \citenamefont
  {Kokkotas}(2018)}]{Papadopoulos:2018nvd}%
  \BibitemOpen
  \bibfield  {author} {\bibinfo {author} {\bibfnamefont {G.~O.}\ \bibnamefont
  {Papadopoulos}}\ and\ \bibinfo {author} {\bibfnamefont {K.~D.}\ \bibnamefont
  {Kokkotas}},\ }\bibfield  {title} {\enquote {\bibinfo {title} {{Preserving
  Kerr symmetries in deformed spacetimes}},}\ }\href {\doibase
  10.1088/1361-6382/aad7f4} {\bibfield  {journal} {\bibinfo  {journal} {Class.
  Quant. Grav.}\ }\textbf {\bibinfo {volume} {35}},\ \bibinfo {pages} {185014}
  (\bibinfo {year} {2018})},\ \Eprint {http://arxiv.org/abs/1807.08594}
  {arXiv:1807.08594 [gr-qc]} \BibitemShut {NoStop}%
\bibitem [{\citenamefont {Papadopoulos}\ and\ \citenamefont
  {Kokkotas}(2021)}]{Papadopoulos:2020kxu}%
  \BibitemOpen
  \bibfield  {author} {\bibinfo {author} {\bibfnamefont {G.~O.}\ \bibnamefont
  {Papadopoulos}}\ and\ \bibinfo {author} {\bibfnamefont {K.~D.}\ \bibnamefont
  {Kokkotas}},\ }\bibfield  {title} {\enquote {\bibinfo {title} {{On Kerr black
  hole deformations admitting a Carter constant and an invariant criterion for
  the separability of the wave equation}},}\ }\href {\doibase
  10.1007/s10714-021-02795-2} {\bibfield  {journal} {\bibinfo  {journal} {Gen.
  Rel. Grav.}\ }\textbf {\bibinfo {volume} {53}},\ \bibinfo {pages} {21}
  (\bibinfo {year} {2021})},\ \Eprint {http://arxiv.org/abs/2007.12125}
  {arXiv:2007.12125 [gr-qc]} \BibitemShut {NoStop}%
\bibitem [{\citenamefont {Chen}(2020)}]{Chen:2020aix}%
  \BibitemOpen
  \bibfield  {author} {\bibinfo {author} {\bibfnamefont {C.-Y.}\ \bibnamefont
  {Chen}},\ }\bibfield  {title} {\enquote {\bibinfo {title} {{Rotating black
  holes without $\mathbb{Z}_2$ symmetry and their shadow images}},}\ }\href
  {\doibase 10.1088/1475-7516/2020/05/040} {\bibfield  {journal} {\bibinfo
  {journal} {JCAP}\ }\textbf {\bibinfo {volume} {05}},\ \bibinfo {pages} {040}
  (\bibinfo {year} {2020})},\ \Eprint {http://arxiv.org/abs/2004.01440}
  {arXiv:2004.01440 [gr-qc]} \BibitemShut {NoStop}%
\bibitem [{\citenamefont {Rezzolla}\ and\ \citenamefont
  {Zhidenko}(2014)}]{Rezzolla:2014mua}%
  \BibitemOpen
  \bibfield  {author} {\bibinfo {author} {\bibfnamefont {L.}~\bibnamefont
  {Rezzolla}}\ and\ \bibinfo {author} {\bibfnamefont {A.}~\bibnamefont
  {Zhidenko}},\ }\bibfield  {title} {\enquote {\bibinfo {title} {{New
  parametrization for spherically symmetric black holes in metric theories of
  gravity}},}\ }\href {\doibase 10.1103/PhysRevD.90.084009} {\bibfield
  {journal} {\bibinfo  {journal} {Phys.Rev.}\ }\textbf {\bibinfo {volume}
  {D90}},\ \bibinfo {pages} {084009} (\bibinfo {year} {2014})},\ \Eprint
  {http://arxiv.org/abs/1407.3086} {arXiv:1407.3086 [gr-qc]} \BibitemShut
  {NoStop}%
\bibitem [{\citenamefont {Konoplya}\ \emph {et~al.}(2016)\citenamefont
  {Konoplya}, \citenamefont {Rezzolla},\ and\ \citenamefont
  {Zhidenko}}]{Konoplya:2016jvv}%
  \BibitemOpen
  \bibfield  {author} {\bibinfo {author} {\bibfnamefont {R.}~\bibnamefont
  {Konoplya}}, \bibinfo {author} {\bibfnamefont {L.}~\bibnamefont {Rezzolla}},
  \ and\ \bibinfo {author} {\bibfnamefont {A.}~\bibnamefont {Zhidenko}},\
  }\bibfield  {title} {\enquote {\bibinfo {title} {{General parametrization of
  axisymmetric black holes in metric theories of gravity}},}\ }\href {\doibase
  10.1103/PhysRevD.93.064015} {\bibfield  {journal} {\bibinfo  {journal} {Phys.
  Rev. D}\ }\textbf {\bibinfo {volume} {93}},\ \bibinfo {pages} {064015}
  (\bibinfo {year} {2016})},\ \Eprint {http://arxiv.org/abs/1602.02378}
  {arXiv:1602.02378 [gr-qc]} \BibitemShut {NoStop}%
\bibitem [{\citenamefont {Konoplya}\ \emph {et~al.}(2018)\citenamefont
  {Konoplya}, \citenamefont {Stuchl\'\i{}k},\ and\ \citenamefont
  {Zhidenko}}]{Konoplya:2018arm}%
  \BibitemOpen
  \bibfield  {author} {\bibinfo {author} {\bibfnamefont {R.~A.}\ \bibnamefont
  {Konoplya}}, \bibinfo {author} {\bibfnamefont {Z.}~\bibnamefont
  {Stuchl\'\i{}k}}, \ and\ \bibinfo {author} {\bibfnamefont {A.}~\bibnamefont
  {Zhidenko}},\ }\bibfield  {title} {\enquote {\bibinfo {title} {{Axisymmetric
  black holes allowing for separation of variables in the Klein-Gordon and
  Hamilton-Jacobi equations}},}\ }\href {\doibase 10.1103/PhysRevD.97.084044}
  {\bibfield  {journal} {\bibinfo  {journal} {Phys. Rev. D}\ }\textbf {\bibinfo
  {volume} {97}},\ \bibinfo {pages} {084044} (\bibinfo {year} {2018})},\
  \Eprint {http://arxiv.org/abs/1801.07195} {arXiv:1801.07195 [gr-qc]}
  \BibitemShut {NoStop}%
\bibitem [{\citenamefont {Konoplya}\ and\ \citenamefont
  {Zhidenko}(2020)}]{Konoplya:2020hyk}%
  \BibitemOpen
  \bibfield  {author} {\bibinfo {author} {\bibfnamefont {R.~A.}\ \bibnamefont
  {Konoplya}}\ and\ \bibinfo {author} {\bibfnamefont {A.}~\bibnamefont
  {Zhidenko}},\ }\bibfield  {title} {\enquote {\bibinfo {title} {{General
  parametrization of black holes: The only parameters that matter}},}\ }\href
  {\doibase 10.1103/PhysRevD.101.124004} {\bibfield  {journal} {\bibinfo
  {journal} {Phys. Rev. D}\ }\textbf {\bibinfo {volume} {101}},\ \bibinfo
  {pages} {124004} (\bibinfo {year} {2020})},\ \Eprint
  {http://arxiv.org/abs/2001.06100} {arXiv:2001.06100 [gr-qc]} \BibitemShut
  {NoStop}%
\bibitem [{\citenamefont {Junior}\ \emph {et~al.}(2020)\citenamefont {Junior},
  \citenamefont {Crispino}, \citenamefont {Cunha},\ and\ \citenamefont
  {Herdeiro}}]{Junior:2020lya}%
  \BibitemOpen
  \bibfield  {author} {\bibinfo {author} {\bibfnamefont {H.~C. D.~L.}\
  \bibnamefont {Junior}}, \bibinfo {author} {\bibfnamefont {L.~C.~B.}\
  \bibnamefont {Crispino}}, \bibinfo {author} {\bibfnamefont {P.~V.~P.}\
  \bibnamefont {Cunha}}, \ and\ \bibinfo {author} {\bibfnamefont {C.~A.~R.}\
  \bibnamefont {Herdeiro}},\ }\bibfield  {title} {\enquote {\bibinfo {title}
  {{Spinning black holes with a separable Hamilton\textendash{}Jacobi equation
  from a modified Newman\textendash{}Janis algorithm}},}\ }\href {\doibase
  10.1140/epjc/s10052-020-08572-w} {\bibfield  {journal} {\bibinfo  {journal}
  {Eur. Phys. J. C}\ }\textbf {\bibinfo {volume} {80}},\ \bibinfo {pages}
  {1036} (\bibinfo {year} {2020})},\ \Eprint {http://arxiv.org/abs/2011.07301}
  {arXiv:2011.07301 [gr-qc]} \BibitemShut {NoStop}%
\bibitem [{\citenamefont {Delaporte}\ \emph {et~al.}(2022)\citenamefont
  {Delaporte}, \citenamefont {Eichhorn},\ and\ \citenamefont
  {Held}}]{Delaporte:2022acp}%
  \BibitemOpen
  \bibfield  {author} {\bibinfo {author} {\bibfnamefont {H.}~\bibnamefont
  {Delaporte}}, \bibinfo {author} {\bibfnamefont {A.}~\bibnamefont {Eichhorn}},
  \ and\ \bibinfo {author} {\bibfnamefont {A.}~\bibnamefont {Held}},\
  }\bibfield  {title} {\enquote {\bibinfo {title} {{Parameterizations of
  black-hole spacetimes beyond circularity}},}\ }\href {\doibase
  10.1088/1361-6382/ac7027} {\bibfield  {journal} {\bibinfo  {journal} {Class.
  Quant. Grav.}\ }\textbf {\bibinfo {volume} {39}},\ \bibinfo {pages} {134002}
  (\bibinfo {year} {2022})},\ \Eprint {http://arxiv.org/abs/2203.00105}
  {arXiv:2203.00105 [gr-qc]} \BibitemShut {NoStop}%
\bibitem [{\citenamefont {Baines}\ and\ \citenamefont
  {Visser}(2023)}]{Baines:2023dhq}%
  \BibitemOpen
  \bibfield  {author} {\bibinfo {author} {\bibfnamefont {J.}~\bibnamefont
  {Baines}}\ and\ \bibinfo {author} {\bibfnamefont {M.}~\bibnamefont
  {Visser}},\ }\bibfield  {title} {\enquote {\bibinfo {title} {{Killing
  Horizons and Surface Gravities for a Well-Behaved Three-Function
  Generalization of the Kerr Spacetime}},}\ }\href {\doibase
  10.3390/universe9050223} {\bibfield  {journal} {\bibinfo  {journal}
  {Universe}\ }\textbf {\bibinfo {volume} {9}},\ \bibinfo {pages} {223}
  (\bibinfo {year} {2023})},\ \Eprint {http://arxiv.org/abs/2303.07380}
  {arXiv:2303.07380 [gr-qc]} \BibitemShut {NoStop}%
\bibitem [{\citenamefont {Aliev}\ and\ \citenamefont
  {Gumrukcuoglu}(2005)}]{Aliev:2005bi}%
  \BibitemOpen
  \bibfield  {author} {\bibinfo {author} {\bibfnamefont {A.~N.}\ \bibnamefont
  {Aliev}}\ and\ \bibinfo {author} {\bibfnamefont {A.~E.}\ \bibnamefont
  {Gumrukcuoglu}},\ }\bibfield  {title} {\enquote {\bibinfo {title} {{Charged
  rotating black holes on a 3-brane}},}\ }\href {\doibase
  10.1103/PhysRevD.71.104027} {\bibfield  {journal} {\bibinfo  {journal} {Phys.
  Rev.}\ }\textbf {\bibinfo {volume} {D71}},\ \bibinfo {pages} {104027}
  (\bibinfo {year} {2005})},\ \Eprint {http://arxiv.org/abs/hep-th/0502223}
  {arXiv:hep-th/0502223 [hep-th]} \BibitemShut {NoStop}%
\bibitem [{\citenamefont {Hayward}(2006)}]{Hayward:2005gi}%
  \BibitemOpen
  \bibfield  {author} {\bibinfo {author} {\bibfnamefont {S.~A.}\ \bibnamefont
  {Hayward}},\ }\bibfield  {title} {\enquote {\bibinfo {title} {{Formation and
  evaporation of regular black holes}},}\ }\href {\doibase
  10.1103/PhysRevLett.96.031103} {\bibfield  {journal} {\bibinfo  {journal}
  {Phys. Rev. Lett.}\ }\textbf {\bibinfo {volume} {96}},\ \bibinfo {pages}
  {031103} (\bibinfo {year} {2006})},\ \Eprint
  {http://arxiv.org/abs/gr-qc/0506126} {arXiv:gr-qc/0506126} \BibitemShut
  {NoStop}%
\bibitem [{\citenamefont {Bardeen}(1968)}]{Bardeen_proceedings}%
  \BibitemOpen
  \bibfield  {author} {\bibinfo {author} {\bibfnamefont {J.}~\bibnamefont
  {Bardeen}},\ }\bibfield  {title} {\enquote {\bibinfo {title} {Conference
  proceedings of gr5},}\ }in\ \href@noop {} {\emph {\bibinfo {booktitle}
  {Conference Proceedings of GR5}}}\ (\bibinfo  {publisher} {Tbilisi},\
  \bibinfo {year} {1968})\ p.\ \bibinfo {pages} {174}\BibitemShut {NoStop}%
\bibitem [{\citenamefont {Ghosh}(2015)}]{Ghosh:2014pba}%
  \BibitemOpen
  \bibfield  {author} {\bibinfo {author} {\bibfnamefont {S.~G.}\ \bibnamefont
  {Ghosh}},\ }\bibfield  {title} {\enquote {\bibinfo {title} {{A nonsingular
  rotating black hole}},}\ }\href {\doibase 10.1140/epjc/s10052-015-3740-y}
  {\bibfield  {journal} {\bibinfo  {journal} {Eur. Phys. J. C}\ }\textbf
  {\bibinfo {volume} {75}},\ \bibinfo {pages} {532} (\bibinfo {year} {2015})},\
  \Eprint {http://arxiv.org/abs/1408.5668} {arXiv:1408.5668 [gr-qc]}
  \BibitemShut {NoStop}%
\bibitem [{\citenamefont {Kumar}\ \emph {et~al.}(2020)\citenamefont {Kumar},
  \citenamefont {Ghosh},\ and\ \citenamefont {Wang}}]{Kumar:2020hgm}%
  \BibitemOpen
  \bibfield  {author} {\bibinfo {author} {\bibfnamefont {R.}~\bibnamefont
  {Kumar}}, \bibinfo {author} {\bibfnamefont {S.~G.}\ \bibnamefont {Ghosh}}, \
  and\ \bibinfo {author} {\bibfnamefont {A.}~\bibnamefont {Wang}},\ }\bibfield
  {title} {\enquote {\bibinfo {title} {{Gravitational deflection of light and
  shadow cast by rotating Kalb-Ramond black holes}},}\ }\href {\doibase
  10.1103/PhysRevD.101.104001} {\bibfield  {journal} {\bibinfo  {journal}
  {Phys. Rev. D}\ }\textbf {\bibinfo {volume} {101}},\ \bibinfo {pages}
  {104001} (\bibinfo {year} {2020})},\ \Eprint
  {http://arxiv.org/abs/2001.00460} {arXiv:2001.00460 [gr-qc]} \BibitemShut
  {NoStop}%
\bibitem [{\citenamefont {Sen}(1992)}]{Sen:1992ua}%
  \BibitemOpen
  \bibfield  {author} {\bibinfo {author} {\bibfnamefont {A.}~\bibnamefont
  {Sen}},\ }\bibfield  {title} {\enquote {\bibinfo {title} {{Rotating charged
  black hole solution in heterotic string theory}},}\ }\href {\doibase
  10.1103/PhysRevLett.69.1006} {\bibfield  {journal} {\bibinfo  {journal}
  {Phys. Rev. Lett.}\ }\textbf {\bibinfo {volume} {69}},\ \bibinfo {pages}
  {1006--1009} (\bibinfo {year} {1992})},\ \Eprint
  {http://arxiv.org/abs/hep-th/9204046} {arXiv:hep-th/9204046} \BibitemShut
  {NoStop}%
\bibitem [{\citenamefont {Randall}\ and\ \citenamefont
  {Sundrum}(1999{\natexlab{a}})}]{Randall:1999ee}%
  \BibitemOpen
  \bibfield  {author} {\bibinfo {author} {\bibfnamefont {L.}~\bibnamefont
  {Randall}}\ and\ \bibinfo {author} {\bibfnamefont {R.}~\bibnamefont
  {Sundrum}},\ }\bibfield  {title} {\enquote {\bibinfo {title} {{A Large mass
  hierarchy from a small extra dimension}},}\ }\href {\doibase
  10.1103/PhysRevLett.83.3370} {\bibfield  {journal} {\bibinfo  {journal}
  {Phys. Rev. Lett.}\ }\textbf {\bibinfo {volume} {83}},\ \bibinfo {pages}
  {3370--3373} (\bibinfo {year} {1999}{\natexlab{a}})},\ \Eprint
  {http://arxiv.org/abs/hep-ph/9905221} {arXiv:hep-ph/9905221 [hep-ph]}
  \BibitemShut {NoStop}%
\bibitem [{\citenamefont {Randall}\ and\ \citenamefont
  {Sundrum}(1999{\natexlab{b}})}]{Randall:1999vf}%
  \BibitemOpen
  \bibfield  {author} {\bibinfo {author} {\bibfnamefont {L.}~\bibnamefont
  {Randall}}\ and\ \bibinfo {author} {\bibfnamefont {R.}~\bibnamefont
  {Sundrum}},\ }\bibfield  {title} {\enquote {\bibinfo {title} {{An Alternative
  to compactification}},}\ }\href {\doibase 10.1103/PhysRevLett.83.4690}
  {\bibfield  {journal} {\bibinfo  {journal} {Phys.Rev.Lett.}\ }\textbf
  {\bibinfo {volume} {83}},\ \bibinfo {pages} {4690--4693} (\bibinfo {year}
  {1999}{\natexlab{b}})},\ \Eprint {http://arxiv.org/abs/hep-th/9906064}
  {arXiv:hep-th/9906064 [hep-th]} \BibitemShut {NoStop}%
\bibitem [{\citenamefont {Will}(2006)}]{will-living}%
  \BibitemOpen
  \bibfield  {author} {\bibinfo {author} {\bibfnamefont {C.~M.}\ \bibnamefont
  {Will}},\ }\bibfield  {title} {\enquote {\bibinfo {title} {The confrontation
  between general relativity and experiment},}\ }\href@noop {} {\bibfield
  {journal} {\bibinfo  {journal} {Living Reviews in Relativity}\ }\textbf
  {\bibinfo {volume} {9}},\ \bibinfo {pages} {3} (\bibinfo {year}
  {2006})}\BibitemShut {NoStop}%
\bibitem [{\citenamefont {Cardoso}\ \emph {et~al.}(2014)\citenamefont
  {Cardoso}, \citenamefont {Pani},\ and\ \citenamefont
  {Rico}}]{Cardoso:2014rha}%
  \BibitemOpen
  \bibfield  {author} {\bibinfo {author} {\bibfnamefont {V.}~\bibnamefont
  {Cardoso}}, \bibinfo {author} {\bibfnamefont {P.}~\bibnamefont {Pani}}, \
  and\ \bibinfo {author} {\bibfnamefont {J.}~\bibnamefont {Rico}},\ }\bibfield
  {title} {\enquote {\bibinfo {title} {{On generic parametrizations of spinning
  black-hole geometries}},}\ }\href {\doibase 10.1103/PhysRevD.89.064007}
  {\bibfield  {journal} {\bibinfo  {journal} {Phys.Rev.}\ }\textbf {\bibinfo
  {volume} {D89}},\ \bibinfo {pages} {064007} (\bibinfo {year} {2014})},\
  \Eprint {http://arxiv.org/abs/1401.0528} {arXiv:1401.0528 [gr-qc]}
  \BibitemShut {NoStop}%
\bibitem [{\citenamefont {Newman}\ and\ \citenamefont
  {Janis}(1965)}]{Newman:1965tw}%
  \BibitemOpen
  \bibfield  {author} {\bibinfo {author} {\bibfnamefont {E.~T.}\ \bibnamefont
  {Newman}}\ and\ \bibinfo {author} {\bibfnamefont {A.~I.}\ \bibnamefont
  {Janis}},\ }\bibfield  {title} {\enquote {\bibinfo {title} {{Note on the Kerr
  spinning particle metric}},}\ }\href {\doibase 10.1063/1.1704350} {\bibfield
  {journal} {\bibinfo  {journal} {J. Math. Phys.}\ }\textbf {\bibinfo {volume}
  {6}},\ \bibinfo {pages} {915--917} (\bibinfo {year} {1965})}\BibitemShut
  {NoStop}%
\bibitem [{\citenamefont {Drake}\ and\ \citenamefont
  {Szekeres}(2000)}]{Drake:1998gf}%
  \BibitemOpen
  \bibfield  {author} {\bibinfo {author} {\bibfnamefont {S.~P.}\ \bibnamefont
  {Drake}}\ and\ \bibinfo {author} {\bibfnamefont {P.}~\bibnamefont
  {Szekeres}},\ }\bibfield  {title} {\enquote {\bibinfo {title} {{Uniqueness of
  the Newman-Janis algorithm in generating the Kerr-Newman metric}},}\ }\href
  {\doibase 10.1023/A:1001920232180} {\bibfield  {journal} {\bibinfo  {journal}
  {Gen. Rel. Grav.}\ }\textbf {\bibinfo {volume} {32}},\ \bibinfo {pages}
  {445--458} (\bibinfo {year} {2000})},\ \Eprint
  {http://arxiv.org/abs/gr-qc/9807001} {arXiv:gr-qc/9807001} \BibitemShut
  {NoStop}%
\bibitem [{\citenamefont {Erbin}(2017)}]{Erbin:2016lzq}%
  \BibitemOpen
  \bibfield  {author} {\bibinfo {author} {\bibfnamefont {H.}~\bibnamefont
  {Erbin}},\ }\bibfield  {title} {\enquote {\bibinfo {title} {{Janis-Newman
  algorithm: generating rotating and NUT charged black holes}},}\ }\href
  {\doibase 10.3390/universe3010019} {\bibfield  {journal} {\bibinfo  {journal}
  {Universe}\ }\textbf {\bibinfo {volume} {3}},\ \bibinfo {pages} {19}
  (\bibinfo {year} {2017})},\ \Eprint {http://arxiv.org/abs/1701.00037}
  {arXiv:1701.00037 [gr-qc]} \BibitemShut {NoStop}%
\bibitem [{\citenamefont {Stephani}\ \emph {et~al.}(2003)\citenamefont
  {Stephani}, \citenamefont {Kramer}, \citenamefont {MacCallum}, \citenamefont
  {Hoenselaers},\ and\ \citenamefont {Herlt}}]{stephani}%
  \BibitemOpen
  \bibfield  {author} {\bibinfo {author} {\bibfnamefont {H.}~\bibnamefont
  {Stephani}}, \bibinfo {author} {\bibfnamefont {D.}~\bibnamefont {Kramer}},
  \bibinfo {author} {\bibfnamefont {M.}~\bibnamefont {MacCallum}}, \bibinfo
  {author} {\bibfnamefont {C.}~\bibnamefont {Hoenselaers}}, \ and\ \bibinfo
  {author} {\bibfnamefont {E.}~\bibnamefont {Herlt}},\ }\href@noop {} {\emph
  {\bibinfo {title} {Exact solutions of Einstein's field equations}}}\
  (\bibinfo  {publisher} {Cambridge Univ. Pr.},\ \bibinfo {address} {Cambridge,
  UK},\ \bibinfo {year} {2003})\BibitemShut {NoStop}%
\bibitem [{\citenamefont {Campanelli}\ \emph {et~al.}(2009)\citenamefont
  {Campanelli}, \citenamefont {Lousto},\ and\ \citenamefont
  {Zlochower}}]{campanelli}%
  \BibitemOpen
  \bibfield  {author} {\bibinfo {author} {\bibfnamefont {M.}~\bibnamefont
  {Campanelli}}, \bibinfo {author} {\bibfnamefont {C.~O.}\ \bibnamefont
  {Lousto}}, \ and\ \bibinfo {author} {\bibfnamefont {Y.}~\bibnamefont
  {Zlochower}},\ }\bibfield  {title} {\enquote {\bibinfo {title} {{Algebraic
  Classification of Numerical Spacetimes and Black-Hole-Binary Remnants}},}\
  }\href {\doibase 10.1103/PhysRevD.79.084012} {\bibfield  {journal} {\bibinfo
  {journal} {Phys. Rev.}\ }\textbf {\bibinfo {volume} {D79}},\ \bibinfo {pages}
  {084012} (\bibinfo {year} {2009})},\ \Eprint {http://arxiv.org/abs/0811.3006}
  {arXiv:0811.3006 [gr-qc]} \BibitemShut {NoStop}%
\bibitem [{\citenamefont {Kinnersley}(1969)}]{Kinnersley:1969zza}%
  \BibitemOpen
  \bibfield  {author} {\bibinfo {author} {\bibfnamefont {W.}~\bibnamefont
  {Kinnersley}},\ }\bibfield  {title} {\enquote {\bibinfo {title} {{Type D
  Vacuum Metrics}},}\ }\href {\doibase 10.1063/1.1664958} {\bibfield  {journal}
  {\bibinfo  {journal} {J. Math. Phys.}\ }\textbf {\bibinfo {volume} {10}},\
  \bibinfo {pages} {1195--1203} (\bibinfo {year} {1969})}\BibitemShut {NoStop}%
\bibitem [{\citenamefont {Guo}\ \emph {et~al.}(2023)\citenamefont {Guo},
  \citenamefont {Nakajima},\ and\ \citenamefont {Lin}}]{Guo:2023wtx}%
  \BibitemOpen
  \bibfield  {author} {\bibinfo {author} {\bibfnamefont {Y.}~\bibnamefont
  {Guo}}, \bibinfo {author} {\bibfnamefont {H.}~\bibnamefont {Nakajima}}, \
  and\ \bibinfo {author} {\bibfnamefont {W.}~\bibnamefont {Lin}},\ }\bibfield
  {title} {\enquote {\bibinfo {title} {{Teukolsky-like equations in a
  non-vacuum axisymmetric type D spacetime}},}\ }\href@noop {} {\  (\bibinfo
  {year} {2023})},\ \Eprint {http://arxiv.org/abs/2309.06237} {arXiv:2309.06237
  [gr-qc]} \BibitemShut {NoStop}%
\bibitem [{\citenamefont {Walker}\ and\ \citenamefont
  {Penrose}(1970)}]{walkerpenrose1970}%
  \BibitemOpen
  \bibfield  {author} {\bibinfo {author} {\bibfnamefont {M.}~\bibnamefont
  {Walker}}\ and\ \bibinfo {author} {\bibfnamefont {R.}~\bibnamefont
  {Penrose}},\ }\bibfield  {title} {\enquote {\bibinfo {title} {{On quadratic
  first integrals of the geodesic equations for type [22] spacetimes}},}\
  }\href {\doibase 10.1007/BF01649445} {\bibfield  {journal} {\bibinfo
  {journal} {Commun. Math. Phys.}\ }\textbf {\bibinfo {volume} {18}},\ \bibinfo
  {pages} {265--274} (\bibinfo {year} {1970})}\BibitemShut {NoStop}%
\bibitem [{\citenamefont {Silva}\ and\ \citenamefont
  {Glampedakis}(2020)}]{Silva:2019scu}%
  \BibitemOpen
  \bibfield  {author} {\bibinfo {author} {\bibfnamefont {H.~O.}\ \bibnamefont
  {Silva}}\ and\ \bibinfo {author} {\bibfnamefont {K.}~\bibnamefont
  {Glampedakis}},\ }\bibfield  {title} {\enquote {\bibinfo {title} {{Eikonal
  quasinormal modes of black holes beyond general relativity. II. Generalized
  scalar-tensor perturbations}},}\ }\href {\doibase
  10.1103/PhysRevD.101.044051} {\bibfield  {journal} {\bibinfo  {journal}
  {Phys. Rev. D}\ }\textbf {\bibinfo {volume} {101}},\ \bibinfo {pages}
  {044051} (\bibinfo {year} {2020})},\ \Eprint
  {http://arxiv.org/abs/1912.09286} {arXiv:1912.09286 [gr-qc]} \BibitemShut
  {NoStop}%
\bibitem [{\citenamefont {Glampedakis}\ \emph {et~al.}(2017)\citenamefont
  {Glampedakis}, \citenamefont {Pappas}, \citenamefont {Silva},\ and\
  \citenamefont {Berti}}]{Glampedakis:2017dvb}%
  \BibitemOpen
  \bibfield  {author} {\bibinfo {author} {\bibfnamefont {K.}~\bibnamefont
  {Glampedakis}}, \bibinfo {author} {\bibfnamefont {G.}~\bibnamefont {Pappas}},
  \bibinfo {author} {\bibfnamefont {H.~O.}\ \bibnamefont {Silva}}, \ and\
  \bibinfo {author} {\bibfnamefont {E.}~\bibnamefont {Berti}},\ }\bibfield
  {title} {\enquote {\bibinfo {title} {{Post-Kerr black hole spectroscopy}},}\
  }\href {\doibase 10.1103/PhysRevD.96.064054} {\bibfield  {journal} {\bibinfo
  {journal} {Phys. Rev. D}\ }\textbf {\bibinfo {volume} {96}},\ \bibinfo
  {pages} {064054} (\bibinfo {year} {2017})},\ \Eprint
  {http://arxiv.org/abs/1706.07658} {arXiv:1706.07658 [gr-qc]} \BibitemShut
  {NoStop}%
\bibitem [{\citenamefont {Glampedakis}\ and\ \citenamefont
  {Silva}(2019)}]{Glampedakis:2019dqh}%
  \BibitemOpen
  \bibfield  {author} {\bibinfo {author} {\bibfnamefont {K.}~\bibnamefont
  {Glampedakis}}\ and\ \bibinfo {author} {\bibfnamefont {H.~O.}\ \bibnamefont
  {Silva}},\ }\bibfield  {title} {\enquote {\bibinfo {title} {{Eikonal
  quasinormal modes of black holes beyond General Relativity}},}\ }\href
  {\doibase 10.1103/PhysRevD.100.044040} {\bibfield  {journal} {\bibinfo
  {journal} {Phys. Rev. D}\ }\textbf {\bibinfo {volume} {100}},\ \bibinfo
  {pages} {044040} (\bibinfo {year} {2019})},\ \Eprint
  {http://arxiv.org/abs/1906.05455} {arXiv:1906.05455 [gr-qc]} \BibitemShut
  {NoStop}%
\bibitem [{\citenamefont {Mishra}\ \emph {et~al.}(2022)\citenamefont {Mishra},
  \citenamefont {Ghosh},\ and\ \citenamefont {Chakraborty}}]{Mishra:2021waw}%
  \BibitemOpen
  \bibfield  {author} {\bibinfo {author} {\bibfnamefont {A.~K.}\ \bibnamefont
  {Mishra}}, \bibinfo {author} {\bibfnamefont {A.}~\bibnamefont {Ghosh}}, \
  and\ \bibinfo {author} {\bibfnamefont {S.}~\bibnamefont {Chakraborty}},\
  }\bibfield  {title} {\enquote {\bibinfo {title} {{Constraining extra
  dimensions using observations of black hole quasi-normal modes}},}\ }\href
  {\doibase 10.1140/epjc/s10052-022-10788-x} {\bibfield  {journal} {\bibinfo
  {journal} {Eur. Phys. J. C}\ }\textbf {\bibinfo {volume} {82}},\ \bibinfo
  {pages} {820} (\bibinfo {year} {2022})},\ \Eprint
  {http://arxiv.org/abs/2106.05558} {arXiv:2106.05558 [gr-qc]} \BibitemShut
  {NoStop}%
\bibitem [{\citenamefont {Bryant}\ \emph {et~al.}(2021)\citenamefont {Bryant},
  \citenamefont {Silva}, \citenamefont {Yagi},\ and\ \citenamefont
  {Glampedakis}}]{Bryant:2021xdh}%
  \BibitemOpen
  \bibfield  {author} {\bibinfo {author} {\bibfnamefont {A.}~\bibnamefont
  {Bryant}}, \bibinfo {author} {\bibfnamefont {H.~O.}\ \bibnamefont {Silva}},
  \bibinfo {author} {\bibfnamefont {K.}~\bibnamefont {Yagi}}, \ and\ \bibinfo
  {author} {\bibfnamefont {K.}~\bibnamefont {Glampedakis}},\ }\bibfield
  {title} {\enquote {\bibinfo {title} {{Eikonal quasinormal modes of black
  holes beyond general relativity. III. Scalar Gauss-Bonnet gravity}},}\ }\href
  {\doibase 10.1103/PhysRevD.104.044051} {\bibfield  {journal} {\bibinfo
  {journal} {Phys. Rev. D}\ }\textbf {\bibinfo {volume} {104}},\ \bibinfo
  {pages} {044051} (\bibinfo {year} {2021})},\ \Eprint
  {http://arxiv.org/abs/2106.09657} {arXiv:2106.09657 [gr-qc]} \BibitemShut
  {NoStop}%
\bibitem [{\citenamefont {Johannsen}(2013{\natexlab{b}})}]{Johannsen:2013vgc}%
  \BibitemOpen
  \bibfield  {author} {\bibinfo {author} {\bibfnamefont {T.}~\bibnamefont
  {Johannsen}},\ }\bibfield  {title} {\enquote {\bibinfo {title} {{Photon Rings
  around Kerr and Kerr-like Black Holes}},}\ }\href {\doibase
  10.1088/0004-637X/777/2/170} {\bibfield  {journal} {\bibinfo  {journal}
  {Astrophys. J.}\ }\textbf {\bibinfo {volume} {777}},\ \bibinfo {pages} {170}
  (\bibinfo {year} {2013}{\natexlab{b}})},\ \Eprint
  {http://arxiv.org/abs/1501.02814} {arXiv:1501.02814 [astro-ph.HE]}
  \BibitemShut {NoStop}%
\bibitem [{\citenamefont {Wei}\ and\ \citenamefont {Liu}(2013)}]{Wei:2013kza}%
  \BibitemOpen
  \bibfield  {author} {\bibinfo {author} {\bibfnamefont {S.-W.}\ \bibnamefont
  {Wei}}\ and\ \bibinfo {author} {\bibfnamefont {Y.-X.}\ \bibnamefont {Liu}},\
  }\bibfield  {title} {\enquote {\bibinfo {title} {{Observing the shadow of
  Einstein-Maxwell-Dilaton-Axion black hole}},}\ }\href {\doibase
  10.1088/1475-7516/2013/11/063} {\bibfield  {journal} {\bibinfo  {journal}
  {JCAP}\ }\textbf {\bibinfo {volume} {11}},\ \bibinfo {pages} {063} (\bibinfo
  {year} {2013})},\ \Eprint {http://arxiv.org/abs/1311.4251} {arXiv:1311.4251
  [gr-qc]} \BibitemShut {NoStop}%
\bibitem [{\citenamefont {Atamurotov}\ \emph {et~al.}(2013)\citenamefont
  {Atamurotov}, \citenamefont {Abdujabbarov},\ and\ \citenamefont
  {Ahmedov}}]{Atamurotov:2013sca}%
  \BibitemOpen
  \bibfield  {author} {\bibinfo {author} {\bibfnamefont {F.}~\bibnamefont
  {Atamurotov}}, \bibinfo {author} {\bibfnamefont {A.}~\bibnamefont
  {Abdujabbarov}}, \ and\ \bibinfo {author} {\bibfnamefont {B.}~\bibnamefont
  {Ahmedov}},\ }\bibfield  {title} {\enquote {\bibinfo {title} {{Shadow of
  rotating non-Kerr black hole}},}\ }\href {\doibase
  10.1103/PhysRevD.88.064004} {\bibfield  {journal} {\bibinfo  {journal} {Phys.
  Rev. D}\ }\textbf {\bibinfo {volume} {88}},\ \bibinfo {pages} {064004}
  (\bibinfo {year} {2013})}\BibitemShut {NoStop}%
\bibitem [{\citenamefont {Ghasemi-Nodehi}\ \emph {et~al.}(2015)\citenamefont
  {Ghasemi-Nodehi}, \citenamefont {Li},\ and\ \citenamefont
  {Bambi}}]{Ghasemi-Nodehi:2015raa}%
  \BibitemOpen
  \bibfield  {author} {\bibinfo {author} {\bibfnamefont {M.}~\bibnamefont
  {Ghasemi-Nodehi}}, \bibinfo {author} {\bibfnamefont {Z.}~\bibnamefont {Li}},
  \ and\ \bibinfo {author} {\bibfnamefont {C.}~\bibnamefont {Bambi}},\
  }\bibfield  {title} {\enquote {\bibinfo {title} {{Shadows of CPR black holes
  and tests of the Kerr metric}},}\ }\href {\doibase
  10.1140/epjc/s10052-015-3539-x} {\bibfield  {journal} {\bibinfo  {journal}
  {Eur. Phys. J. C}\ }\textbf {\bibinfo {volume} {75}},\ \bibinfo {pages} {315}
  (\bibinfo {year} {2015})},\ \Eprint {http://arxiv.org/abs/1506.02627}
  {arXiv:1506.02627 [gr-qc]} \BibitemShut {NoStop}%
\bibitem [{\citenamefont {Abdujabbarov}\ \emph {et~al.}(2016)\citenamefont
  {Abdujabbarov}, \citenamefont {Amir}, \citenamefont {Ahmedov},\ and\
  \citenamefont {Ghosh}}]{Abdujabbarov:2016hnw}%
  \BibitemOpen
  \bibfield  {author} {\bibinfo {author} {\bibfnamefont {A.}~\bibnamefont
  {Abdujabbarov}}, \bibinfo {author} {\bibfnamefont {M.}~\bibnamefont {Amir}},
  \bibinfo {author} {\bibfnamefont {B.}~\bibnamefont {Ahmedov}}, \ and\
  \bibinfo {author} {\bibfnamefont {S.~G.}\ \bibnamefont {Ghosh}},\ }\bibfield
  {title} {\enquote {\bibinfo {title} {{Shadow of rotating regular black
  holes}},}\ }\href {\doibase 10.1103/PhysRevD.93.104004} {\bibfield  {journal}
  {\bibinfo  {journal} {Phys. Rev. D}\ }\textbf {\bibinfo {volume} {93}},\
  \bibinfo {pages} {104004} (\bibinfo {year} {2016})},\ \Eprint
  {http://arxiv.org/abs/1604.03809} {arXiv:1604.03809 [gr-qc]} \BibitemShut
  {NoStop}%
\bibitem [{\citenamefont {Younsi}\ \emph {et~al.}(2016)\citenamefont {Younsi},
  \citenamefont {Zhidenko}, \citenamefont {Rezzolla}, \citenamefont
  {Konoplya},\ and\ \citenamefont {Mizuno}}]{Younsi:2016azx}%
  \BibitemOpen
  \bibfield  {author} {\bibinfo {author} {\bibfnamefont {Z.}~\bibnamefont
  {Younsi}}, \bibinfo {author} {\bibfnamefont {A.}~\bibnamefont {Zhidenko}},
  \bibinfo {author} {\bibfnamefont {L.}~\bibnamefont {Rezzolla}}, \bibinfo
  {author} {\bibfnamefont {R.}~\bibnamefont {Konoplya}}, \ and\ \bibinfo
  {author} {\bibfnamefont {Y.}~\bibnamefont {Mizuno}},\ }\bibfield  {title}
  {\enquote {\bibinfo {title} {{New method for shadow calculations: Application
  to parametrized axisymmetric black holes}},}\ }\href {\doibase
  10.1103/PhysRevD.94.084025} {\bibfield  {journal} {\bibinfo  {journal} {Phys.
  Rev. D}\ }\textbf {\bibinfo {volume} {94}},\ \bibinfo {pages} {084025}
  (\bibinfo {year} {2016})},\ \Eprint {http://arxiv.org/abs/1607.05767}
  {arXiv:1607.05767 [gr-qc]} \BibitemShut {NoStop}%
\bibitem [{\citenamefont {Mizuno}\ \emph {et~al.}(2018)\citenamefont {Mizuno},
  \citenamefont {Younsi}, \citenamefont {Fromm}, \citenamefont {Porth},
  \citenamefont {De~Laurentis}, \citenamefont {Olivares}, \citenamefont
  {Falcke}, \citenamefont {Kramer},\ and\ \citenamefont
  {Rezzolla}}]{Mizuno:2018lxz}%
  \BibitemOpen
  \bibfield  {author} {\bibinfo {author} {\bibfnamefont {Y.}~\bibnamefont
  {Mizuno}}, \bibinfo {author} {\bibfnamefont {Z.}~\bibnamefont {Younsi}},
  \bibinfo {author} {\bibfnamefont {C.~M.}\ \bibnamefont {Fromm}}, \bibinfo
  {author} {\bibfnamefont {O.}~\bibnamefont {Porth}}, \bibinfo {author}
  {\bibfnamefont {M.}~\bibnamefont {De~Laurentis}}, \bibinfo {author}
  {\bibfnamefont {H.}~\bibnamefont {Olivares}}, \bibinfo {author}
  {\bibfnamefont {H.}~\bibnamefont {Falcke}}, \bibinfo {author} {\bibfnamefont
  {M.}~\bibnamefont {Kramer}}, \ and\ \bibinfo {author} {\bibfnamefont
  {L.}~\bibnamefont {Rezzolla}},\ }\bibfield  {title} {\enquote {\bibinfo
  {title} {{The Current Ability to Test Theories of Gravity with Black Hole
  Shadows}},}\ }\href {\doibase 10.1038/s41550-018-0449-5} {\bibfield
  {journal} {\bibinfo  {journal} {Nature Astron.}\ }\textbf {\bibinfo {volume}
  {2}},\ \bibinfo {pages} {585--590} (\bibinfo {year} {2018})},\ \Eprint
  {http://arxiv.org/abs/1804.05812} {arXiv:1804.05812 [astro-ph.GA]}
  \BibitemShut {NoStop}%
\bibitem [{\citenamefont {Narang}\ \emph {et~al.}(2020)\citenamefont {Narang},
  \citenamefont {Mohanty},\ and\ \citenamefont {Kumar}}]{Narang:2020bgo}%
  \BibitemOpen
  \bibfield  {author} {\bibinfo {author} {\bibfnamefont {A.}~\bibnamefont
  {Narang}}, \bibinfo {author} {\bibfnamefont {S.}~\bibnamefont {Mohanty}}, \
  and\ \bibinfo {author} {\bibfnamefont {A.}~\bibnamefont {Kumar}},\ }\bibfield
   {title} {\enquote {\bibinfo {title} {{Test of Kerr-Sen metric with black
  hole observations}},}\ }\href@noop {} {\  (\bibinfo {year} {2020})},\ \Eprint
  {http://arxiv.org/abs/2002.12786} {arXiv:2002.12786 [gr-qc]} \BibitemShut
  {NoStop}%
\bibitem [{\citenamefont {Konoplya}\ and\ \citenamefont
  {Zhidenko}(2021)}]{Konoplya:2021slg}%
  \BibitemOpen
  \bibfield  {author} {\bibinfo {author} {\bibfnamefont {R.~A.}\ \bibnamefont
  {Konoplya}}\ and\ \bibinfo {author} {\bibfnamefont {A.}~\bibnamefont
  {Zhidenko}},\ }\bibfield  {title} {\enquote {\bibinfo {title} {{Shadows of
  parametrized axially symmetric black holes allowing for separation of
  variables}},}\ }\href {\doibase 10.1103/PhysRevD.103.104033} {\bibfield
  {journal} {\bibinfo  {journal} {Phys. Rev. D}\ }\textbf {\bibinfo {volume}
  {103}},\ \bibinfo {pages} {104033} (\bibinfo {year} {2021})},\ \Eprint
  {http://arxiv.org/abs/2103.03855} {arXiv:2103.03855 [gr-qc]} \BibitemShut
  {NoStop}%
\bibitem [{\citenamefont {Zubair}\ \emph {et~al.}(2023)\citenamefont {Zubair},
  \citenamefont {Raza},\ and\ \citenamefont {Maqsood}}]{Zubair:2023cor}%
  \BibitemOpen
  \bibfield  {author} {\bibinfo {author} {\bibfnamefont {M.}~\bibnamefont
  {Zubair}}, \bibinfo {author} {\bibfnamefont {M.~A.}\ \bibnamefont {Raza}}, \
  and\ \bibinfo {author} {\bibfnamefont {E.}~\bibnamefont {Maqsood}},\
  }\bibfield  {title} {\enquote {\bibinfo {title} {{Rotating black hole in
  Kalb\textendash{}Ramond gravity: Constraining parameters by comparison with
  EHT observations of Sgr A* and M87*}},}\ }\href {\doibase
  10.1016/j.dark.2023.101334} {\bibfield  {journal} {\bibinfo  {journal} {Phys.
  Dark Univ.}\ }\textbf {\bibinfo {volume} {42}},\ \bibinfo {pages} {101334}
  (\bibinfo {year} {2023})},\ \Eprint {http://arxiv.org/abs/2310.12325}
  {arXiv:2310.12325 [gr-qc]} \BibitemShut {NoStop}%
\bibitem [{\citenamefont {Tsukamoto}\ \emph {et~al.}(2014)\citenamefont
  {Tsukamoto}, \citenamefont {Li},\ and\ \citenamefont
  {Bambi}}]{Tsukamoto:2014tja}%
  \BibitemOpen
  \bibfield  {author} {\bibinfo {author} {\bibfnamefont {N.}~\bibnamefont
  {Tsukamoto}}, \bibinfo {author} {\bibfnamefont {Z.}~\bibnamefont {Li}}, \
  and\ \bibinfo {author} {\bibfnamefont {C.}~\bibnamefont {Bambi}},\ }\bibfield
   {title} {\enquote {\bibinfo {title} {{Constraining the spin and the
  deformation parameters from the black hole shadow}},}\ }\href {\doibase
  10.1088/1475-7516/2014/06/043} {\bibfield  {journal} {\bibinfo  {journal}
  {JCAP}\ }\textbf {\bibinfo {volume} {06}},\ \bibinfo {pages} {043} (\bibinfo
  {year} {2014})},\ \Eprint {http://arxiv.org/abs/1403.0371} {arXiv:1403.0371
  [gr-qc]} \BibitemShut {NoStop}%
\bibitem [{\citenamefont {Tsukamoto}(2018)}]{Tsukamoto:2017fxq}%
  \BibitemOpen
  \bibfield  {author} {\bibinfo {author} {\bibfnamefont {N.}~\bibnamefont
  {Tsukamoto}},\ }\bibfield  {title} {\enquote {\bibinfo {title} {{Black hole
  shadow in an asymptotically-flat, stationary, and axisymmetric spacetime: The
  Kerr-Newman and rotating regular black holes}},}\ }\href {\doibase
  10.1103/PhysRevD.97.064021} {\bibfield  {journal} {\bibinfo  {journal} {Phys.
  Rev. D}\ }\textbf {\bibinfo {volume} {97}},\ \bibinfo {pages} {064021}
  (\bibinfo {year} {2018})},\ \Eprint {http://arxiv.org/abs/1708.07427}
  {arXiv:1708.07427 [gr-qc]} \BibitemShut {NoStop}%
\bibitem [{\citenamefont {Shaikh}(2019)}]{Shaikh:2019fpu}%
  \BibitemOpen
  \bibfield  {author} {\bibinfo {author} {\bibfnamefont {R.}~\bibnamefont
  {Shaikh}},\ }\bibfield  {title} {\enquote {\bibinfo {title} {{Black hole
  shadow in a general rotating spacetime obtained through Newman-Janis
  algorithm}},}\ }\href {\doibase 10.1103/PhysRevD.100.024028} {\bibfield
  {journal} {\bibinfo  {journal} {Phys. Rev. D}\ }\textbf {\bibinfo {volume}
  {100}},\ \bibinfo {pages} {024028} (\bibinfo {year} {2019})},\ \Eprint
  {http://arxiv.org/abs/1904.08322} {arXiv:1904.08322 [gr-qc]} \BibitemShut
  {NoStop}%
\bibitem [{\citenamefont {Bardeen}\ and\ \citenamefont
  {DeWitt}(1973)}]{Bardeen_1973}%
  \BibitemOpen
  \bibfield  {author} {\bibinfo {author} {\bibfnamefont {J.}~\bibnamefont
  {Bardeen}}\ and\ \bibinfo {author} {\bibfnamefont {B.~S.}\ \bibnamefont
  {DeWitt}},\ }\href@noop {} {\emph {\bibinfo {title} {Black holes}}}\
  (\bibinfo  {publisher} {Gordon and Breach},\ \bibinfo {year}
  {1973})\BibitemShut {NoStop}%
\bibitem [{\citenamefont {Cardenas-Avendano}\ \emph {et~al.}(2020)\citenamefont
  {Cardenas-Avendano}, \citenamefont {Nampalliwar},\ and\ \citenamefont
  {Yunes}}]{Cardenas-Avendano:2019zxd}%
  \BibitemOpen
  \bibfield  {author} {\bibinfo {author} {\bibfnamefont {A.}~\bibnamefont
  {Cardenas-Avendano}}, \bibinfo {author} {\bibfnamefont {S.}~\bibnamefont
  {Nampalliwar}}, \ and\ \bibinfo {author} {\bibfnamefont {N.}~\bibnamefont
  {Yunes}},\ }\bibfield  {title} {\enquote {\bibinfo {title}
  {{Gravitational-wave versus X-ray tests of strong-field gravity}},}\ }\href
  {\doibase 10.1088/1361-6382/ab8f64} {\bibfield  {journal} {\bibinfo
  {journal} {Class. Quant. Grav.}\ }\textbf {\bibinfo {volume} {37}},\ \bibinfo
  {pages} {135008} (\bibinfo {year} {2020})},\ \Eprint
  {http://arxiv.org/abs/1912.08062} {arXiv:1912.08062 [gr-qc]} \BibitemShut
  {NoStop}%
\bibitem [{\citenamefont {Carson}\ and\ \citenamefont
  {Yagi}(2020{\natexlab{b}})}]{Carson:2020iik}%
  \BibitemOpen
  \bibfield  {author} {\bibinfo {author} {\bibfnamefont {Z.}~\bibnamefont
  {Carson}}\ and\ \bibinfo {author} {\bibfnamefont {K.}~\bibnamefont {Yagi}},\
  }\bibfield  {title} {\enquote {\bibinfo {title} {{Probing beyond-Kerr
  spacetimes with inspiral-ringdown corrections to gravitational waves}},}\
  }\href@noop {} {\  (\bibinfo {year} {2020}{\natexlab{b}})},\ \Eprint
  {http://arxiv.org/abs/2003.02374 (submitted to Phys. Rev. D)}
  {arXiv:2003.02374 (submitted to Phys. Rev. D) [gr-qc]} \BibitemShut {NoStop}%
\bibitem [{\citenamefont {Shashank}\ and\ \citenamefont
  {Bambi}(2022)}]{Shashank:2021giy}%
  \BibitemOpen
  \bibfield  {author} {\bibinfo {author} {\bibfnamefont {S.}~\bibnamefont
  {Shashank}}\ and\ \bibinfo {author} {\bibfnamefont {C.}~\bibnamefont
  {Bambi}},\ }\bibfield  {title} {\enquote {\bibinfo {title} {{Constraining the
  Konoplya-Rezzolla-Zhidenko deformation parameters III: Limits from
  stellar-mass black holes using gravitational-wave observations}},}\ }\href
  {\doibase 10.1103/PhysRevD.105.104004} {\bibfield  {journal} {\bibinfo
  {journal} {Phys. Rev. D}\ }\textbf {\bibinfo {volume} {105}},\ \bibinfo
  {pages} {104004} (\bibinfo {year} {2022})},\ \Eprint
  {http://arxiv.org/abs/2112.05388} {arXiv:2112.05388 [gr-qc]} \BibitemShut
  {NoStop}%
\bibitem [{\citenamefont {Kumar}\ \emph {et~al.}(2023)\citenamefont {Kumar},
  \citenamefont {Chowdhuri},\ and\ \citenamefont
  {Bhattacharyya}}]{Kumar:2023bdf}%
  \BibitemOpen
  \bibfield  {author} {\bibinfo {author} {\bibfnamefont {S.}~\bibnamefont
  {Kumar}}, \bibinfo {author} {\bibfnamefont {A.}~\bibnamefont {Chowdhuri}}, \
  and\ \bibinfo {author} {\bibfnamefont {A.}~\bibnamefont {Bhattacharyya}},\
  }\bibfield  {title} {\enquote {\bibinfo {title} {{Prospects of detecting
  deviations to Kerr geometry with radiation reaction effects in EMRIs}},}\
  }\href@noop {} {\  (\bibinfo {year} {2023})},\ \Eprint
  {http://arxiv.org/abs/2311.05983} {arXiv:2311.05983 [gr-qc]} \BibitemShut
  {NoStop}%
\bibitem [{\citenamefont {Contreras}\ \emph {et~al.}(2021)\citenamefont
  {Contreras}, \citenamefont {Ovalle},\ and\ \citenamefont
  {Casadio}}]{Contreras:2021yxe}%
  \BibitemOpen
  \bibfield  {author} {\bibinfo {author} {\bibfnamefont {E.}~\bibnamefont
  {Contreras}}, \bibinfo {author} {\bibfnamefont {J.}~\bibnamefont {Ovalle}}, \
  and\ \bibinfo {author} {\bibfnamefont {R.}~\bibnamefont {Casadio}},\
  }\bibfield  {title} {\enquote {\bibinfo {title} {{Gravitational decoupling
  for axially symmetric systems and rotating black holes}},}\ }\href {\doibase
  10.1103/PhysRevD.103.044020} {\bibfield  {journal} {\bibinfo  {journal}
  {Phys. Rev. D}\ }\textbf {\bibinfo {volume} {103}},\ \bibinfo {pages}
  {044020} (\bibinfo {year} {2021})},\ \Eprint
  {http://arxiv.org/abs/2101.08569} {arXiv:2101.08569 [gr-qc]} \BibitemShut
  {NoStop}%
\bibitem [{\citenamefont {Vigeland}\ \emph
  {et~al.}(2011{\natexlab{b}})\citenamefont {Vigeland}, \citenamefont {Yunes},\
  and\ \citenamefont {Stein}}]{Vigeland:2011ji}%
  \BibitemOpen
  \bibfield  {author} {\bibinfo {author} {\bibfnamefont {S.}~\bibnamefont
  {Vigeland}}, \bibinfo {author} {\bibfnamefont {N.}~\bibnamefont {Yunes}}, \
  and\ \bibinfo {author} {\bibfnamefont {L.}~\bibnamefont {Stein}},\ }\bibfield
   {title} {\enquote {\bibinfo {title} {{Bumpy Black Holes in Alternate
  Theories of Gravity}},}\ }\href {\doibase 10.1103/PhysRevD.83.104027}
  {\bibfield  {journal} {\bibinfo  {journal} {Phys.Rev.}\ }\textbf {\bibinfo
  {volume} {D83}},\ \bibinfo {pages} {104027} (\bibinfo {year}
  {2011}{\natexlab{b}})},\ \Eprint {http://arxiv.org/abs/1102.3706}
  {arXiv:1102.3706 [gr-qc]} \BibitemShut {NoStop}%
\end{thebibliography}%

\end{document}